\documentclass[10pt, twocolumn,article]{IEEEtran}
\usepackage{setspace}
\usepackage{cite}  
\usepackage[latin1]{inputenc}
\usepackage{epsf}\epsfverbosetrue
\usepackage{graphics,epsfig,color}
\usepackage{graphicx,epsfig}
\usepackage{epsfig}
\usepackage{multirow}
\usepackage{slashbox}
\usepackage{alltt}
\usepackage{subfigure}
\usepackage{color}
\usepackage{float}
\usepackage{url}
\usepackage{graphicx}
\usepackage{verbatim} 
\usepackage{footnote} 
\usepackage{latexsym} 
\usepackage{amsmath}
\usepackage{sidecap} 
\usepackage{wrapfig}
\usepackage{eso-pic}
\usepackage{fix-cm}

\usepackage{layout}
\usepackage{amsmath}
\usepackage{textcomp}
\usepackage{multirow}
\usepackage[table,usenames,dvipsnames]{xcolor} 

\usepackage{soul}
\usepackage{amsmath}
\usepackage{verbatim}

\usepackage[normalem]{ulem}	

\usepackage{pifont} 

\usepackage{ltxtable} 	
\usepackage{longtable}
\usepackage{afterpage}
\usepackage{ulem}

\newcommand{\rst}[1]{} 

\begin{document}

\title{Internet of Things-aided Smart Grid: Technologies, Architectures, Applications, Prototypes, and Future Research Directions}

\author{Yasir Saleem, {\it Student Member, IEEE}, Noel Crespi, {\it Senior Member, IEEE}, Mubashir Husain Rehmani, {\it Senior Member, IEEE}, and Rebecca Copeland
\thanks{Y. Saleem, N. Crespi and R. Copeland are with the Department of Wireless Networks and Multimedia Services, Telecom SudParis, Institut Mines-Telecom, 91011 Evry, France (e-mail: yasir-saleem\_shaikh@telecom-sudparis.eu; noel.crespi@it-sudparis.eu; rebecca.copeland@coreviewpoint.com).}
\thanks{M.H. Rehmani is with Cork Institute of Technology, Cork, Ireland (e-mail: mshrehmani@gmail.com).}
}
\maketitle

\begin{abstract}
Traditional power grids are being transformed into Smart Grids (SGs) to address the issues in existing power system due to uni-directional information flow, energy wastage, growing energy demand, reliability and security. SGs offer bi-directional energy flow between service providers and consumers, involving power generation, transmission, distribution and utilization systems. SGs employ various devices for the monitoring, analysis and control of the grid, deployed at power plants, distribution centers and in consumers' premises in a very large number. Hence, an SG requires connectivity, automation and the tracking of such devices. This is achieved with the help of Internet of Things (IoT). IoT helps SG systems to support various network functions throughout the generation, transmission, distribution and consumption of energy by incorporating IoT devices (such as sensors, actuators and smart meters), as well as by providing the connectivity, automation and tracking for such devices. In this paper, we provide a comprehensive survey on IoT-aided SG systems, which includes the existing architectures, applications and prototypes of IoT-aided SG systems. This survey also highlights the open issues, challenges and future research directions for IoT-aided SG systems.
\end{abstract}

\begin{IEEEkeywords}
Home Area Network (HAN), Internet of Things (IoT), Neighborhood Area Network (NAN), Smart Grid (SG), Wide Area Network (WAN)
\end{IEEEkeywords}

\section{Introduction}
\label{sec:introduction}

\subsection{Motivation}
A traditional power grid consists of a large number of loosely interconnected synchronous Alternate Current (AC) grids. It performs three main functions: generation, transmission and distribution of electrical energy \cite{collier2017emerging}, in which electric power flows only in one direction, i.e., from a service provider to the consumers. Firstly in power generation, a large number of power plants generate electrical energy, mostly from burning carbon and uranium based fuels. Secondly in power transmission, the electricity is transmitted from power plants to remote load centers through high voltage transmission lines. Thirdly in power distribution, the electrical distribution systems distribute electrical energy to the end consumers at reduced voltage. Each grid is centrally controlled and monitored to ensure that the power plants generate electrical energy in accordance with the needs of the consumers within the constraints of power systems. Nearly, all the generation, transmission and distribution of electrical energy is owned by the utility companies who provide electrical energy to consumers and bill them accordingly to recover their costs and earn profit. 

The traditional power grid worked very well from its inception in 1870 until 1970 \cite{collier2017emerging}. Even though the consumers' demand for energy grew exponentially, it was still rather predictable. However, there has been a dramatic change in the nature of electrical energy consumption since 1970, as the load of electronic devices has become the fastest growing element of the total electricity demand and new sources of high electricity consumption have been developed, such as Electric Vehicles (EVs). The power grids endure a significant wastage of energy due to a number of factors, such as consumers' inefficient appliances and lack of smart technology, inefficient routing and dispensation of electrical energy, unreliable communication and monitoring, and most importantly, lack of a mechanism to store the generated electrical energy \cite{deng2015survey, temel2014routing, ma2013smart}. Furthermore, power grids face some other challenges as well, including growing energy demand, reliability, security, emerging renewable energy sources and aging infrastructure problems to name a few. 

In order to solve these challenges, the Smart Grid (SG) paradigm has appeared as a promising solution with a variety of information and communication technologies. Such technologies can improve the effectiveness, efficiency, reliability, security, sustainability, stability and scalability of the traditional power grid \cite{wang2011survey}. SG differs from traditional power grids in many aspects. For instance, SG offers a bi-directional communication flow between service providers and consumers, while a traditional power grid only offers uni-directional communication from the service provider to the consumer. SG provides Advanced Metering Infrastructure (AMI), smart meters, fault tolerance, unauthorize usage detection, and load balancing \cite{yaacoub2014automatic, yigit2014cloud, sun2012relaying}, as well as self-healing, i.e., detection and recovery from faults \cite{bush2014network}.

SG solves the problem of electrical energy wastage by generating electrical energy which closely matches the demand \cite{deng2015survey, temel2014routing, ma2013smart}. SG helps to make important decisions according to the demand of energy, such as real-time pricing, self-healing, power consumption scheduling and optimized electrical energy usage. Such decisions can significantly improve the power quality as well as the efficiency of the grid by maintaining a balance between power generation and its usage \cite{yaacoub2014automatic}.

SG deploys various types of devices for monitoring, analyzing and controlling the grid. Such monitoring devices are deployed at power plants, transmission lines, transmission towers, distribution centers and consumers premises, and their numbers amount to the hundreds of millions or even billions \cite{collier2017emerging}. One of the main concerns for SG is the connectivity, automation and tracking of such large number of devices, which requires distributed monitoring, analysis and control through high speed, ubiquitous and two-way digital communications. It requires distributed automation of SG for such devices or ``things". This is already being realized in the real world through the Internet of Things (IoT) technology. 
The IoT is defined as a network that can connect any object with the Internet based on a protocol for exchanging information and communication among various smart devices in order to achieve monitoring, tracking, management and location identification objectives \cite{wang2014research}. Over the past few years, the IoT technology has gained significant attention in various applications, and has allowed for the interconnection of the Internet to various network-embedded devices used in daily life \cite{chen2012application}.

Initially, the Internet served as a connectivity of people-to-people and people-to-things. However by 2008, the number of things connected to the Internet surpassed the number of people in the world \cite{collier2017emerging}, thus the impact of IoT technology keeps rising. IoT is a network of physical objects or things connected to the Internet. Such objects are equipped with embedded technology to interact with their internal and external environments. These objects sense, analyze, control and decide individually or in collaboration with other objects through high speed and two-way digital communications in a distributed, autonomous and ubiquitous manner. This is exactly what is required for the SG. Hence, IoT technology can help SGs by supporting various network functions throughout the power generation, storage, transmission, distribution and consumption by incorporating IoT devices (such as sensors, actuators and smart meters), as well as by providing connectivity, automation and tracking for such devices \cite{meng2014smart}. 

SG is considered as one of the largest applications of the IoT \cite{smartgridnews}. 
Today, although a lot of domestic devices that use electricity are connected to the Internet, however still there is a large number of domestic devices that are not connected to the Internet. For example, throughout the globe, the number of microwave ovens and washing machines that are connected to the Internet is much lower than the ones that are not connected to the Internet. Essentially, everything that uses electricity could be made more useful by connecting it to the Internet (such as, microwave ovens and washing machines, that are connected to the Internet can be operated remotely and at off-peak times, thus saving cost, as well as provide comfort to human through automation). Hence in the future, we can predict that the SG integrated with IoT would be larger than the SG today, and the modern and intelligent grid (i.e., SG) will not be possible without the IoT technology. 

By making the IoT technology as a global standard for the communications and the basis for SG, new doors will be opened for maximizing the prospects for future innovations. Based on the need of integration of IoT and SG, recently a new conference dedicated to SG and IoT \cite{sgiot_conf}, as well as a special issue on Smart Grid Internet of Things \cite{fadlullah2017smart} have been initiated. Such initiatives also shows the need and importance of the integration of IoT and SG.

\subsection{Comparison with Existing Survey Articles}
Our current survey on IoT-aided SG systems is significantly different from previous surveys on IoT, SGs and IoT-aided SGs, as we comprehensively cover the domain of IoT-aided SG systems which has not been covered so far in that comprehensive manner. There is a number of attempts in the literature that covered IoT and SGs, and some overviewed IoT-aided SG systems. In this section, we first present the previous surveys on IoT-aided SG systems, IoT and SGs, and subsequently, we justify how we are novel and different from them and what are our contributions.

\subsubsection{Surveys on IoT-aided SG Systems}
There is a number of attempts in the literature that have overviewed various aspects of IoT-aided SG systems. For instance, \cite{collier2017emerging} is not a survey paper rather a short magazine paper in which Collier presented the convergence of SG with the IoT. However, since it is mainly a short magazine paper which presented the convergence, therefore, it did not survey the existing works on IoT-aided SG systems.
In \cite{alturjman2019iot}, the authors focused and surveyed AMI and smart metering technologies for the monitoring of reliability and power quality in IoT-aided SG systems. However, its main focus is only on AMI and smart metering, rather than covering the other important aspects of IoT-aided SG systems, such as applications, architectures, prototypes etc.
There are two recent surveys on security in IoT-aided SG systems. In \cite{gupta2019prevailing}, Gupta et al., mainly focused on security in IoT-aided SG systems and surveyed the vulnerabilities, security threads and their countermeasures in IoT-aided SG systems. In \cite{de2019implementation}, the authors focused on deep packet inspection as a security tool for SG and industrial IoT. The authors studied the existing models and recommendations from academia and governmental bodies (e.g., NIST) and evaluated deep packet inspection as a security tool for SG. However, their complete focus in mainly on deep packet inspection. Both surveys \cite{gupta2019prevailing, de2019implementation} have mainly focused on security aspects in IoT-aided SG systems.
Furthermore, in \cite{bedi2018review}, Bedi et al., focused on how IoT can be applied to electric power and energy systems, such as the value and importance of IoT in electric power network, economic, environmental and societal impacts of IoT-aided electric power systems, challenges and recommendation related to IoT. It provides just one application of IoT in electric power and energy systems. Similarly, Reka et al.,  \cite{reka2018future} focused on providing the awareness to researchers in the field of IoT and SG systems. Both \cite{bedi2018review, reka2018future} are not comprehensive surveys like ours, rather they are very similar to \cite{collier2017emerging} that provides the importance of IoT in SG and a generic overview of IoT-aided SG systems. There is another review \cite{sohraby2017review}, but it only focuses on satellite and wireless-based M2M services applied to the SG.
Jain et al. \cite{jain2014survey} also presents a survey on SG technologies, including smart metering, IoT and energy management system. However, firstly it is a very short survey and secondly, it focuses separately on the technologies, rather than surveying them combinely as IoT-aided SG system. 
Al-Ali and Aburukba \cite{al2015role} presents a conceptual model for SG within the IoT context. However, its main focus is on the SG communication layer based on IPv6 as the backbone and it presents a conceptual model rather than surveying existing articles. 
Viswanath et al. \cite{viswanath2016system} focuses on residential SG, specifically smart home. They developed a testbed for smart home that controls the energy based on dynamic pricing and performs an energy management system. The authors also developed an Android application for allowing remote access to consumers.
Yang \cite{yang2019internet} presents the application of IoT in SG and provides brief overview of opportunities, challenges and future directions. However, it does not survey existing work in the domain of IoT-aided SG systems.

\subsubsection{Surveys on IoT}
There is a number of surveys on the IoT that have focused on the overview, principles, vision, applications, challenges and architectures of the IoT \cite{alfuqaha2015internet, borgia2014internet, mashal2015choices, nitti2015the}. Many surveys have also focused on security \cite{granjal2015security_comst, granjal2015security_ahn, sicari2015security} and cloud computing for the IoT \cite{botta2015integration}. Moreover, social IoT has become a very hot topic of IoT \cite{atzori2012social, ortiz2014cluster}. Other groups of surveys on the IoT have focused on software defined networking (SDN) \cite{sood2016software}, data mining \cite{tsai2014data}, LTE uplink scheduling \cite{mehaseb2016classification} and context-aware computing \cite{perera2014context}. The standardization of the IoT is also surveyed in \cite{palattella2013standardized, sheng2013survey}. Another survey \cite{gluhak2011survey} has focused on experimentation with the IoT.
Moreover, there are several books on the IoT \mbox{\cite{sinclair2017iot, rossman2016the, kranz2016building, waher2015learning, kellmereit2013the, mcewen2013designing}} for the readers who want to understand in more details.

\subsubsection{Surveys on SG}
Surveys focused on the SG have presented the general overview of the SG concept \cite{ma2013smart}, including challenges and standardization activities \cite{fan2013smart, fang2012smart}, as well as applications and architectures \cite{wang2011survey, alanbagi2016survey, erol2015energy, gao2012survey}. Moreover, information management schemes for SG have been surveyed in \cite{liang2014stochastic}. Security issues, which are important factors to be considered in the SG, are surveyed in \cite{yan2012survey, komninos2014survey}, and privacy-aware metering is reviewed in \cite{finster2015privacy}. Other areas which have been surveyed in the domain of SG include energy efficiency \cite{sun2016comprehensive}, power demand forecasting \cite{hernandez2014survey}, load balancing \cite{rahman2014survey}, optimization and pricing methods \cite{vardakas2015survey}, and renewable energy \cite{reddy2014review}. Additionally, some surveys focus on solving the problem of spectrum scarcity in SG through Cognitive Radio (CR) technology \cite{khan2016cognitive, yigit2014cloud, ma2013smart, meng2014smart}. A survey on cloud-assisted IoT-based SCADA systems security has very recently been published in \cite{anam2016cloud}.
Our this survey differs from previous individual surveys on IoT and SG systems because it combines IoT and SG systems together and it specifically covers IoT-aided SG systems.
Additionally, there are several books on the SG \mbox{\cite{babu2018smart, mouftah2018transportation, borlase2017smart, buchholz2014smart, momoh2012smart, sioshansi2011smart}} for the readers who want to understand in more details.

\begin{table*}[!th]\scriptsize
\centering
\caption{Summary of related surveys of IoT, SG and IoT-aided SG.}
\label{tab:comparison}
\begin{tabular}{|p{1.25cm}|p{2.25cm}|p{1cm}|p{8cm}|}
\hline
\bfseries Considered Network & \bfseries Survey & \bfseries  Publication Year &  \bfseries Focus  \\
\hline

\multirow{12}{*}{IoT-aided SG} & Our survey & - & Our survey focuses on providing a review on existing works on the integration of IoT and SG systems, e.g., vision, architectures, applications, prototypes and future research directions. It differs from previous individual surveys on IoT and SG systems in a way that it combines IoT and SG systems together and specifically covers IoT-aided SG systems \\ \cline{2-4}

& Collier \cite{collier2017emerging} & 2017 & A short magazine paper on the convergence of IoT with the IoT. \\ \cline{2-4}

& Al-Turjman and Abujubbeh \cite{alturjman2019iot} & 2019 & AMI and Smart Metering technologies for the monitoring of reliability and power quality in IoT-aided SG systems. \\ \cline{2-4}

& Gupta et al. \cite{gupta2019prevailing} & 2019 & Security in IoT-aided SG systems, specifically vulnerabilities, security threads and their countermeasures. \\ \cline{2-4}

& De La Torre Parra et al. \cite{de2019implementation} & 2019 & Evaluating deep packet inspection as a security tool for IoT-aided SG systems. \\ \cline{2-4}

& Bedi et al. \cite{bedi2018review} & 2018 & Electric power and energy systems, mainly how IoT can be applied to eletric power and energy systems, e.g., the value of IoT in electric power network. Also presents an example of IoT in electric power and energy systems \\ \cline{2-4}

& Reka et al. \cite{reka2018future} & 2018 & Providing awareness to researchers in the field of IoT and SG systems \\ \cline{2-4}

& Sohraby et al. \cite{sohraby2017review} & 2017 & Satellite and wireless-based M2M services applied to SG \\ \cline{2-4}

& Jain et al. \cite{jain2014survey} & 2014 & A short survey on SG technologies, including smart metering, IoT and energy management system with separate focus on the technologies, rather than surveying them combinely as IoT-aided SG system \\ \cline{2-4}

& Al-Ali and Aburukba \cite{al2015role} & 2015 & Presents a conceptual model for SG within the IoT context with the main focus on SG communication layer based on IPv6 as the backbone \\ \cline{2-4}

& Viswanath et al. \cite{viswanath2016system} & 2016 & Developed a testbed for residential SG, i.e., smart home that controls the energy based on dynamic pricing and performs an energy management system. The authors also developed an Android application for allowing remote access to consumers \\ \cline{2-4}

& Yang \cite{yang2019internet} & 2019 & Presents the application of IoT in SG with a focus on brief overview of opportunities, challenges and future directions. However, it does survey existing work in the domain of IoT-aided SG systems \\ 

\hline


\multirow{16}{*}{IoT} & Al-Fuqaha et al. \cite{alfuqaha2015internet} & 2015 & \multirow{4}{8cm}{Overview, principles, vision, applications, challenges and architectures of IoT} \\ \cline{2-3}
& Borgia et al. \cite{borgia2014internet} & 2014 & \\ \cline{2-3}
& Mashal et al. \cite{mashal2015choices} & 2015 &  \\ \cline{2-3}
& Nitti et al. \cite{nitti2015the} & 2015 & \\ \cline{2-4}

& Granjal et al. \cite{granjal2015security_comst} & 2015 & \multirow{3}{8cm}{Security in IoT} \\ \cline{2-3}
& Granjal et al.  \cite{granjal2015security_ahn} & 2015 & \\ \cline{2-3}
& Sicari et al. \cite{sicari2015security} & 2015 & \\ \cline{2-4}

& Botta et al. \cite{botta2015integration} & 2015 & Cloud computing for IoT \\ \cline{2-4}

& Atzori et al. \cite{atzori2012social} & 2012 & \multirow{2}{8cm}{Social IoT} \\ \cline{2-3}
& Ortiz et al. \cite{ortiz2014cluster} & 2014 & \\ \cline{2-4}

& Sood et al. \cite{sood2016software} & 2015 & SDN for IoT \\ \cline{2-4}

& Tsai et al. \cite{tsai2014data} & 2014 & Data mining in IoT \\ \cline{2-4}

& Mehaseb et al. \cite{mehaseb2016classification} & 2015 & LTE uplink scheduling in IoT \\ \cline{2-4}

& Perara et al. \cite{perera2014context} & 2014 & Context-aware computing in IoT \\ \cline{2-4}

& Palattella et al. \cite{palattella2013standardized} & 2013 & \multirow{2}{8cm}{Standardization in IoT} \\ \cline{2-3}
& Sheng et al. \cite{sheng2013survey} & 2013 & \\ \cline{2-4}

& Gluhak et al. \cite{gluhak2011survey} & 2011 & Experimentation in IoT \\ \hline

\multirow{21}{*}{SG} & Ma et al. \cite{ma2013smart} & 2013 & General overview of SG \\ \cline{2-4}

& Fan et al. \cite{fan2013smart},  & 2013 & \multirow{2}{8cm}{Challenges and standardization activities in SG} \\ \cline{2-3}
& Fang et al. \cite{fang2012smart} & 2012 & \\ \cline{2-4}

& Wang et al. \cite{wang2011survey} & 2011 &  \multirow{4}{8cm}{Applications and architectures of SG} \\ \cline{2-3}
& Al-Anbagi et al. \cite{alanbagi2016survey} & 2016 & \\ \cline{2-3}
& Erol et al. \cite{erol2015energy} & 2015 & \\ \cline{2-3}
& Gao et al. \cite{gao2012survey} & 2012 & \\ \cline{2-4}

& Liang et al. \cite{liang2014stochastic} & 2014 & Information management schemes for SG \\ \cline{2-4}

& Yan et al. \cite{yan2012survey} & 2012 & \multirow{2}{8cm}{Security in SG} \\ \cline{2-3}
& Komninos et al. \cite{komninos2014survey} & 2014 & \\ \cline{2-4} 

& Finster et al. \cite{finster2015privacy} & 2015 & Privacy-aware metering \\ \cline{2-4}

& Sun et al. \cite{sun2016comprehensive} & 2016 & Energy efficiency in SG \\ \cline{2-4}

& Hernandez et al. \cite{hernandez2014survey} & 2014 & Power demand forecasting \\ \cline{2-4}

& Rahman et al. \cite{rahman2014survey} & 2014 & Load balancing \\ \cline{2-4}

& Vardakas et al. \cite{vardakas2015survey} & 2015 & Optimization and pricing methods \\ \cline{2-4}

& Reddy et al. \cite{reddy2014review} & 2014 & Renewable energy \\ \cline{2-4}

& Khan et al. \cite{khan2016cognitive} & 2016 &  \multirow{4}{8cm}{Solving the problem of spectrum scarcity in SG through CR technology} \\ \cline{2-3}
& Yigit et al. \cite{yigit2014cloud} & 2014 & \\ \cline{2-3}
& Ma et al. \cite{ma2013smart} & 2013 & \\ \cline{2-3}
& Meng et al. \cite{meng2014smart} & 2014 \\ \cline{2-4}

& Anam et al. \cite{anam2016cloud} & 2016 & Cloud-assisted IoT-based SCADA security \\ \hline

\end{tabular}
\end{table*}

\subsubsection{Novelty and Contributions of This Paper}
We have presented the surveys on IoT-aided SG systems, IoT and SG, however, as presented above and to the best of our knowledge, none of the above attempts have comprehensively covered all major aspects of IoT-aided SG systems, such as integration, applications, architectures, prototypes, taxonomies,  an overview of IoT and non-IoT communications technologies, as well as detailed and extensive open issues and future research directions. Table \ref{tab:comparison} provides the summary of the above references in a comparative manner which also supports our claim of lack of comprehensive survey like our survey paper. In order to fill this gap, our intention and effort is to provide a comprehensive survey that covers all the major aspects of IoT-aided SG systems. 

The contributions of this survey are summarized as follows:
\begin{itemize}
\item A detailed discussion with taxonomy on the applications of IoT-aided SG systems;
\item A detailed discussion on existing architectures for IoT-aided SG systems;
\item A discussion on the current prototypes for IoT-aided SG systems;
\item A discussion on big data analytics and cloud in IoT-aided SG systems;
\item An overview of IoT and non-IoT communication technologies for SG systems;
\item Authors' insights (as subsections) under each topic (e.g., applications, architecture, prototypes, big data analytics and communications technologies); and
\item A layered-wise detailed and extensive presentation of the open issues, challenges and future research directions of IoT-aided SG systems.
\end{itemize}

\begin{figure*}[!th]
\centering
\includegraphics[width=1.0\textwidth]{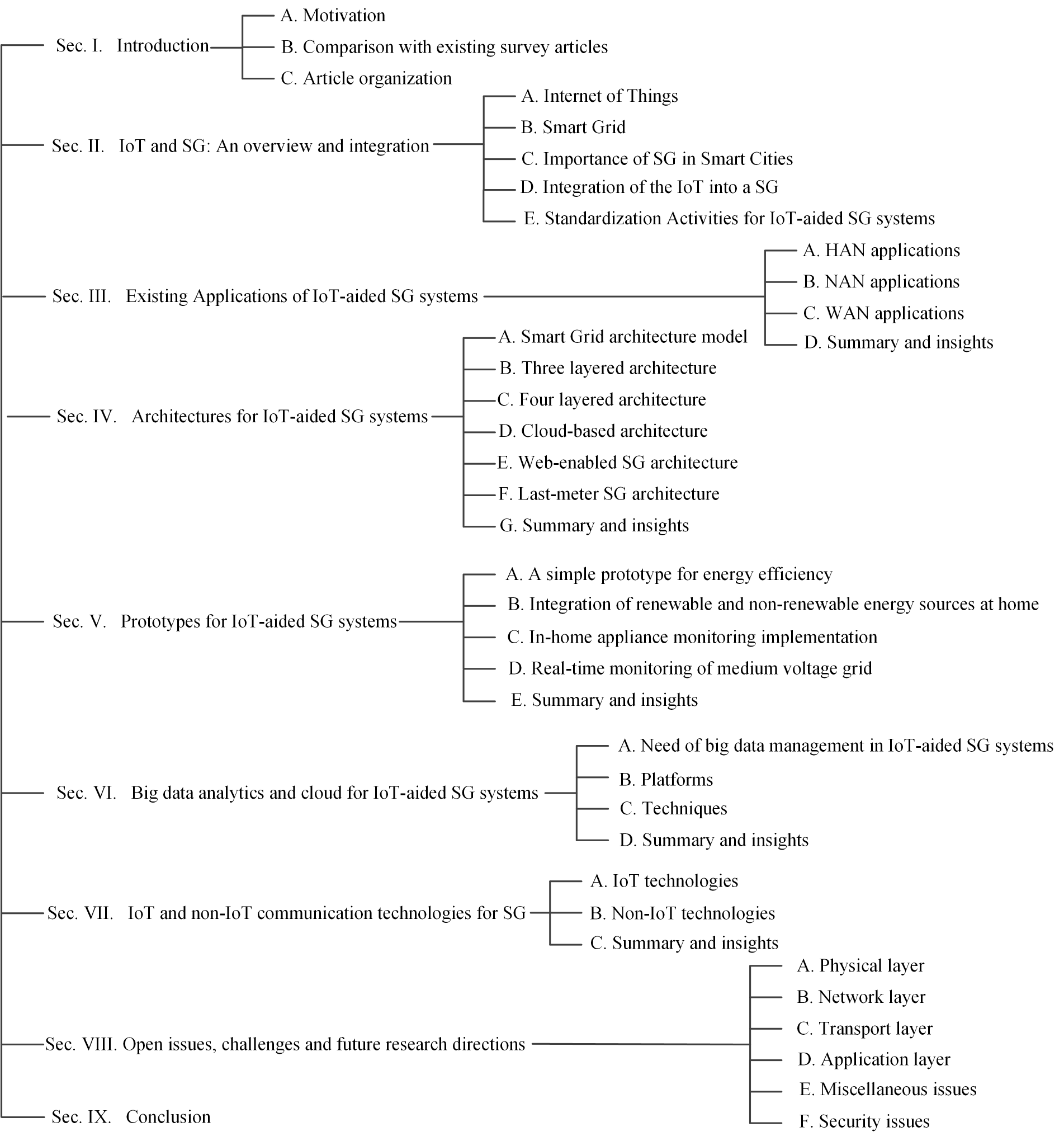}
\caption{Article organization.}
\label{fig:organization}
\end{figure*}

\subsection{Article Organization}
The organization of the paper is presented in Fig. \ref{fig:organization} and is as follows: 
In Section \ref{sec:overview_iot_sg}, we provide an overview of IoT technology and SG systems, as well as the integration of IoT technology into SG systems. 
In Section \ref{sec:applications_iot_sg}, we present the current applications of IoT-aided SG systems, followed by a detailed analysis of the existing architectures of IoT-aided SG systems in Section \ref{sec:architectures_iot_sg}. 
In Section \ref{sec:prototypes_experimentation_iot_sg}, we present prototypes and experimentations for IoT-aided SG systems. We discuss big data management in IoT-aided SG systems in Section \ref{sec:big_data_mgmt_iot_sg}. 
Section \ref{sec:iot_communication_tech_sg} presents an overview of IoT and non-IoT communication technologies for SG systems. 
In Section \ref{sec:open_issues}, we highlight the open issues, challenges and future research directions for IoT-aided SG systems. 
Finally, we conclude in Section \ref{sec:conclusion}.

\section{Internet of Things and Smart Grid: An Overview and Integration}
\label{sec:overview_iot_sg}

\begin{figure*}[t]
\centering
\includegraphics[width=0.9\textwidth]{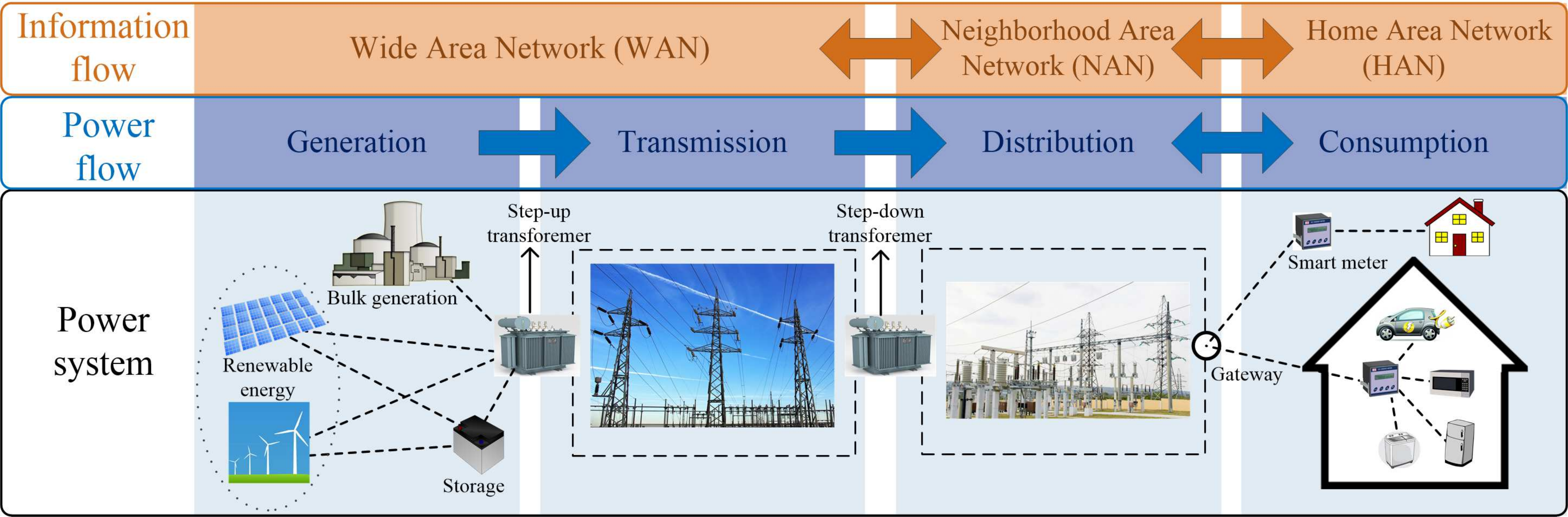}
\caption{Smart grid (SG) architecture presenting power systems, power flow and information flow. The SG is comprised of four main subsystems (power generation, transmission, distribution and utilization) and three types of networks (a wide area network (WAN), a neighborhood area network (NAN) and a home area network (HAN). The power flows through the subsystems while the information flows through the networks.}
\label{fig:architecture_sg}
\end{figure*}

\subsection{Internet of Things}
\label{sec:iot_overview}

The IoT is defined as a network that can connect any object with the Internet based on a protocol for exchanging information and communication among various smart devices in order to achieve monitoring, tracking, management and location identification objectives \cite{wang2014research}. The IoT focuses on the realization of three main concepts, namely things-oriented, Internet-oriented and semantic-oriented. The things oriented concept involves smart devices, such as RFID tags, sensors, actuators, cameras, laser scanners, the Global Positioning System (GPS) and NFC. The Internet oriented concept enables communication among smart devices through various communication technologies, such as ZigBee, WiFi, Bluetooth and cellular communications. and connects them to the Internet. The semantic oriented concept realizes a variety of applications with the help of smart devices.

Over the past few years, the IoT technology has gained significant attention in various applications, and has allowed for the interconnection of the Internet to various network-embedded devices used in daily life \cite{chen2012application}. It has automated the operation of various systems, such as health care, transportation, military, home appliances, security, surveillance, agriculture and power grids. IoT devices are normal objects that are equipped with transceivers, micro-controller and protocol stacks, enabling their communication with other devices, as well as with external entities (e.g., humans) to allow the realization of completely automated systems that make them an essential part of the Internet \cite{zanella2014internet}.

\subsection{Smart Grid}
\label{sec:sg_overview}

The SG has been promoted as a promising solution for minimizing the wastage of electrical energy and as a means to solve the problems of traditional power grids, making possible advances in efficiency, effectiveness, reliability, security, stability, and increasing demand of electrical energy \cite{iot3}. The main SG attributes are that it offers self-healing, improved electricity quality, distributed generation and demand response, mutual operation and user participation, and effective asset management. 

As presented in Fig. \ref{fig:architecture_sg}, the SG completely revolutionizes the energy generation, transmission, distribution and consumption in four sub-systems. It is comprised of three types of networks, a Home Area Network (HAN), a Neighborhood Area Network (NAN) and a Wide Area Network (WAN). HAN is the first layer; it manages the consumers' on-demand power requirements and consists of smart devices, home appliances (including washing machines, televisions, air conditioners, refrigerators and ovens), electrical vehicles, as well as renewable energy sources (such as solar panels). HAN is deployed within residential units, in industrial plants and in commercial buildings and connects electrical appliances with smart meters. NAN, also known as Field Area Network (FAN), is the second layer of an SG and consists of smart meters belonging to multiple HANs. NAN supports communication between distribution substations and field electrical devices for power distribution systems. It collects the service and metering information from multiple HANs and transmits it to the data collectors which connect NANs to a WAN. WAN is the third layer of an SG and it serves as a backbone for communication between network gateways or aggregation points. It facilitates the communication among power transmission systems, bulk generation systems, renewable energy sources and control centers \cite{deng2015survey}. Additionally, video cameras have been used in SG management to build video surveillance system for assets safety, fire alarm and safe operations. Zhang and Huo \cite{zhang2014research} developed a video surveillance system to assist safety operations in smart substation. They developed integrated SCADA system with video cameras embedded in supervisory graph for improving the efficiency. The authors first developed a system and layered communication architectures. Secondly, for software reuse, they present component-based software development. Finally, they present the communication protocol between SCADA system and video surveillance system.

EVs are of paramount importance when we talk about SG. EVs are considered as an effective tool for reducing the gas emissions and oil demands, as well as for increasing the energy conversion \cite{liu2013opportunities}. The emergence of SG has opened new opportunity for EVs. Now EVs are used to exchange energy with the power grid \cite{liu2013opportunities}. They not only consume energy from the power grid, in fact through their bidirectional charger, they also distribute the energy back to the power grid. There are three main emerging concepts which are based on the capability of charging/discharging of EVs: Vehicle-to-Grid (V2G), Vehicle-to-Vehicle (V2V) and Vehicle-to-Home (V2H) \cite{liu2013opportunities}. In V2G, EVs are connected to the power grid. They can obtain energy, as well as deliver energy back to the power grid. One way of earning money is to buy energy during off-peak hours at low price from the grid. Then during on-peak hours, the EV can deliver the energy back to the grid at higher price. In V2V, EVs distribute energy among other EVs. Using bidirectional chargers, the EVs first transfer their energy using a local grid, and subsequently, by using a controller (also known as aggregator), the energy is distributed among EVs \cite{liu2013opportunities}. In V2H, the EVs supply energy to the homes. The EVs charge their batteries during off-peak hours at low price from the grid. Then at on-peak hours, when the energy price is higher, the houses consumes energy from EV batteries fulfilling the whole or partial demand of the house and avoids buying the expensive energy during on-peak hours \cite{dufo2014can}. For more  details on V2G, V2V and V2H, the readers are referred to \cite{liu2013opportunities}.

\subsection{Importance of SG in Smart Cities}
\label{sec:imp_sg_smart_cities}
Smart cities are comprised of several entities, such as governance, buildings, security, healthcare, economy, transportation and energy. Among them, energy is the most important sector for moving towards a more endurable urban life, as well as for integrating various stakeholders and sensitive infrastructure \cite{atasoy2015analysis}. In other words, smart cities are tightly coupled with the modernization of traditional power grid, i.e., SG because if the power is unavailable for a certain period of time, all other functions of smart cities will be halted. SG provides three main functions which are highly required by a smart city. Firstly, the traditional power grid is transitioned into SG by automation, remote controlling and monitoring. Secondly, SG enables the consumers to be aware about their energy consumption, costs, as well as allow them to adjust their energy accordingly. Thirdly, the integration of renewable and distributed energy resources is enabled by the SG. Therefore, we can say that without these functionalities, smart cities cannot fully exist \cite{atasoy2015analysis, curiale2014from}.







\subsection{Integration of the IoT into a SG}
\label{sec:integration_iot_sg}

The SG has already achieved wide adoption in information sensing, transmission and processing, and now IoT technology plays a significant role in grid construction. 
The driving force behind the initiative of SG is to improve planning, maintenance and operations by ensuring that each component of the power grid is able `listen' and `talk', and to enable automation in SG \cite{what_is_sg}. For example, in traditional power grid, the utility company only knows about the disruption of service when a customer informs himself/herself. In SG, the utility company will automatically know about the disruption of service because certain components of SG (such as smart meters in the affection region) will cease sending the collected sensor data. Here, the IoT plays the key role in enabling this scenario because all the components of the grid (see Fig. \ref{fig:architecture_sg}) must have IP addresses and should be capable of two-way communication. This is enabled by the IoT.
IoT technology provides interactive real-time network connection to the users and devices through various communication technologies, power equipment through various IoT smart devices, and the cooperation required to realize real-time, two-way and high-speed data sharing across various applications, enhancing the overall efficiency of a SG \cite{yun2010research}. The application of the IoT in SGs can be classified into three types based on the three-layered IoT architecture \cite{yaqoob2017internet, alfuqaha2015internet}. Firstly, IoT is applied for deploying various IoT smart devices for the monitoring of equipment states (i.e., at perception layer of IoT). Secondly, IoT is applied for information collection from equipment with the help of its connected IoT smart devices through various communication technologies (see Section \ref{sec:iot_communication_tech_sg}) (i.e., at network layer of IoT). Thirdly, IoT is applied for controlling the SG through application interfaces (i.e., at application layer of IoT). 

IoT sensing devices are generally comprised of wireless sensors, RFIDs, M2M (machine-to-machine) devices, cameras, infrared sensors, laser scanners, GPSs and various data collection devices. The information sensing in an SG can be highly supported and improved by IoT technology. The IoT technology also plays an essential role in the infrastructure deployment of data sensing and transmission for the SG, assisting in network construction, operation, safety management, maintenance, security monitoring, information collection, measurement, user interaction etc. Moreover, the IoT also enables the integration of information flow, power flow and distribution flow in a SG \cite{liu2011applications},\cite{zaveri2016collaborative}. Additionally, existing SG architectures mainly focus on the needs of power distributors to manage the complete power grid \cite{samarakoon2013reporting}. The consumers are accessed with a network of smart meters by means of General Packet Radio Service (GPRS) or other mobile networks. The new reality where consumers may already have other smart home infrastructures (such as WiFi) has not yet been incorporated in the network communications of existing SG architectures \cite{khan2011web, benzi2011electricity}. While some architectures do consider existing smart home infrastructures, they are not designed for scalability in large deployments \cite{hu2013hardware, viani2013wireless}. The protocols specific to IoT and SG systems therefore cannot be directly applied to IoT-aided SG systems, as they only consider the individual characteristics of either the IoT or the SG systems, which is not sufficient for an integrated IoT-aided SG system.

A SG is comprised of four main subsystems, power generation, power transmission, power distribution and power utilization. IoT can be applied to all these subsystems and appears as a promising solution for enhancing them, making the IoT a key element for SG. In the area of power generation, the IoT can be used for the monitoring and controlling of energy consumption, units, equipment, gas emissions and pollutants discharge, power use/production prediction, energy storage and power connection, as well as for managing distributed power plans, pumped storage, wind power, biomass power and photo-voltaic power plants \cite{shu2011research, wang2012research,tsg5}. In the area of power transmission, the IoT can be used for the monitoring and control of transmission lines and substations, as well as for transmission tower protection \cite{shu2011research, wang2012research}. In the area of power distribution, IoT can be used for distributed automation, as well as in the management of operations and equipment. In the area of power utilization, the IoT can be used for smart homes, automatic meter reading, electric vehicle charging and discharging, for collecting information about home appliances' energy consumption, power load controlling, energy efficiency monitoring and management, power demand management and multi network consumption \cite{shu2011research, wang2012research}. Fig. \ref{fig:applications_iot_sg} presents the existing (see Section \ref{sec:applications_iot_sg}) and potential applications of IoT-aided SG systems. 

In the rest of this section, we describe the suitability of IoT technology and preferred communication technologies for various functions of the three SG layers, i.e., for HAN, NAN and WAN. 

\subsubsection{Home Area Networks (HANs)}
HAN is the first layer; it manages the consumers' on-demand power requirements and consists of smart devices, home appliances (including washing machines, televisions, air conditioners, refrigerators and ovens), electrical vehicles, as well as renewable energy sources (such as solar panels). HAN is deployed within residential units, in industrial plants and in commercial buildings and connects electrical appliances with smart meters \cite{deng2015survey}.
HANs may have either a star topology or a mesh topology. The preferred communication technologies for HANs are powerline communications (wired technology), ZigBee, Bluetooth and WiFi (wireless technologies). A HAN is comprised of a variety of IoT smart devices, such as a home gateway, smart meters, sensor and actuator nodes, smart home appliances and electric vehicles. A home gateway connects to smart meters and periodically collects power consumption data of the home appliances. 

HANs perform two-fold functions, commissioning and control. The commissioning function identifies new devices and manages the devices. The control function enables communication among smart devices by establishing the links and performs reliable operation for the various SG layers. A HAN uses two-way communications for demand response management services \cite{mashima2016residential, aburukba2016iot, mortaji2016smart}. In the forward communication direction, the smart meters' load and real-time power consumption information of the home equipment, connected to IoT smart devices, are collected by home gateways and transmitted from the consumer side (the HAN) to the NAN to be forwarded to a utility center. In the backward communication direction, the home gateway acts as a central node and receives dynamic electricity pricing information from the NAN, which is then provided to smart meters or IoT smart devices for triggering the required action for home appliances. Mosrello et al. \cite{morello2017smart} present the role of IoT and advance sensing in smart power meter for monitoring the energy flow in SG.

\subsubsection{Neighbor Area Networks (NANs)}
NAN is the second layer of an SG and consists of smart meters belonging to multiple HANs. NAN supports communication between distribution substations and field electrical devices for power distribution systems. It collects the service and metering information from multiple HANs and transmits it to the data collectors which connect NANs to a WAN \cite{deng2015survey}.

The communication technologies for NANs need to cover a radius of a thousand meters. The communication channels between smart meters and data aggregation points therefore must be interference free \cite{zhu2012overview}. A gateway in the NAN collects consumers' energy consumption data from smart meters in HANs and transmits the collected data to the utility companies through either private or public WANs. Basically, the topology of a NAN is comprised of two types of gateways, NAN gateways and HAN gateways. A NAN gateway connects various HAN gateways and serves as an access point to provide a single hop connection to HAN gateways in a hybrid access manner. The HAN gateways transmit their energy consumption data to NAN gateways through either wired (e.g., PLC, DSL) or wireless (e.g., cellular, mobile broadband wireless access or digital microwave technology) communication technologies. 

\subsubsection{Wide Area Networks (WANs)}
The WAN is the third layer of an SG and it serves as a backbone for communication between network gateways, NANs, distributed grid devices, utility control centers and substations. It facilitates the communication among power transmission systems, bulk generation systems, renewable energy sources and control centers \cite{deng2015survey}. It is comprised of two interconnected networks, core networks and backhaul networks. The core network provides communication to utility control centers with low latency and high data rate through fiber optics or cellular communications. The backhaul networks provide broadband connections and monitoring devices to NANs through wired (e.g., optical networks, DSL), wireless (e.g., cellular network, mobile broadband wireless access) or hybrid fiber-wireless networks.

\subsection{Standardization Activities for IoT-aided SG systems}
For the practical realization of technologies, standardization activities play an important role. Though efforts have taken place on both IoT and SG systems from the perspective of standardization at the individual level, there is still a need to put joint efforts towards standardization of IoT-aided SG systems. Below, we summarize standardization activities in IoT and SG systems, as well as our insights on standardization of IoT-aided SG systems. 

\subsubsection{Standardization Activities in IoT}

In IoT, several standardization activities are going on. A very detailed discussion on standardization activities from the perspective of IoT and different standardization bodies can be found in \cite{palattella2013standardized, bandyopadhyay2011internet}. The oneM2M \cite{swetina2014toward, onem2m2016} is standard for M2M and IoT, which is widely adopted. OMA LWM2M (Lightweight M2M) \cite{oma2016} is another standard for IoT which is gaining popularity due to its greater simplicity. Additionally, the Internet Engineering Task Force (IETF) has developed a set of protocols and standards for connecting the IoT devices to the Internet \cite{sheng2013survey}. For instance, the Routing Protocol for Low Power and Lossy Networks (RPL) and the Constrained Application Protocol (CoAP) are examples of two protocols developed by IETF to connect lower power devices to the Internet. Machine-to-Machine (M2M) communication is an enabling technology for IoT. M2M-based communication networks and their protocol stack is discussed in \cite{aijaz2015cognitive}, wherein authors briefly discuss protocol layers and standardization efforts to connect M2M devices to the Internet. A security protocol is key requirement for the successful operation of IoT. In this regard, IETE's DICE working group (WG) has put some effort to define and adapt the Datagram Transport Layer Security (DTLS) protocol \cite{keoh2014securing}.

\subsubsection{Standardization Activities in SG}

In SG, there are several worldwide standardization efforts. IEEE, ANSI, NIST, and IETF are the key organizations which are involved in the standardization process of SG systems. In Europe, various Expert Groups have been established which look at the standardization efforts. 
In March 2011, European Commission (EC) and European Free Trade Association (EFTA) published a SG mandate M/490 \cite{m490} which has been accepted by three European Standardization Organizations (ESOs): the European Committee for Standardization (CEN), the European Committee for Electrotechnical Standardization (CENELEC) and the European Telecommunications Standards Institute (ETSI) in June 2011 \cite{cenelec_sg_standardization}. The mandate M/490 requested CEN, CENELEC and ETSI to design a framework enabling ESOs to develop and enhance SG standard continuously. The mandate M/490 presents the following points: need of fast action, a need to consider the large number of stakeholders and to operate in an international environment. In July 2011, the ESOs are established together with CEN-CENELEC-ETSI SG Coordination Group (SG-CG) in order to perform the requested mandated work that coordinates the ESOs reply to mandate M/490 \cite{cenelec_sg_standardization, cenelec_sg_standardization2}. By the end of 2014, SG-CG finalized four mandated reports: i) extended set of standards \cite{sgcg_m490_standards_set}, ii) overview methodology \cite{sgcg_m490_methodologies} and its annexes: general market model development \cite{sgcg_m490_annex1}, SG architecture model user manual \cite{sgcg_m490_annex2} and flexibility management \cite{sgcg_m490_annex3}, iii) SG interoperability \cite{sgcg_m490_interop} and its tool \cite{sgcg_m490_interop_tool}, and iv) SG information security \cite{sgcg_m490_security}. These reports were approved in December 2014 by CEN, CENELEC and ETSI board \cite{cenelec_sg_standardization}. 
IEEE-SA has been closely working on SG systems and in this regard, the IEEE P2030 standard has been proposed, which deals with the interoperability issues of different power systems. Another effort by IEEE is the approval of IEEE's 1547 standard which deals with the interconnection of distributed energy resources. To deal with the challenges of NAN communications, the IEEE 802.15.4g standard was introduced. This standard deals with the physical layer issues associated with NAN communications \cite{nanstandard}. The IEEE P1901 working group is developing standards for multimedia communication in cognitive radio-based SG systems \cite{multimediasg}.

For EVs, the standardization is a critical issue. SG led to new European policies for the deployment of new infrastructures to recharge/replace the batteries of EVs. As an initial step, CENELEC and CEN have created a coordination group, known as European Electro-Mobility Coordinate Group (eM-CG) \cite{standardization_electric_vehicles}, in October 2011, in response to Mandate M/468 \cite{m468}, an standardization mandate related to the charging of electric vehicles. The eM-CG ensures that the required standards for eMobility are seriously dealt by the appropriate technical bodies in a rational way.

\subsubsection{Standardization of IoT-aided SG Systems}
IoT-aided SG system is a complex system that needs heterogeneous communication technologies to fulfil the diverse requirements of the system. Hence, unlike traditional telecommunication standardization (such as 3GPP LTE or IEEE 802.11n), the communication standardization of IoT-aided SG systems involves enabling interoperability among interfaces, workflows and messages. Therefore, rather than one particular communication technology for IoT-aided SG system, it is imperative to reach a consensus on the usage and interpretation of messages and interfaces for the seamless bridging of different technologies and standards. In short, the primary objective of communication standardization of IoT-aided SG system is to achieve interoperability among different components (e.g., devices, meters and protocols) of SG instead of defining them \cite{smcgm4412009}.

In summary, there are still many issues which need to be addressed by the standardization community to deal with interoperability issues of IoT-aided SG systems.

\begin{figure*}[t]
\centering
\includegraphics[width=1.0\textwidth]{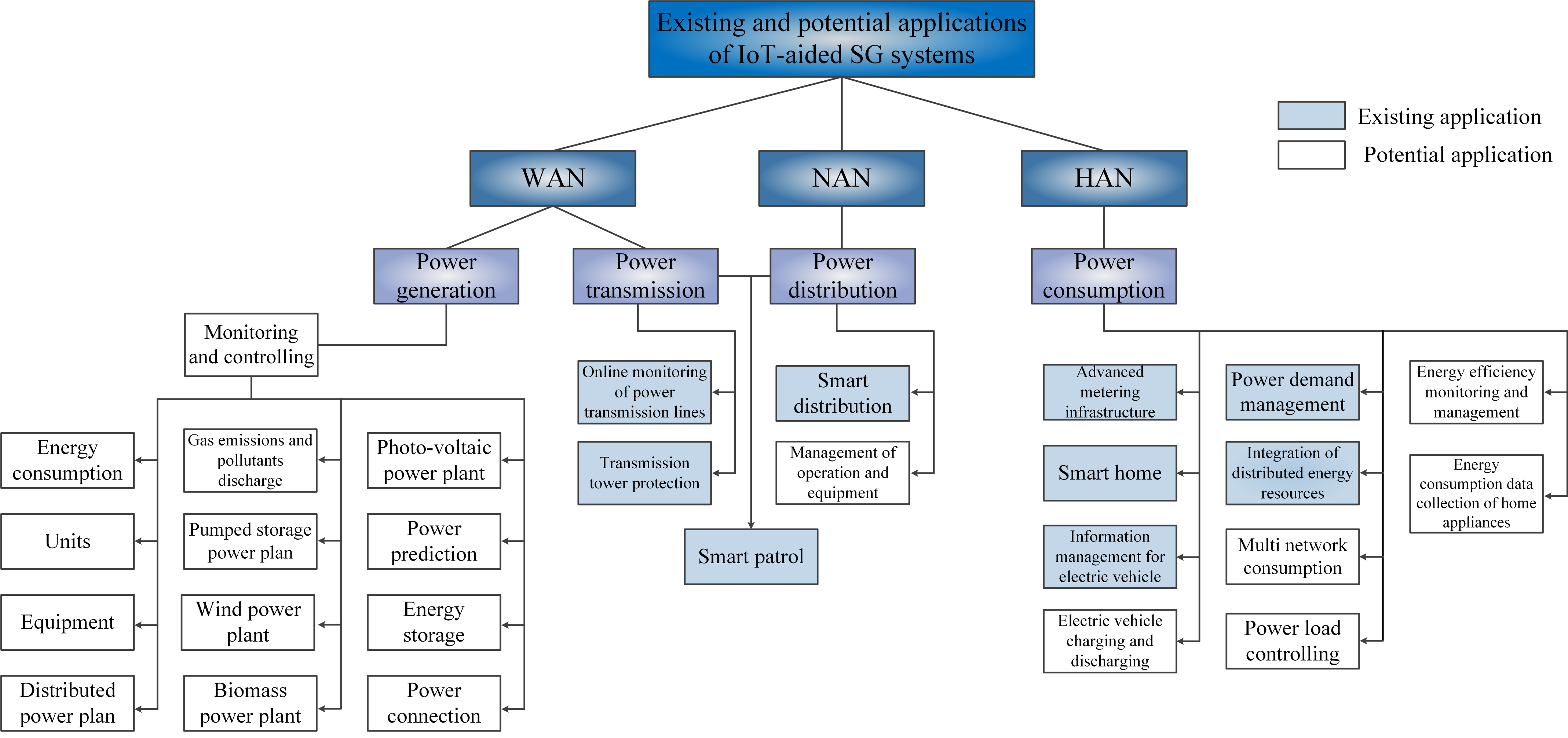}
\caption{Existing and potential applications of IoT-aided SG systems classified into WAN, NAN and HAN. These are further classified into subsystems, i.e., power generation, transmission, distribution and utilization. The blue (shaded) boxes represent existing applications while the white (empty) boxes represent potential applications. The existing applications are discussed in Section \ref{sec:applications_iot_sg}.}
\label{fig:applications_iot_sg}
\end{figure*}

\begin{table*}[hbtp]\scriptsize
\centering
\caption{Summary of applications of IoT-aided SG systems.}
\label{tbapplications}
\begin{tabular}{|p{4.5cm}|p{2cm}|p{1.5cm}|p{2.5cm}|}
\hline
\bfseries Applications & \bfseries References & \bfseries SG Area & \bfseries SG Sub-Area \\
\hline
Transmission Tower Protection & \cite{zhen2012transmission} & WAN  & Power Transmission \\
\hline
Monitoring of Power Transmission Lines & \cite{chen2011integration} & WAN  & Power Transmission \\
\hline
Smart Patrol & \cite{liu2011applications} & NAN  & Power Distribution \\
\hline
Smart Home & \cite{santoso2015securing, jaradat2015internet, li2011smart} & HAN  & Power Consumption \\
\hline
Information Management Systems for EVs & \cite{liu2011applications} & HAN  & Power Consumption \\
\hline
Automatic Meter Reading & \cite{chen2011integration} & HAN & Power Consumption \\
\hline
Distributed Energy Resources & \cite{yu2014carbon, jaradat2015internet, rana2015microgrid} &  HAN & Power Consumption \\
\hline
Power Demand Management & \cite{balijepalli2011review,tsg3,siano2014demand,jaradat2014integration} & HAN  & Power Consumption \\
\hline
\end{tabular}
\end{table*}

\section{Existing Applications of IoT-aided SG Systems}
\label{sec:applications_iot_sg}

There are many existing applications of IoT-aided SG systems, and many more have been proposed, as represented in Fig. \ref{fig:applications_iot_sg}. Table \ref{tbapplications} presents the work done on the applications of IoT-aided SG systems. In this section, we discuss existing applications of IoT-aided SG systems in the literature \cite{zhen2012transmission, santoso2015securing, jaradat2015internet, li2011smart, liu2011applications, chen2011integration, yu2014carbon, rana2015microgrid, mohanty2014implementation, balijepalli2011review, tsg3, siano2014demand, jaradat2014integration}.

\subsection{HAN Applications}

\subsubsection{Smart Home}
IoT technology plays a major role in the SG for the realization of smart homes and appliances \cite{santoso2015securing, khan2016internet, mukhopadhyay2016home}, such as smart TVs, home security systems, smart refrigerators, washing machines, fire detection, lighting control and temperature monitoring. The smart home includes sensor and actuator nodes for environmental monitoring that transmit the surveillance data to the home's control unit \cite{jaradat2015internet}. The control unit enables users to monitor and control the appliances remotely from anywhere and at any time. The smart home is thus a key element of the SG to realize real-time interaction between users and the grid, improve the quality of services and enhance the capacity of integrated grid services, as well as to fulfill users' energy demands in a most efficient possible way. Optimizing daily power consumption makes broad use of smart home services. For instance, users can turn on heaters or air conditioners before coming home in order to enjoy their desired environment without waiting. Additionally, users can switch on their power-intensive electrical appliances, such as their washing machine, in the middle of the night when the electricity price is lower. The control unit also uses surveillance data to detect suspicious activities and inform users to take appropriate actions. All these functions can only able to be realized, thanks to the IoT technology. Viswanath et al. \cite{viswanath2016system} have designed a system and developed a testbed for smart home with a house having one living and three bedrooms that can accommodate six to nine persons. This system controls the energy based on dynamic pricing and performs as an energy management system. Hence, it avoids home appliances usage during the peak hours. The authors developed an Android application for allowing remote access to consumers as well. Additionally, Shah et al. \cite{shah2016enhancing} proposed a model for security enhancement in smart home by implementing Reed Solomon Codes for error detection and correction.

The IoT is applied to various aspects of the SG in smart homes, for instance, in a smart home's sensor LAN protocol to control smart appliances, multi-meter reading, information gathering of power consumption (including electricity, water and gas), load monitoring and control and user interaction with smart appliances. IoT technology also provides its services to NANs, linking a group of smart homes in a neighborhood through a NAN to form a smart community \cite{li2011smart}. Smart homes in the smart community can thus share the results of outdoor surveillance cameras to detect any accident or suspicious activity and inform the appropriate emergency centers and police stations autonomously. The concept of the smart community could be extended to form a smart city. Similarly, a comprehensive surveillance system could be developed in a smart city to monitor various activities within an entire city or even a country. Some IoT requirements and considerations for smart buildings are discussed in \cite{minoli2017iot} for building management systems and energy optimization.

\subsubsection{Information Management System for Electric Vehicles}

EVs provide an eco-friendly transportation option by reducing carbon dioxide emissions \cite{liu2011applications, yu2016distributed}. This is an interesting opportunity for IoT aided SG systems. The charging system for EV is composed of a power supply system, charging equipment and a monitoring system. The power supply system is responsible for the output and management of electricity. The charging equipment charges and discharges the EVs, and includes both AC and DC chargers. AC chargers provide slow charging and are generally implemented in home. Rapid EV charging requires DC chargers, generally implemented at public charging stations. Both types of chargers also include a billing function. The monitoring system is responsible for real-time monitoring of the charging system and its security. IoT technology plays a leading role in this monitoring system, providing an information management system that integrates different components of the charging system \cite{liu2011applications}. For instance, the IoT technology enables the power supply and real-time monitoring systems to send their information to the information management system, which in turns passes information to the control station, so that necessary actions can be taken. EVs are equipped with GPS, which let the IoT to help drivers to manage their batteries more efficiently by locating the nearest, most suitable charging station with the shortest waiting time, as well as providing traffic and parking information \cite{liu2011applications}.

\subsubsection{Advanced Metering Infrastructure (AMI)}
Traditionally, power consumption information was collected manually on site at specific time intervals. This practice inevitably led to inadequacies in terms of accuracy and timeliness. IoT enables AMI \cite{ghasempour2016optimizing} or remote meter reading systems based on WSN and \cite{rashid2016applications} PLC and Optical PLC (OPLC) by using public or private communication networks. An AMI system is one of the most important functions of the SG, collecting the real-time electricity consumption data with high reliability, processing this information and hence providing real-time monitoring, statistics and power consumption analysis. 
AMI meters (or smart meters) are enhanced and digital versions of the traditional electric meters that are deployed outside the home. Besides measuring the electricity consumption, AMI meters also transmit energy and pricing information from utility companies to the consumer premises, thus allowing two-way communication, thanks to the IoT \cite{amr_vs_ami}. Hence, in this manner, the consumers can adjust their energy usage based on the energy and pricing information received through AMI, and hence can save their money.
The importance of this system lies in the timeliness, efficiency and accuracy in the power consumption data. By using this system, IoT technology could help users to save money by adjusting their electricity usage behavior based on the analysis of their power consumption \cite{chen2011integration}.

The European Commission set certain directives (e.g., Directive 2009/72/EC \cite{union2009adirective} and Directive 2009/73/EC \cite{directive2009bdirective} of the European Parliament) which oblige each member state to implement smart metering systems in its own legislation in order to enable consumers to have active participation in the electricity and gas supply markets \cite{eu2012smartmeter}.
Following these European Commission directives \cite{union2009adirective, directive2009bdirective}, France has begun its AMI rollout in March 2010 with a smart metering pilot project called `Linky' \cite{linky}. The French grid manager called Electricite Reseau Distribution France (ERDF) is managing the deployment of 28 million smart meters, the Linky, in France. Currently, 35 million Linky meters have been deployed in France which costs around 4.5 billion euros. Based on the estimates, the Linky smart meter installation cost is around 150-200 euros or 1-2 euros per month per household over 10 years \cite{smartmeter101}. The estimated savings it will generate on average are around 50 euros per annum which is much lower than the installation cost. 
For all EU member states, the Joint Research Centre (JRC) of European Commission \cite{jrc} prepared a report on benchmarking the deployment of smart meters in EU member states focusing on the electricity \cite{benchmarking_smartmeter}. This report performs a detailed analysis of each EU member state's roll-out plans by relating it with cross-country indicators and metrics. Additionally, this report also presents the lessons learned and best practices from EU member states which have successfully roll-out the smart metering (or AMI) \cite{eu_consumers_smart_meter_2020}. For complete information, the readers are referred to the report \cite{benchmarking_smartmeter}.

\subsubsection{Integration of Distributed Energy Resources (DERs)}

Renewable energy generators, such as solar cells, photovoltaic cells and wind turbines, are progressively integrated into today's power grid. They have recently attracted considerable attention in SG studies due to climate changes and environmental pressures. Renewable energy has a positive impact on the global environment by generating electricity without carbon emissions \cite{yu2014carbon}. Reducing or holding back the annual increase in greenhouse gas emissions contributes to limiting the Earth's increasing temperature. In the past few years, governments and organizations have installed a substantial number of solar cells and wind turbines to satisfy part of their power requirements \cite{jaradat2015internet}. Power generation patterns of renewable energy sources (solar and wind), which are distributed over the grid, are intermittent in nature and dependent on location and climate, so they pose significant challenges for the predictability and reliability of the power supply. Such problems are addressed using the seamless interoperability and connectivity provided by the IoT technology. Furthermore, the IoT technology uses sensors to collect real-time weather information which helps in forecasting energy availability in the near future. A Kalman Filter-based state estimation and discrete-time linear quadratic regulation method for controlling the state deviations is proposed in \cite{rana2015microgrid}, utilizing IoT aided SG systems. It uses IoT technology and WSNs to sense, estimate and control the states of DERs. Furthermore, a web of things-based SG architecture is proposed in \cite{mohanty2014implementation} for the remote monitoring and controlling of renewable energy sources.

\subsubsection{Power Demand Management}

Power demand management, also known as demand-side energy management, is defined as the change in the energy consumption profiles of consumers according to the time-varying electricity prices from utility companies \cite{balijepalli2011review,tsg3}. It is used to minimize the consumer's electricity bill, and to reduce the operational cost of the power grid and energy losses, as well as to shift the demand load from peak times \cite{siano2014demand}. IoT devices collect the energy consumption requirements of various home appliances and transmit them to the home control units. Subsequently, the SG control unit schedules the energy consumption of home appliances based on the users' defined preferences so that each consumer's electricity bill is minimized. Demand-side energy management can be performed at different levels of the SG. For instance, it can be performed at a home-level to maintain consumers' privacy. It can also be performed at higher levels to not only benefit the consumers but also the utility companies by generating a more effective scheduling plan \cite{jaradat2014integration}.

\subsection{NAN Applications}

\subsubsection{Smart Distribution}

Smart distribution grid is introduced and defined in \cite{ma2016key} and is characterized by high reliability, improved power quality, improved compatibility, better interaction ability, higher exploitation rate of power grid assets and visualization management platform. Its main functional requirements include reliable and flexible grid structure, operational control, open communication architecture and software component \cite{ma2016key}. The detailed discussion of smart distribution grid is provided in \cite{ma2016key}.
Smart distribution grid is based on advanced automated IoT technology and is one of the important component of SG \cite{ma2016key}. It is the component that is directly connected to users in smart grid. Smart distribution grid consists of communication system, power distribution remote unit, master unit and station unit \cite{ma2016key}. With the help of IoT technology, the smart distribution grid can immediately identify the faults in case of any disorder and can overcome the fault instantly. IoT technology helps smart distribution grid by providing various types of sensors for collecting data about temperature, humidity, noise etc. which ensures monitoring and secure operation of distribution grid \cite{jingchao2016implementation}. The first demonstration of smart distribution by using IoT technology in distribution network of smart grid was performed in Henan Hebi IoT demonstration project in Hebi, China \cite{jingchao2016implementation}. It uses temperature, noise and tower tilt sensors, as well as ZigBee, GPRS, 3G and power fiber. It realizes online monitoring, online inspection and life cycle management. This project was performed using 10 kV underground and overhead laying mixed line which covers 45 utility distribution transformers, 68 units of surge arresters, circuit breakers on 20 pillars, a ring counter, and four cable branch box. Readers are referred to \cite{jingchao2016implementation} for detailed description and results of this project.

\subsubsection{Smart Patrol}

The patrolling of power generation, transmission and distribution used to be mainly a manual task, performed regularly at specific time intervals. However, due to climate conditions and both human and environmental factors, the quality and quantity of patrolling is not always as desired. Furthermore, it is usually not easy for power workers to patrol unattended substation equipment. The IoT technology offers a promising solution to this problem by introducing Smart patrol \cite{liu2011applications}. Smart patrol is comprised of WSN and RFID tags which are connected to the power substation with the help of IoT technology, and are used to locate the power equipment in order to improve the quality of patrolling, as well as to enhance the stability, efficiency and reliability of a power system and its supply. Smart patrol can be used for a number of applications, such as patrol staff positioning, equipment status reports, environment monitoring, state maintenance and standard operations guidance \cite{liu2011applications}.

\subsection{WAN Applications}

\subsubsection{Transmission Tower Protection}

An integral part of power transmission, the transmission tower protection is a WAN application of IoT-aided SG systems, developed to enhance the safety of transmission towers from physical damage by plundering of components, natural disasters, unsafe construction and growing trees under the foundations \cite{zhen2012transmission}. Burglary and intentional damage by people are the major causes of transmission tower damage. Natural disasters, such as typhoons, strong thunderstorms and global warming affects also can cause transmission towers to collapse. Additionally, key infrastructure projects, such as highways and high speed railways, are often constructed near the transmission towers and must sometimes cross high voltage transmission lines. Often the construction companies are not fully aware of the risks involved in operating near high voltage transmission towers. They use a number of large construction machines which not only pose serious dangers to their crew, but can also damage transmission lines and towers. Such construction contractors sometimes do not inform the relevant power transmission departments which makes it impossible for power transmission employees to inspect and monitor all the power transmission facilities, which could lead to risks to transmission towers. 

\begin{figure*}[th]
\centering
\includegraphics[width=0.8\textwidth]{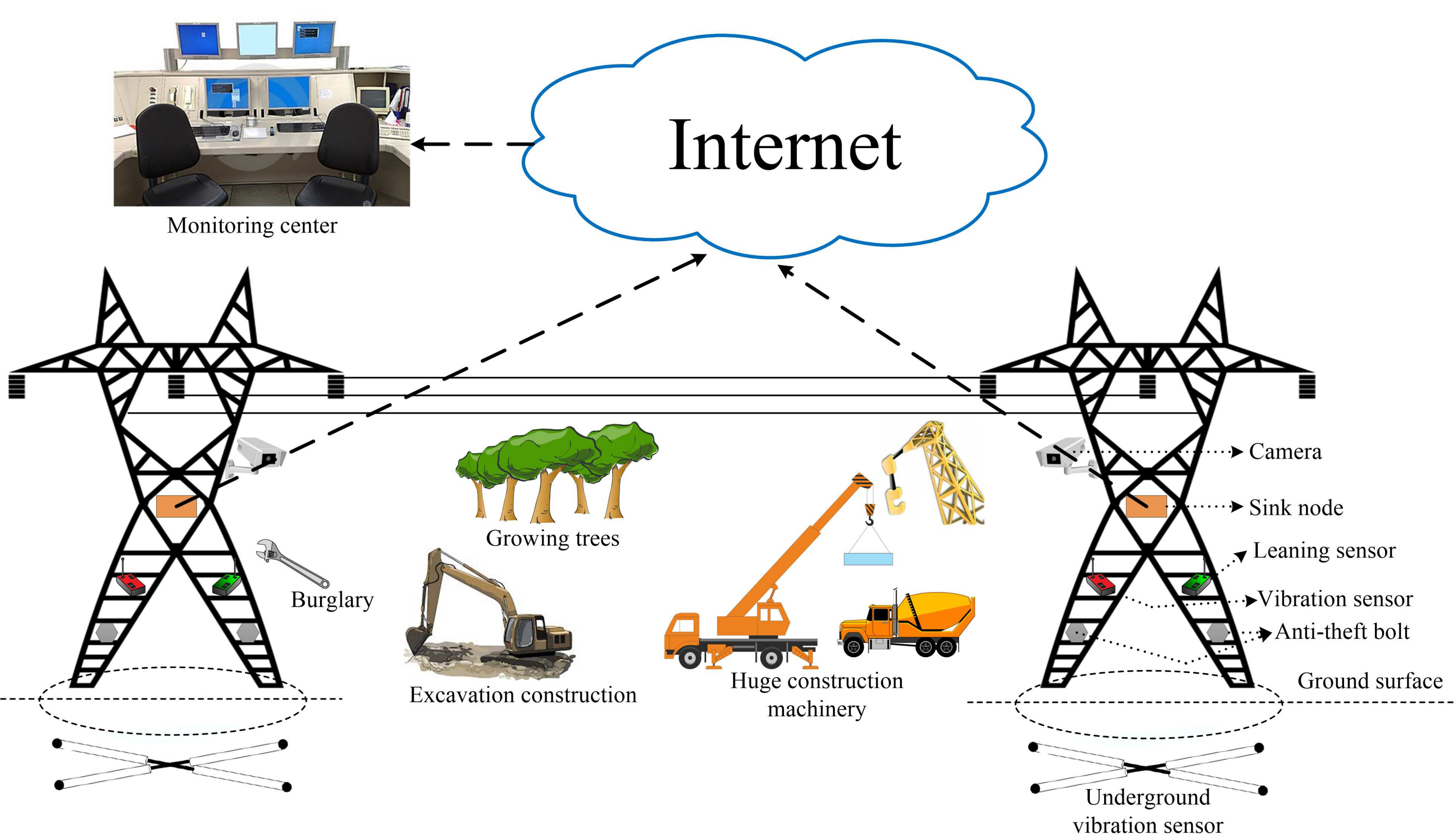}
\caption{IoT-aided transmission tower protection system for the safety of transmission towers from the threats of buglary, natural disasters, barbaric construction and growing trees. This system is comprised of a sink node and various sensors which generate early warnings to the monitoring centers about the threats to high voltage transmission towers. The sensors include two vibration sensors (one underground in the base of tower and other on the tower), anti-theft bolts, leaning sensor and video cameras \cite{zhen2012transmission}.}
\label{fig:transmission_tower_protection}
\end{figure*}

Currently, the main method of transmission towers protection is manual patrol by the staff. However, regular manual patrol of high voltage transmission lines and towers by staff is very difficult due to manpower realities, divisions of responsibility and the level of knowledge of the staff. Furthermore, some transmission towers are difficult to approach due to their physical positioning. Hence, the patrolling quality cannot be guaranteed. The patrolling period varies from 1-10 weeks, which means insufficient monitoring and higher security risks. While it is true that some equipment, such as cameras and infrared alarms, are installed on transmission towers to monitor burglary and other potential damage, the accuracy and stability of this equipment is not yet satisfactory \cite{zhen2012transmission}.

With the help of WSNs, IoT technology can provide remote monitoring in addressing these security threats. The IoT-aided transmission tower protection system contains various sensors which generate early warnings of threats to high voltage transmission towers, enabling quick responses. The sensors include vibration sensors, anti-theft bolts, a leaning sensor and a video camera. These sensors and the sink node form a WSN \cite{zhen2012transmission}. The sensors detect any threat, and send the relevant signals to the sink node. The sink node receives these signals from the sensors, processes them into data and transmits the data to the monitoring center through the Internet or any other public/private communication network.  

The sensors' deployment is presented in Fig. \ref{fig:transmission_tower_protection} and is as follows. The anti-theft bolts are deployed on the lower part of the tower. One vibration sensor is deployed underground in the base of the transmission tower and the other is deployed on the tower, about 3-5 meters above the ground. The leaning sensor is deployed close to the vibration sensor on the tower. The camera is installed on the tower about 6-8 meters high, directed towards the transmission line. Finally, the sink node is deployed in the middle of the transmission tower. 


The vibration sensors monitor the vibration signals of the ground and tower. When they detect signals that indicate excavation construction close to the transmission tower, they transmit those signals to the sink node. The vibration sensor on the tower, the leaning sensor and the anti-theft bolts can detect the signals of a burglary or vandalism in progress, and subsequently transmit those signals immediately to the sink node. In both cases, the sink node combines and processes the signals, and  triggers the video cameras to send real-time images to the sink if the threat is identified. For huge construction machinery and trees that get too close to the transmission lines, the video cameras directed towards the transmission lines identify the threat and transmit real-time images to the sink. In all of the threat cases, the sink generates an alarm and forwards the images to the monitoring center through the Internet. The monitoring center handles the real-time data of a series of transmission towers. The alarm signal and images from the sink inform the monitoring center staff about any threat to a transmission tower and the staff then takes appropriate actions to handle such threats.

\subsubsection{Online Monitoring of Power Transmission Lines}

The online monitoring of power transmission lines is one of the most important applications of the IoT in the SG, specifically for disaster prevention and mitigation. In recent years, natural disasters have highlighted the challenges of security, reliability and stability inherent to high voltage power transmission lines. Traditionally, high voltage transmission line monitoring has been performed manually. Sensors measuring conductor galloping, wind vibration, conductor temperature, micro-meterology and icing can now  be used to  achieve real-time online monitoring of power transmission lines \cite{chen2011integration}. This new online power transmission line monitoring system is comprised of two parts. In the first part, the sensors are installed on the power transmission lines between transmission towers to monitor the states of power transmission lines. The second part has sensors installed on the transmission towers in order to monitor their states and their environmental parameters. IoT enables the communication between the power transmission line sensors and the transmission tower sensors.


\subsection{Summary and Insights}
We have comprehensively surveyed the existing applications of IoT-aided SG systems. While there could be many applications of IoT-aided SG systems, as presented in Fig. \ref{fig:applications_iot_sg}, to date there is very little work on these applications. Some applications of IoT-aided SG systems are already deployed, but many more are yet to materialize, when the full capabilities of instantaneous knowledge and massive data processing are exploited. Applications today tackle several areas of interest, such as:

\begin{itemize}
\item Surveillance of premises or of power equipment installations (towers and power transmission lines);
\item Adjusting home consumption by dynamic scheduling, which takes advantage of fluctuating pricing;
\item Meter reading and consumption monitoring, residential and commercial;
\item Electric Vehicle charging and parking;
\item Power demand and supply management, including integrated renewable energy sources; and
\item Maintenance of power supply systems, by detecting line faults and failures.  
\end{itemize}

This survey reveals that there is little literature on the application of IoT-aided SG systems in the area of power generation, transmission and utilization. The potential of renewable energy may depend, for example, on IoT-based prediction of weather conditions that can regulate energy flow between different regions, and on monitoring the efficiency of the involved equipment (solar panels, wind turbines).  

\section{Architectures for IoT-aided SG Systems}
\label{sec:architectures_iot_sg}

A number of architectures are available for IoT-aided SG systems. We survey such architectures in this section, and their classification is represented in Fig. \ref{fig:architectures_iot_sg}. We start by presenting and discussing SGAM (Smart Grid Architecture Model) which is designed for SG planning and is the SG reference architecture by EU Mandate M/490 \cite{m490}. Then we present existing IoT-aided SG architectures.

\subsection{Smart Grid Architecture Model (SGAM)}
The SGAM is a reference architecture which aim is to illustrate the use cases of SG in an architectural point of view. It is an outcome of the EU Mandate M/490's reference architecture working group \cite{bruinenberg2012cen, trefke2013smart}. The SGAM is mainly comprised of five layers: business, function, information, communication and component layers. These are referred as interoperability layers. Each interoperability layer is comprised of smart grid plane, which covers electrical domains and information management zones. The main purpose of this model is to represent which domains interact with each other over which zones of information management \cite{bruinenberg2012cen}. For better understanding, we can divide SGAM into SGAM SG planes and SGAM interoperability layers. 

\subsubsection{SGAM SG Planes}
In power management system, it is important to differentiate between electrical process and information management. This differentiation can be achieved by dividing into the physical domains of the electrical energy conversion chains and the hierarchical zones for the management of electrical processes. Hence, the SG plane represents which domains interact with each other over which levels (or hierarchical zones) of information management. The domains or energy conversion chains includes: generation, transmission, distribution and DER. The customer premises include the end users, as well as the producers of electricity. The hierarchical zones for the management of electrical processes include process, field, station, operation, enterprise and market \cite{trefke2013smart}.

\subsubsection{SGAM Interoperability Layers}
The SGAM is divided into five interoperability layers for better presentation and handling of the architecture model. The rest of this section discusses each interoperability layer \cite{bruinenberg2012cen}.  

\paragraph{Business Layer}
It describes the business-related aspects to SG on the information exchange and deals with organizational entities, business processes, business capabilities and regulatory conditions. In this manner, it helps business executives in decision making of new or existing business models, projects or use cases, as well as defining new market models through regulatory conditions.

\paragraph{Function Layer}
It represents functions, services and their relation from an architectural point of view. The functions are defined independent of the physical implementations and actors in systems, applications and components. The functions are mainly extracted from the use case functionality.

\paragraph{Information Layer}
It deals with the information exchanged among services, functions and components. It is comprised of the underlying data models and the information objects. The underlying data models and the information objects enable interoperable information exchange by providing the common semantics for services and functions via communication means.

\paragraph{Communication Layer}
It provides the mechanisms and protocols for interoperable information exchange between the components (e.g., the underlying service, function or use case) and the data models or information objects.

\paragraph{Component Layer}
It deals with the physical distribution of all the participating components in the SG which involves applications, actors, equipment, devices, communication infrastructure and servers. They communicate among themselves through the information objects and the communication protocols.

\subsection{Three-layered Architecture}

\begin{figure}[!b]
\centering
\includegraphics[width=0.5\textwidth]{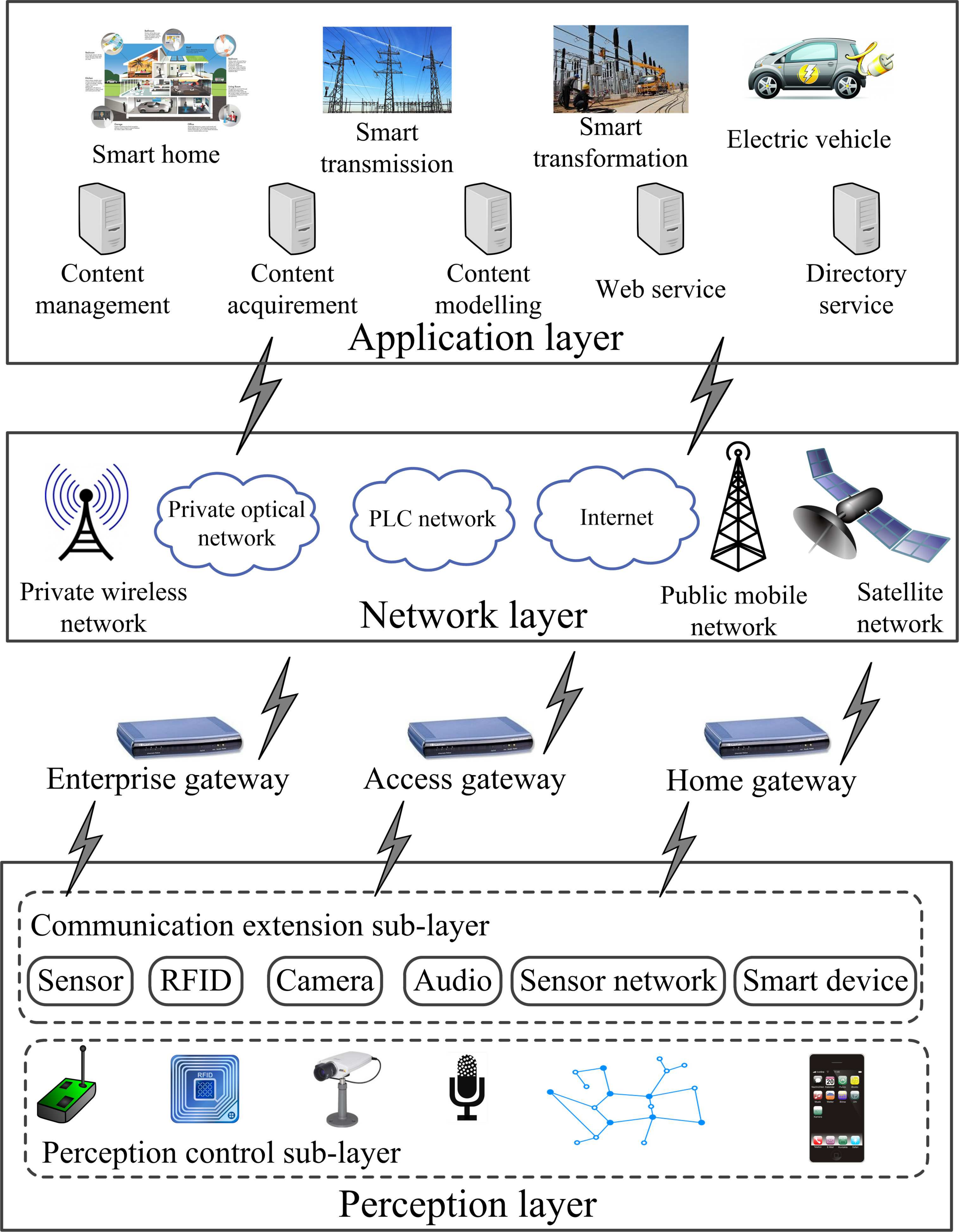}
\caption{Three-layered architecture of IoT-aided SG systems, comprised of a perception layer, a network layer and an application layer. The perception layer consists of two sub-layers, a communication extension sub-layer and a perception control sub-layer.}
\label{fig:architecture_three_layered}
\end{figure}

A three layered architecture for IoT-aided SG systems, based on the characteristics of the IoT, has been widely used in the literature and it has been proposed in \cite{song2015research, wang2014research, chen2012application, liu2011applications, chen2011integration, ou2012application, zheng2012research}. As presented in Fig. \ref{fig:architecture_three_layered}, this architecture is comprised consists of three layers, a perception layer, a network layer and an application layer.

\subsubsection{Perception Layer} 
This layer enables the main objectives of sensing and collecting information in IoT-aided SG systems using a variety of devices. It is comprised of various kinds of IoT sensing devices, such as RFID tags, cameras, WSN, GPS and M2M devices, in order to collect data in a SG. It is categorized into two sub-layers, a perception control sub-layer and a communication extension sub-layer. The perception control sub-layer realizes the perception of the physical world by processing IoT devices, information acquisition, monitoring and control, while the communication extension sub-layer has a communication module which connects IoT devices with the network layer. 

\subsubsection{Network Layer} 

The network layer is comprised of the converged network formed by various telecommunication networks and the Internet \cite{ou2012application}. The network layer has been widely accepted due to its mature technologies. Its function is to map the information collected by the IoT devices in the perception layer to the telecommunication protocols \cite{song2015research}. Subsequently, it transmits the mapped data to the application layer through the relevant telecommunication network. The core network, i.e., the Internet is responsible for the routing, information transmission, and control. The access network will be based on other telecommunication networks. The IoT management and information centers also belong to the network layer. The network layer can rely on public as well as on industry-specific communication networks. 

\subsubsection{Application Layer}
The application layer is the integration of IoT technology and industry expertise for the realization of a broad set of IoT-aided SG applications \cite{ou2012application}. Its function is to process the information received from the network layer, and based on this information, it monitors and troubleshoots the IoT devices and SG environment in real-time. It provides various applications of IoT-aided SG systems which are presented in Fig. \ref{fig:applications_iot_sg}. It is comprised of application infrastructure/middleware and various types of servers related to content, web services and directory services. The application infrastructure/middleware provides computing, processing and resources for IoT technology. The key elements provided by the application layer are information sharing and security. The application layer is set to grow, especially for SGs that can provide much richer data sets. In turns, these applications dictate what data is required, at what time intervals, from the sensors.

\subsection{Four-layered Architecture}
\begin{figure}[!t]
\centering
\includegraphics[width=0.4\textwidth]{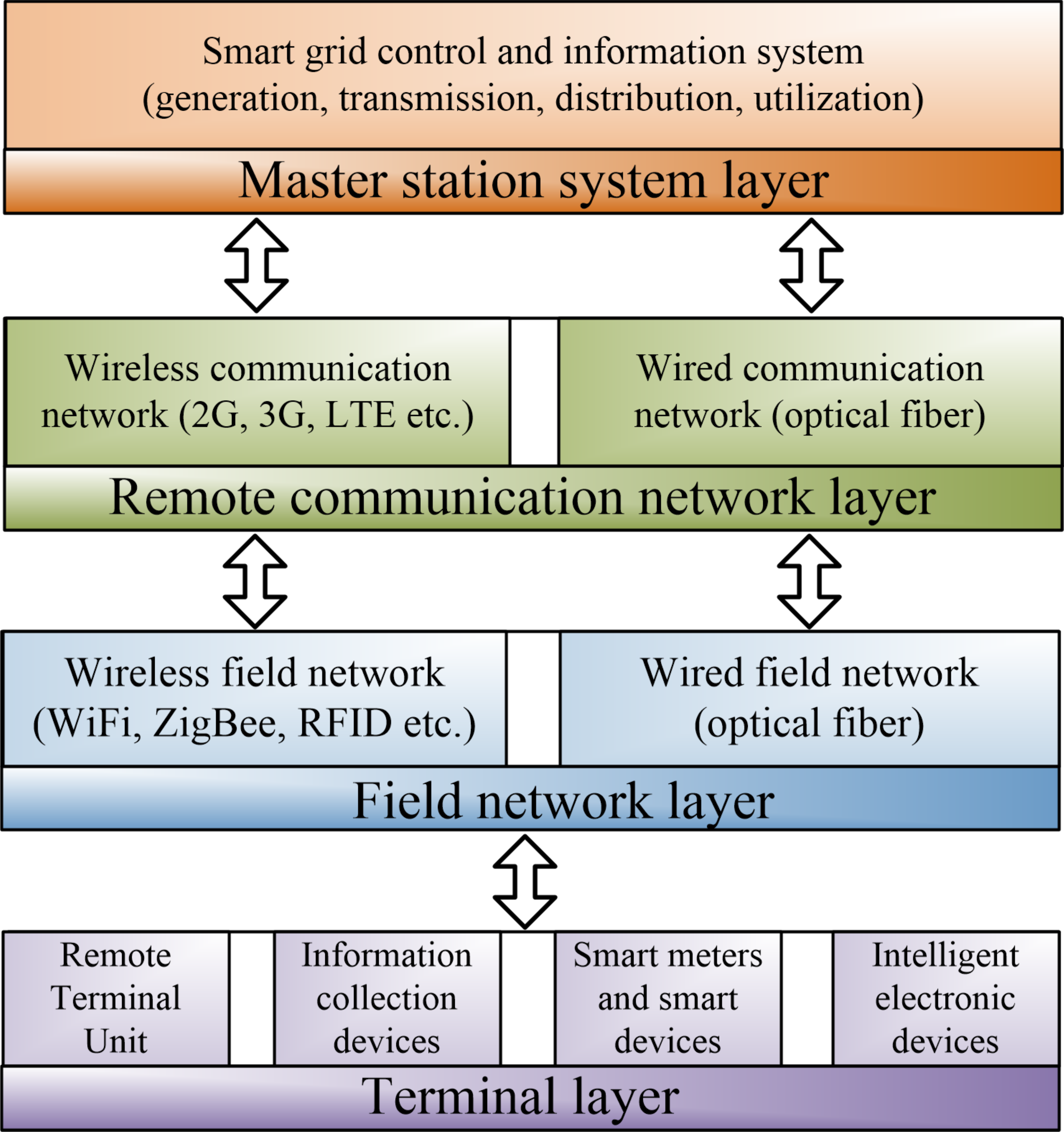}
\caption{Four-layered architecture of IoT-aided SG systems based on SG characteristics. It is comprised of a terminal layer, a field network layer, a remote communication layer and a master station system layer \protect\cite{wang2012research}.}
\label{fig:architecture_four_layered}
\end{figure}

A four-layered architecture for IoT-aided SG systems based on the characteristics of SG information and communication systems was proposed in \cite{wang2012research}. As presented in Fig. \ref{fig:architecture_four_layered}, this architecture is comprised of a terminal layer, a field network layer, a remote communication layer and a master station system layer. According to the three-layered IoT hierarchical model, the terminal and field network layers in this architecture correspond to the IoT perception layer, the remote communication layer corresponds to the IoT network layer, and the master station system layer corresponds to the IoT application layer. 

The terminal layer is comprised of IoT devices deployed in various SG functions, such as power generation, transmission, distribution and utilization. The IoT devices include remote terminal units, information collection devices, smart meters, smart devices and intelligent electronic devices. This layer collects information from IoT devices and transmits the collected data to a field network layer. The field network layer can be wired or wireless. Depending on the type of IoT devices, the appropriate communication network is used. As an example, ZigBee is used by sensor nodes in order to transmit the collected data to a remote communication network layer. 

The remote communication network layer can also be wired or wireless. It is comprised of various communication networks which provide connectivity to the Internet, such as 2G, 3G, and LTE as wireless networks, and optical networks as a wired network. This layer serves as a middleware between IoT devices and the master station system layer. The master station system layer is the control and information system of a SG. It controls and manages all the SG functions. It can also be considered as an interface to the IoT-aided SG applications.

\subsection{Cloud-based Architecture}
Improving a building's energy efficiency is an important aspect of the SG, an aspect that is required for global sustainability. Thus, smart energy is an important research area of the IoT. In the United States, buildings are responsible for around 71\% of the total electrical energy consumption \cite{usgbc2007national}, a major motivation for designing green buildings to be energy efficient. However, thus far, such green buildings have not been as energy efficient as expected \cite{pan2015internet}. This may be due to the ill-fitting of rigid and static schedules to the lifestyle of consumers, or to the unpredictability of business requirements. Now, with the help of IoT technology, the ability to change energy settings can be entrusted to the users, using their smartphones or PCs. This allows users to suit their own schedules at will, and respond to events instantaneously by adjusting the policies as and when necessary. In 2015, an energy efficient location-based automated energy control IoT framework was proposed, using smartphones and cloud computing technologies \cite{pan2015internet}. That proposed framework changes the currently static energy management and centralized control modes to dynamic and distributed energy control at the consumer-side SG, which is composed of usually various buildings.

\begin{figure}[!b]
\centering
\includegraphics[width=0.5\textwidth]{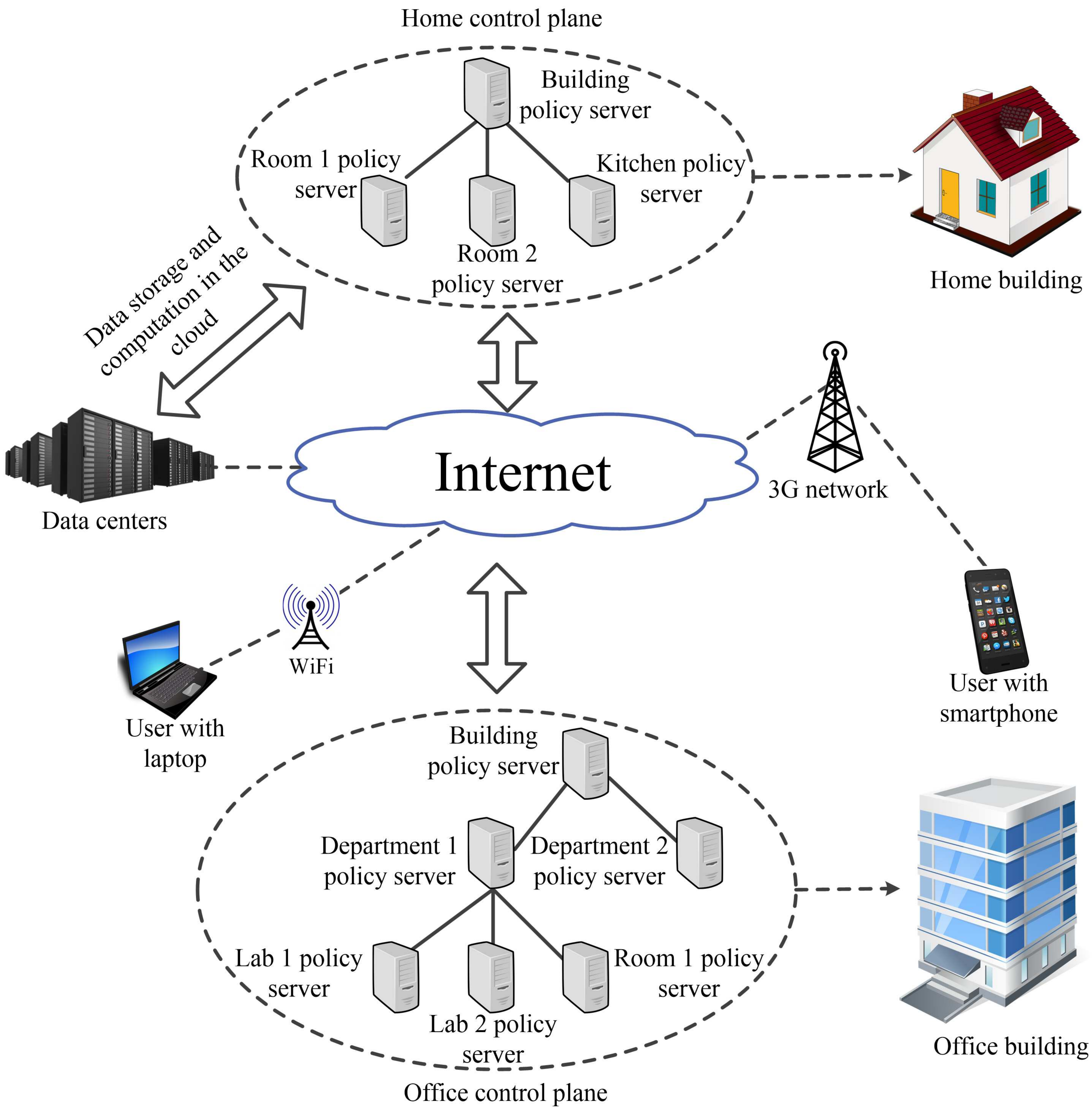}
\caption{Smart location-based automated energy control IoT framework for energy efficiency, changing the current static and centralized energy control to dynamic and distributed energy control. It is comprised of four main components, i.e., multi-source energy saving policies, monitoring and control via mobile devices, location-based automatic control, and a cloud computing platform for data storage and computation \protect\cite{pan2015internet}.}
\label{fig:framework_energy_efficiency}
\end{figure}

This framework is represented in Fig. \ref{fig:framework_energy_efficiency} and is comprised of four main components, (i) multi-source energy saving policies, (ii) monitoring and control via mobile devices, (iii) location-based automatic control, and (iv) a cloud computing platform for data storage and computation. An organization's premise consists of various parts, such as a campus, buildings, departments, labs and rooms/offices, each with different energy requirements and policies that are important to take into account while managing energy consumption. Even in a single home, each family member has his/her own energy consumption preference which needs to be considered. Hence, this framework has a tree-like structure control plane as presented in Fig. \ref{fig:framework_energy_efficiency}, comprised of several layers of energy saving policies at different levels (e.g., building, department, lab and room). 

Smartphones are equipped with multiple networking interfaces, such as Wireless Fidelity (WiFi), 3G, Long Term Evolution (LTE), bluetooth and WiMAX, as well as GPS, which helps to obtain precise location positioning. Thanks to the global Internet access, smartphones are the ideal choice for monitoring and controlling energy control systems remotely from any place at any time. After an initial authentication and authorization, consumers can dynamically modify their energy saving policies through their smartphones by interacting with the policy servers of their home and their office buildings. The consumer can be connected to the Internet through smartphones, tablets, or laptops with WiFi or with 3G networks.

Furthermore, the location information of smartphones is used in designing automatic control policies which can 
switch energy consuming devices ON or OFF in homes and office buildings by detecting the location and direction of user movements. These dynamic adjustment policies also enable the coordination between buildings' policies. For example, as presented in Fig. \ref{fig:framework_energy_efficiency}, when a user has moved out of a predefined distance range from the home premises and is moving into a predefined distance range of an office building, a message will be sent to the policy server to trigger the policy control process, e.g., start the heating. The now deserted home will transit into an energy saving mode while the office building will start the processes based on the user's preferences, such as cooling/heating to the user's desired temperature.

Finally, this framework uses the cloud computing platform for data storage and modeling computation and analysis. The cloud provides the basic data storage and retrieval services for each building's energy consumption data. Most of the computation-intensive modeling and analysis jobs are performed in the cloud. Moreover, the cloud provides security, reliability and configurability for the network communication between the cloud and the user. The cloud has an application layer which contains a user friendly web interface for managing the buildings' energy systems, so that a remote user can easily configure and manage the system. 

\subsection{Web-enabled SG Architecture}
\begin{figure}[!thbp]
\centering
\includegraphics[width=0.5\textwidth]{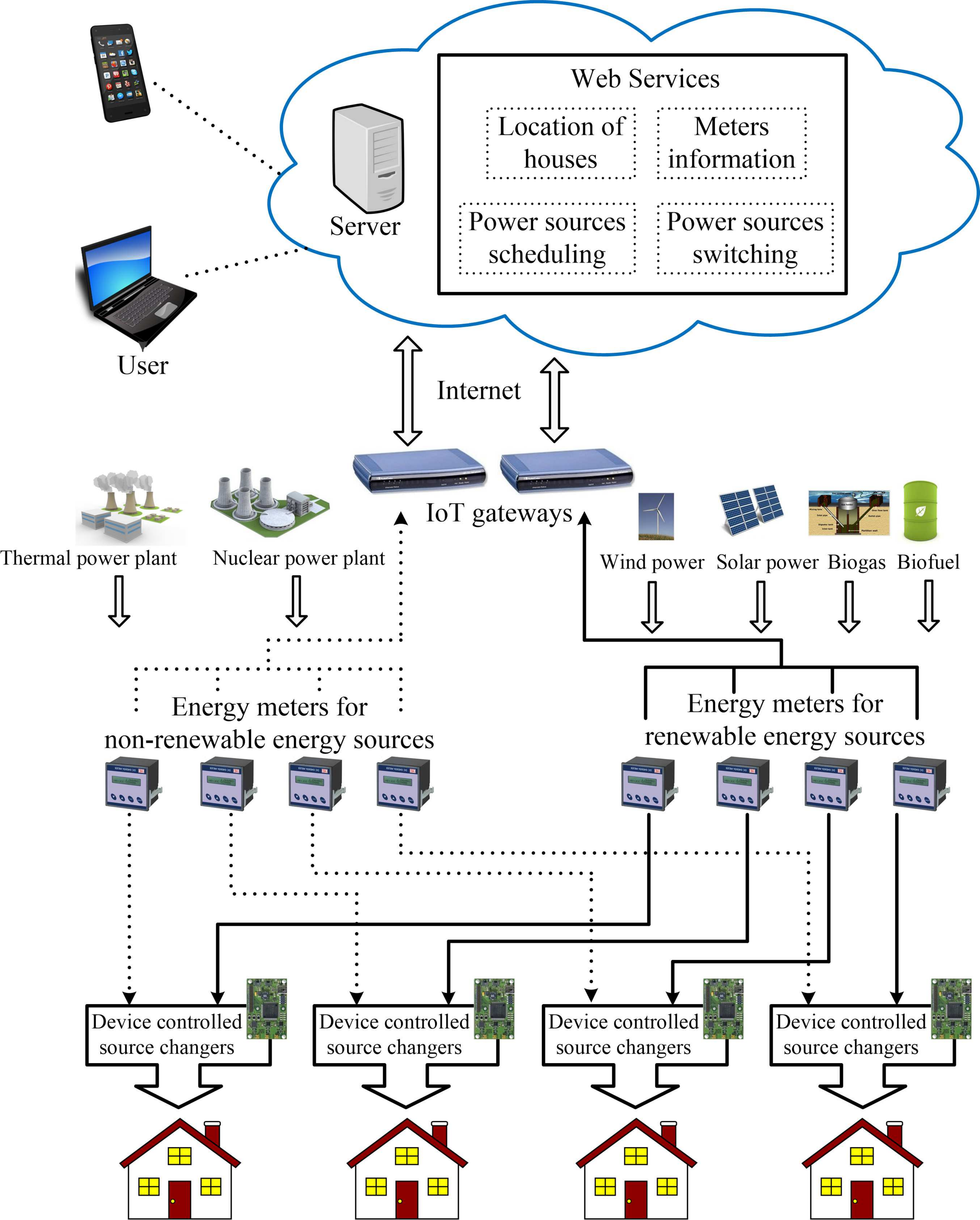}
\caption{Web-enabled SG architecture comprised of web services on the top of IoT devices \protect\cite{mohanty2014implementation}. It has two types of energy sources, renewable and non-renewable, both of which are connected to digital energy meters. The digital energy meters collect household energy consumption data which is further collected by IoT gateways to communicate to the server containing the web services.}
\label{fig:web_enabled_sg}
\end{figure}

An architecture for IoT-aided SG systems is proposed in \cite{mohanty2014implementation} based on the web of things (Fig. \ref{fig:web_enabled_sg}). The web of things is comprised of a number of web services provided on top of the IoT devices in which the web browser acts as an interface to these web services. There are two types of energy sources, non-renewable and renewable energy sources. Non-renewable energy sources include thermal power plants (combusting coal or oil) that release carbon emissions to the environment, as well as nuclear power plants. The renewable energy sources are environmental friendly and are comprised of hydropower, wind farms, solar, biogas plants and biofuel sources, as well as geothermal and tidal/wave sources. 

The energy sources in this architecture are connected to individual digital energy meters. These digital energy meters are responsible for collecting household energy consumption data. The meter readings from energy meters of non-renewable and renewable sources are collected by separate IoT gateways which communicate regularly with these meters. The collected data from IoT gateways are updated to the server periodically, and the server provides web services on top of these IoT devices. These web services include the locations of houses connected through the SG, and the meter information. Moreover, for each home, the scheduling of power sources and the controlling of the energy sources by switching the source controllers remotely through IoT devices are provided as web services. Through the Internet, by connecting to any device, a user can access these services. The energy sources for each household are switched through source changers which are controlled through IoT devices; these IoT devices switch the energy sources upon receiving instruction from the user via the server. 

\subsection{Last-meter SG Architecture}
\begin{figure}[h]
\centering
\includegraphics[width=0.5\textwidth]{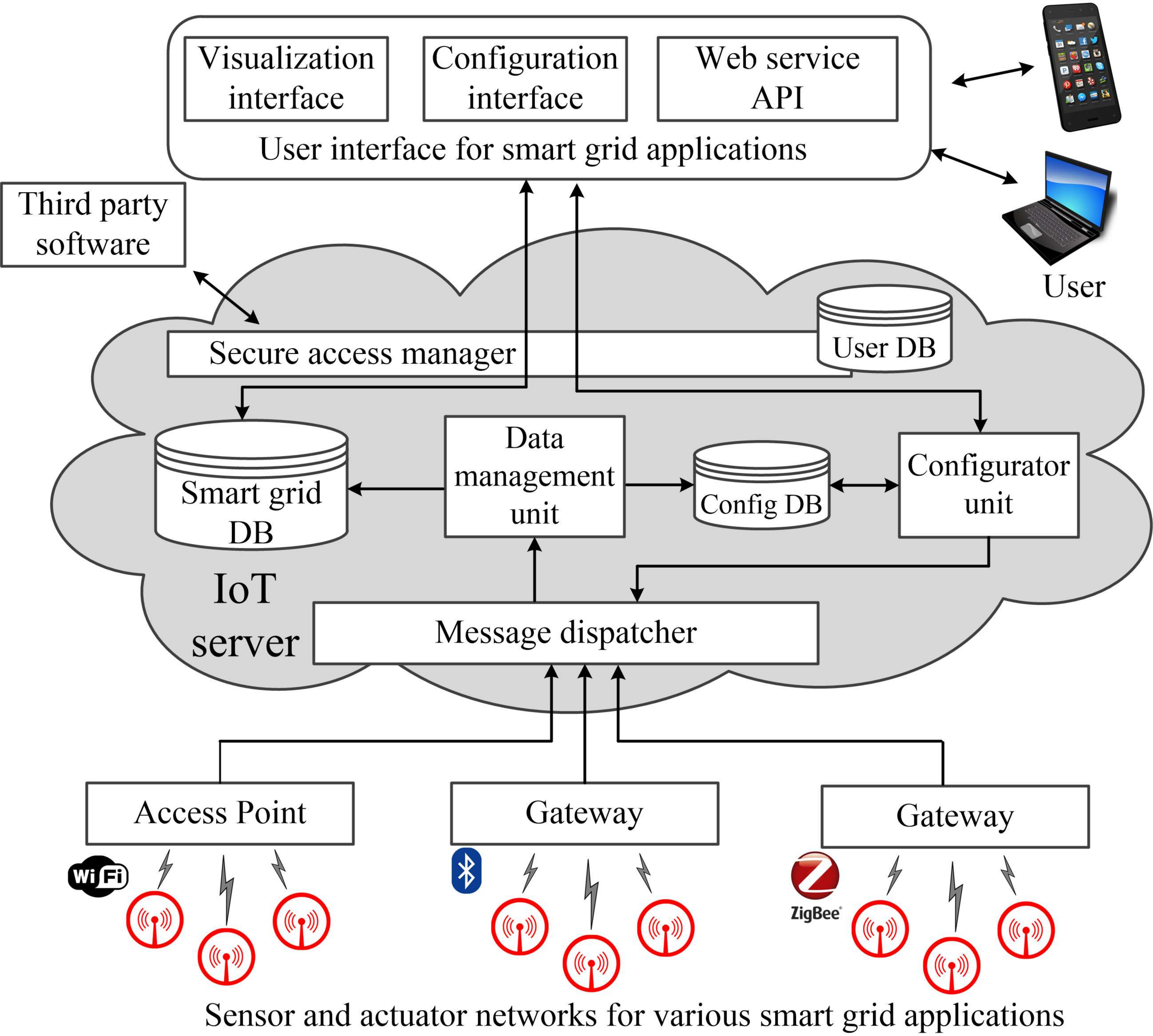}
\caption{Last-meter SG architecture embedded in the IoT \protect\cite{spano2015last, spano2013intragrid}. It is comprised of three main components, sensor and actuator networks, an IoT server and user interfaces. The sensor and actuator networks consist of sensor and actuator nodes, and IP gateways. The IoT server consists of a message dispatcher, an SG database, a data management unit, a configuration database, a configurator unit, a user database and a secure access manager. The user interfaces consist of a visualization interface, a configuration interface and a web service API.}
\label{fig:last_meter_architecture}
\end{figure}

The last-meter SG is a part of the SG that is closer to the home, i.e., the part with which the consumers interact. An architecture for the last meter SG embedded in the IoT is proposed in \cite{spano2015last, spano2013intragrid} and is presented in Fig. \ref{fig:last_meter_architecture}. This architecture is comprised of three main components, sensor and actuator networks, an IoT server and user interfaces.

\subsubsection{Sensor and Actuator Networks for Various SG applications}
The sensor and actuator networks enable the realization of various SG applications and contain two main components: sensor and actuator nodes, and IP gateways. 

\paragraph{Sensor and actuator nodes}
The sensor and actuator nodes can be part of wired or wireless networks. This architecture can accommodate heterogeneous sensor networks, as the data management unit in the IoT server translates information into the format required by the SG database. The bi-directional node communication enables an IoT server to configure, program and interrogate the sensor and actuator nodes. This architecture supports the possibility of adding or removing network components in real time, even if specific node characteristics depend upon the network implementation. Hence, any node can join the network without changing the network implementation and the newly added node is automatically identified and accessible from the network administration interface for registration and configuration. 

Each sensor and actuator node has to be uniquely identified in order to ensure global accessibility. However, the node addresses in sensor and actuator networks are generally only unique within a single network and may change over time. Therefore, the IoT server assigns a unique ID to each node of the network and maintains the mapping of each ID and its corresponding network address, provided by the local network coordinator. Subsequently, when a node sends a message to the IoT server, the gateway node translates its network address into a unique ID, and vice versa.

\paragraph{IP gateway}
The IP gateway connects sensor and actuator nodes to the IoT server via an IP link if there is no IP capability. For uplink communication, the gateway collects data from sensor/actuator nodes, performs encapsulation/reformatting (if required) and forwards the data to the message dispatcher in the IoT server over a secure TCP/IP link. For downlink communication, it forwards the commands received from the IoT server to the intended receiver nodes. In general, the gateway performs a conversion of data into a universal format. However in this architecture, such conversion is performed by the IoT server, and therefore the gateway sends the messages over a TCP/IP link in the native format. This procedure has two main advantages. Firstly, the gateway will have low computational complexity and hardware requirements. It only has to ensure an IP connection, encapsulate the messages into TCP/IP packets and ensure the security level required by specific applications. Secondly, new functionalities and applications can be developed and added without modifying the gateway. 

\subsubsection{IoT Server}
The IoT server has four main components: the first one is a message dispatcher, the second one is a data management unit and database storage, the third one is a configurator unit and database, while the fourth one is a secure access manager and user database.

\paragraph{Message dispatcher}
To perform bi-directional communication between each gateway and the rest of the system, the message dispatcher will be used. The message dispatcher will listen to new connections from IP nodes that want to join the system for uplink communication. It decrypts incoming packets for every connection and forwards them to the data management unit for storage and interpretation. For downlink communication, it encrypts the messages from the configurator unit into a TCP message and then forwards them to the destination gateway. 

\paragraph{Data management unit and database storage}
The data management unit is a collection of software modules. In this unit, each software module is able to manage the messages of a specific sensor/actuator network. These modules receive messages in their native format and then extract their payload. Two storing mechanisms are performed based on the payload. Firstly, the messages are stored in a unique format in the SG database if the payload contains either measurement data from a sensor node or a notification from an actuator node. Secondly, the message is stored in the original format into the configurator database if the payload contains specific network messages (such as configuration, management information, a node address or a communication channel).

The SG database decouples the data collection from the data processing and visualization so that the users do not have to interrogate the nodes directly. The decoupling allows the nodes to stay in sleep mode most of the time and to only wake up periodically to receive commands and configuration messages, as well as to send measurement and status data. This approach is very useful when the sensor networks are heterogeneous, as well as when nodes are battery powered. The sensors' data in a SG database is represented by a unique format that is associated with the physical nodes through the unique node ID, independent of the local sensor network protocol. In this manner, data can be processed and visualized independently of the characteristics of the physical source.

\paragraph{Configurator unit and database}
The configurator unit is a collection of software modules, where each module is dedicated to a specific type of sensor/actuator network. This unit configures the nodes and networks according to the inputs from applications and users, as well as according the stored system status in the configurator database. 

\paragraph{Secure access manager and user database}
This component coordinates all the communication between the end users and the IoT server by ensuring privacy and data protection. It only allows authorized users and third-party applications to access the stored information and the network configuration based on the user database, and the users' access permission for each resource. The network owners have administrative privileges over their networks by default.

\subsubsection{User interfaces}
The user interfaces allow users, application developers and service providers to interact with an IoT server. The user interfaces offer different functionalities to standard and administrative users. Standard users can only access the energy consumption data pertaining to their own households. Administrative users, meanwhile, have higher access privileges and they can also view the configuration and status of the IoT devices, and configure them dynamically. The user interface is divided into three main components: a visualization interface, a configuration interface and a web service API.

\paragraph{Visualization interface}
It displays the current and past information of household energy consumption. Additionally, it allows authorized users to send commands to actuator nodes attached to home appliances.

\paragraph{Configuration interface}
It enables the users to remotely manage and configure their networks. Additionally, users can set the data visibility of their household appliances as well as third-party applications access. This interface also allows administrators to remotely register new gateways and configure new network connections. To establish a new connection, each gateway requires the IP address and network name of the message dispatcher, the port number on which the message dispatcher can accept the connections and the network AES security key. 

\begin{figure*}[!th]
\centering
\includegraphics[width=0.8\textwidth]{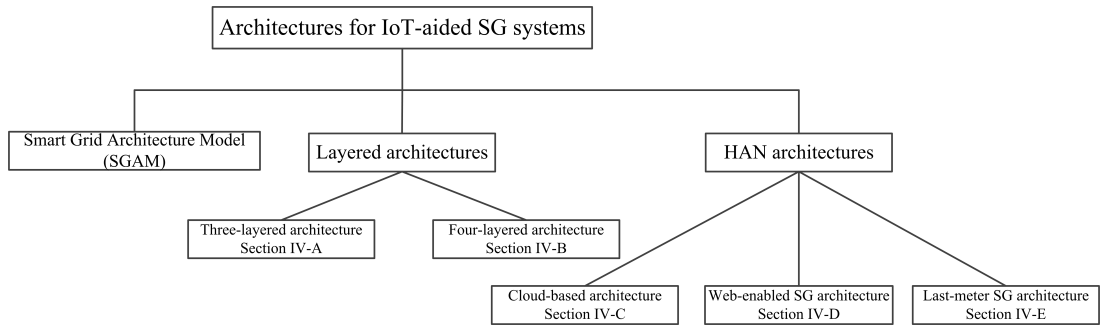}
\caption{Architectures for IoT-aided SG systems classified into layered architectures and HAN architectures.}
\label{fig:architectures_iot_sg}
\end{figure*}

\paragraph{Web service API}
It opens the platform (IoT server) to service providers and new client applications. 
APIs offer an easy and unified way to retrieve information collected from heterogeneous sources. Utilities, service providers and third-parties use web service APIs to obtain single, multiple or aggregated measurement data, as well as to notify consumers about dynamic changes in tariffs, of weather related data and alarms. For security and privacy purposes, only registered users and authorized third-party applications can access energy consumption data from an SG database through web service APIs.

\subsection{Summary and Insights}

We have surveyed the existing architectures for IoT-aided SG systems as summarized in Table \ref{tab:architectures_iotsg}. As presented in Fig. \ref{fig:architectures_iot_sg}, the major focus of the existing studies is either generic layered architectures or the HAN architectures specifically for controlling and managing home appliances remotely. 
However, the layered architectures (three-layered and four-layered) are very generic architectures and they do not cover many specific aspects of SG, such as all the networks (HAN, NAN and WAN) and systems (power generation, transmission, distribution and consumption). Hence, it is very important to consider these specific aspects in an architecture for IoT-aided SG systems. Additionally, four-layered architecture is an extension of three-layered architecture by dividing the network layer into field network layer and remote communication network layer. It is very important to analyze why there is a need of four-layered architecture and why we should not have five-layered or six-layered architectures.

The other three architectures (cloud-based, last meter SG and web-enabled SG) are mainly designed for HAN. The cloud-based architecture saves energy by using smart location-based automated energy control framework, the last-meter SG architecture aims at automating the HAN of SG using IoT technology and the web-enabled SG architecture also connects HAN to the web by offering a number of web services on the top of IoT devices. However, these current studies on architectures for IoT-aided SG systems are very limited; there are still no architectures for NAN and WAN of IoT-aided SG systems, and other areas of HAN are also yet lacking architectures. Hence, we strongly recommend future research on architectures for NAN, WAN and the other HAN areas, which should focus on the individual characteristics and requirements of HANs, NANs and WANs. Subsequently, the individual architectures for HAN, NAN and WAN will need to be combined as a whole for the realization of complete IoT-aided SG systems. Since, IoT and non-IoT communication technologies play a vital role in the realization of IoT-aided SG systems in order to provide connectivity, it is very important to consider communication technologies and topologies in the overall architectures for IoT-aided SG systems.

\begin{table*}[hbtp]\scriptsize
\centering
\caption{Summary of architectures of IoT-aided SG systems.}
\label{tab:architectures_iotsg}
\begin{tabular}{|p{2.5cm}|p{1.25cm}|p{2.25cm}|p{4cm}|p{6.5cm}|}
\hline
\bfseries Architecture & 
\bfseries Reference & 
\bfseries Category & 
\bfseries Advantages &
\bfseries Characteristics \\
\hline

Smart Grid Architecture Model (SGAM) & 
\cite{bruinenberg2012cen, trefke2013smart} & 
Reference architecture & 
Offers a reference architecture for future development.  &
A reference architecture and an outcome of EU Mandate M490 reference architecture working group. Comprised of five layers (or interoperability layers): business, function, information, communication and component layers. SGAM is divided into SGAM SG planes and SGAM interoperability layers. \\
\hline

Three-layered architecture & 
\cite{song2015research, wang2014research, chen2012application, liu2011applications, chen2011integration, ou2012application, zheng2012research} & 
Layered architecture &  
Offers an architecture based on the characteristics of IoT.  &
Based on the characteristics of the IoT. Comprised of three layers: perception, network and application layers. \\
\hline

Four-layered architecture & 
\cite{wang2012research} & 
Layered architecture & 
Offers an architecture based on the characteristics of SG. &
Based on the characteristics of SG information and communication systems. Comprised of: terminal, field network, remote communication and master station layers. Terminal and field network layers correspond to perception layer in three-layered architecture; remote communication layer corresponds to network layer; and master station system layer corresponds to application layer. \\
\hline

Cloud-based architecture & 
\cite{pan2015internet} & 
HAN architecture &
Saves energy by using smart location-based automated energy control framework. & 
Comprised of four main components: multi-source energy saving policies, monitoring and control via mobile devices, location-based automatic control, and a cloud computing platform for data storage and computation. Offers tree-like structure for control plane. \\
\hline

Web-enabled SG architecture & 
\cite{mohanty2014implementation} & 
HAN architecture & 
Connects HAN to the web by offering a number of web services on the top of IoT devices.  &
Based on the web of things and comprised of a number of web services on top of IoT devices. Deals with two types of energy sources: non-renewable and renewable. \\
\hline

Last-meter SG architecture & 
\cite{spano2015last, spano2013intragrid} & 
HAN architecture & 
Aims at automating the HAN of SG using IoT technology.  &
Part of the SG that is closer to the home with which the consumers interact. Comprised of three main components: sensor and actuator networks, IoT server and user interfaces. \\
\hline

\end{tabular}
\end{table*}

\section{Prototypes for IoT-aided SG Systems}
\label{sec:prototypes_experimentation_iot_sg}

Prototypes play an important role in the development of a system in order to test various functions and verify the operation before the actual commercial implementation. Some prototypes have been developed for IoT-aided SG systems as well, however they are very few and there is a need to develop more prototypes for IoT-aided SG systems. In this section, we discuss the existing prototypes of IoT-aided SG systems.

\subsection{A Simple Prototype for Energy Efficiency}
\label{sec:prototype_energy_efficiency}

A simple prototype for smart energy involving one user was designed in a Leadership in Energy and Environmental Design (LEED)-gold-certificated green office building. It linked electrical appliances in two locations, an apartment and an office in an office building \cite{pan2015internet}. This prototype is a simplified scenario of the framework presented in Fig. \ref{fig:framework_energy_efficiency}. The objective of this prototype was to enable a user to dynamically control and manage appliances in two locations. The prototype enables the server to trigger energy policy control process by turning electrical appliances ON or OFF in the two locations after detecting changes in the user's location. This helps users to effortlessly implement and control their own energy policies in real time, making their energy consumption directly proportional to their actual utilization. Although this simple prototype only involves one user with control devices in two locations, this system could be easily scaled up with multiple users controlling their devices at different locations simultaneously.

\begin{figure}[!thbp]
\centering
\includegraphics[width=0.5\textwidth]{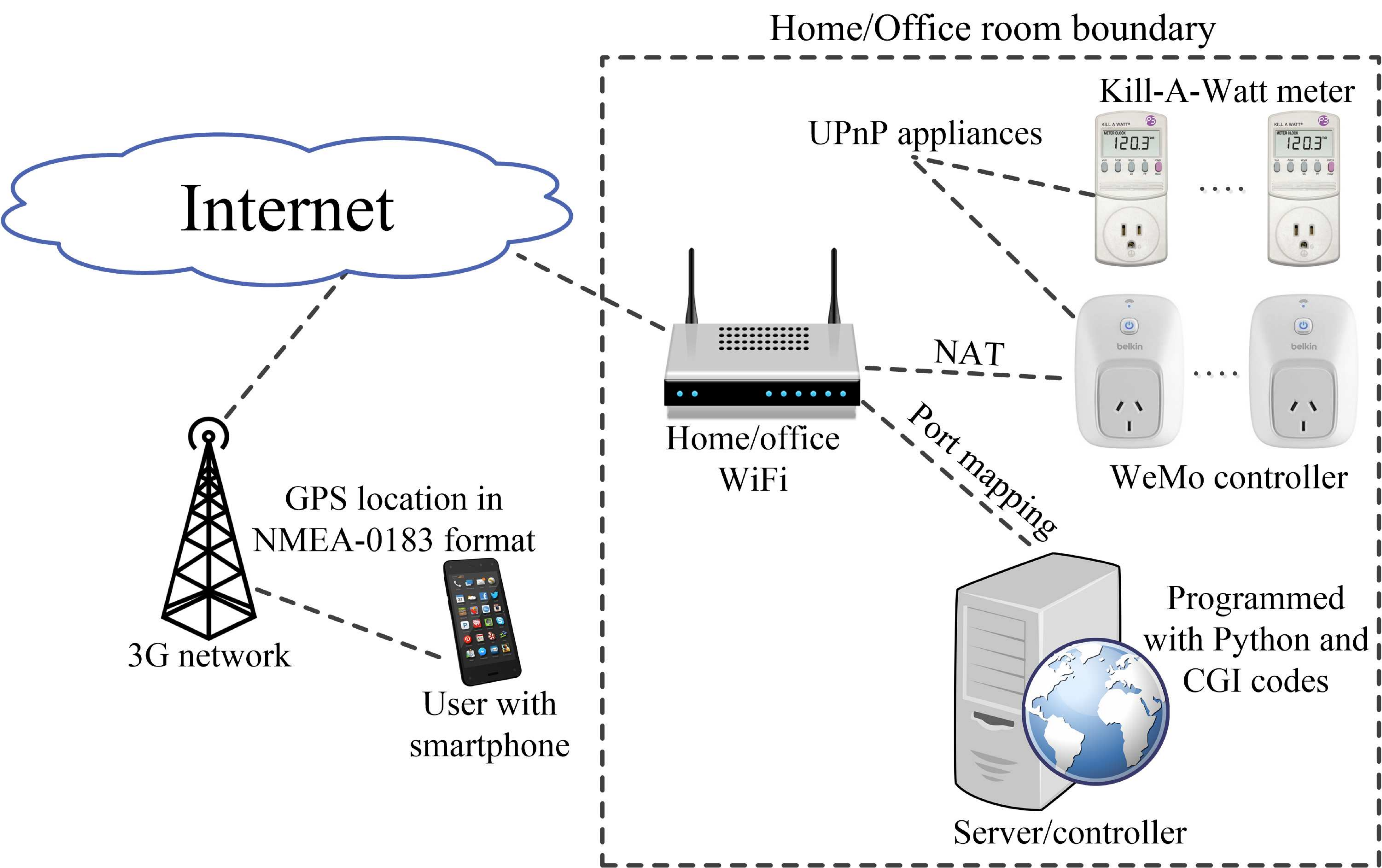}
\caption{Smart location-based automated energy control IoT framework for energy efficiency with the objective of enabling users to dynamically control and manage the appliances in residential and office buildings. This is a very simple prototype in which one user controls devices in two locations. The hardware includes Kill-A-Watt electric meters, WeMo controllers, WiFi routers, one server in each location and smart devices with GPS modules. Its software includes a GPS location data recording software, and a configuration and management software for WiFi routers \protect\cite{pan2015internet}.}
\label{fig:prototype_energy_efficiency}
\end{figure}

Fig. \ref{fig:prototype_energy_efficiency} presents the structure of this prototype. The hardware includes Kill-A-Watt electrical meters \cite{killawatt}, WeMo controllers \cite{wemo}, WiFi routers, one server in each location, and smart devices with location sensors that have a GlobalSat GPS module. This prototype requires two software packages. One package handles the data recording of GPS locations and sends that information to the server in National Marine Electronics Association (NMEA) 0183 compliant format. The other package is a configuration and management software for WiFi routers that provides a port mapping service for accessing the server from outside the Network Address Translation (NAT). The server executes Python codes programmed with Common Gateway Interface (CGI) scripts to control the WeMo devices through the Universal Plug and Play (UPnP) protocol. The smart device is able to control the electrical appliances in both locations through the inter-operation of hardware and software.

In this prototype, a smart mobile device (smartphone, tablet or laptop), with a location sensor, sends its location at predefined intervals to two servers in two locations (i.e., the apartment and the office room). The servers inside the NAT or firewall are accessed from outside by means of port mapping technology. The servers calculate the distance of the mobile device from themselves, and if the distance is within the predefined threshold, they trigger the energy policy control process that instructs the controller to turn the electrical appliances ON or OFF in the two locations according to pre-defined energy policies. This simple example of automation demonstrates the ambition to move from explicit (but hit-or-miss) user actions to intuitive and sensor-based activation that are more reliable. Further sensing (e.g., weather) and other data (calendar, habitual behavior) can be combined to refine the automatic activation control.

\subsection{Integration of Renewable and Non-renewable energy Sources at Home}
\label{sec:prototype_web_of_things}
A prototype for integrating renewable and non-renewable energy sources at home using IoT is presented in \cite{mohanty2014implementation}. This prototype is designed for the HAN architecture presented in Fig. \ref{fig:web_enabled_sg}. The electric meters are connected to non-renewable and renewable energy sources to record the current and voltage readings. This configuration has a direct connection to IoT embedded devices, and hence the IoT embedded devices change the energy source by controlling the source changers connected to the SG supplies of the houses. The web services allow consumers to track energy consumption on various scales, such as daily, weekly, monthly and yearly, as well as to compare the consumption data for different times, and thus enable consumers to configure and schedule the switching of energy sources (e.g., non-renewable or renewable) in advance. The consumer also has the ability to reconfigure the current energy source instantly in case of an emergency. 

The pilot hardware and software components are as follows:

The hardware includes an ARM cortex M3 processor to design the IoT device. An LPC1768 processor from NXP \cite{lpc1768} is used as the ARM processor, interfaced with an Ethernet port, an LCD and an RS232 port. CMSIS \cite{cmsis2016} is used as a real-time operating system for task optimization, and LwIP protocol stack is implemented to support TCP/IP functionality on the board. The ARM processor interfaces its Universal Asynchronous Serial Transmission (UART) port with MAX232 ICI in order to communicate with its RS232 port. However, the collected data from electric meters is in the form of RS485 port out. Hence, the output from RS232 is converted to RS485 for compatibility and the RS484 MODBUS protocols can then transmit serial data up to 1.2km. Several electric meters can thus be accommodated by a single processor board within a 1km radius. Current and voltage transformers are required if the the current is higher than 5A and the voltage is higher than 450V. The transformers also help to isolate the electric meters from the high current and voltage of the input supply. The electrical source changers are basically DC voltage-controlled, interfaced with the embedded controller board through relay controllers (such as H-bridge drivers).

To incorporate the Internet with this prototype, an Ethernet port RJ45 is interfaced to the LPC1768 processor. The Internet connection is established with the help of the LwIP protocol suite in three steps. Firstly, initializing the Internet connection, i.e., mapping a MAC address to the IP address with access to the world wide web (www). Secondly, connecting to the Internet in order to transmit or to receive data. Thirdly, terminating the connection upon completion of a transmission or reception. 

The software includes a Graphical User Interface (GUI) for accessing the web services and user account management. A consumer can access web services via the web browser (any sort) of any device connected to the Internet. A consumer initially registers himself, and after the required verification and authentication, the administrative staff adds the consumer's home to the system. Subsequently, the consumer can monitor his home's average power consumption, making it possible to keep track of energy consumption and plan the scheduling of power sources.

\subsection{In-Home Appliance Monitoring Implementation}
\label{sec:prototype_in_home_iot_sg}
An in-home IoT-aided SG prototype is implemented in \cite{spano2015last, spano2013intragrid} for the architecture presented in Fig. \ref{fig:last_meter_architecture}. This prototype includes a ZigBee network connected to the IoT server through a ZigBee IP gateway. In order to collect the real time energy consumption of home appliances, smart plugs are deployed between each home appliance and its wall socket to collect the loading information of the each home appliance. The smart plug is also an actuator node that can turn the load ON and OFF, configured through the user interface. 

The communication with the ZigBee network is provided through a Freescale MC13224 SoC. The board includes an ARM7 processor with 128KB of flash, 96KB of RAM and 80KB of ROM memory. For energy measurement, an analog device ADE7953 was used. Load control was implemented using a single-pole bistable 12V relay which can support up to 16A. The power supply unit in the board can supply 12V for the relay and 3V for the ADE7953 IC and MC13224 SoC. BeeStack, a Freescale ZigBee stack, is used to implement the firmware, running on the smart plug. 

Since the smart meter is a ZigBee device, a ZigBee/IP gateway is used for the communication with IoT server. The ZigBee/IP gateway is comprised of a micro controller, an Ethernet interface and a ZigBee RF transceiver. The Freescale Kinetis K60 is used as a micro controller which is based on an ARM Cortex M4 processor with hardware encryption. The message dispatcher is implemented as a multi process application on a Linux machine. It continuously listens to new connections from the gateways. 

The data collection unit is implemented using CoMo software \cite{iannaccone2006fast}. It is designed for fast prototyping of network data mining applications and it is used in large testbed deployments, e.g., PlanetLab \cite{peterson2006planetlab}. Hence, it is scalable to very large systems with high data rates. The CoMo architecture provides an abstraction between the IoT server and the network interface. It follows a modular approach and it is comprised of several modules. Each CoMo module interprets the packets of a specific type of sensor network that belongs to the households and extracts data for inclusion in either the SG database or a configuration database. The configurator and user databases are implemented with MySQL. After the required authentication, the Tornado implements direct access to the home appliances data using HTTP. 

The user interface is implemented as a web interface. The web interface, the secure access module and the configurator unit are based on an open source non-blocking web server, known as Tornado \cite{tornado}. The graphical interface is based on Twitter Bootstrap \cite{otto2013bootstrap}. The web interface mainly consists of five functionalities, home appliances' energy consumption data visualization, registration and configuration of new networks, privileges management, registration of new sensors for home appliances, and sending commands. Firstly, the visualization of energy consumption of the home appliances provides a time range selector and a graph where energy consumption data is plotted as a function of time. The time option allows the desired measurement period to be selected. Secondly, there are dedicated web interfaces that administrators use to register and configure new networks. The network configuration is stored in the configurator database. Thirdly, the administrators can manage privileges by assigning or removing different privileges to different users and third-parties. These privileges include the modification of specific network options (such as the communication channel and the security level), as well as managing sensors' visibility of home appliances and security. Fourthly, any unregistered home appliances that are sending data through a registered network or gateway are automatically detected by the system. The system then informs administrators about such home appliances, and they decide whether to register these new home appliances or not. Finally, authorized users send commands to the actuator nodes in home premises so that the required actions can be performed.

\subsection{Real-time Monitoring of Medium Voltage Grid}
\label{sec:prototype_real_time_state_estimation}

A prototype of real-time monitoring of medium voltage grid has been implemented at Ecole polytechnique federale de Lausanne (EPFL) campus in Switzerland \cite{pignati2015real}. This monitoring system estimates the network state for a 20kV active distribution system through Phasor Measurement Unit (PMU) for achieving significantly low latency and high state estimation frame rates. It offers short lines, high volatile load demand, 6MW of combined power and heat generation units and active power injections generated by 2MW photo-voltaic panels. This prototype validates the accuracy and time latency of a real-time 3-ph state estimation process deployed in a real Active Distribution Network (ADN). There are three main components of this system: i) dedicated PMUs which are connected on the medium-voltage side of secondary network substations through specific current transducers, ii) a communication network for supporting time limits, and iii) an state estimation process for real-time monitoring that considers state estimation process and phasor-data concentration. 

PMU offers fast and synchronized measurements of amplitude, phase and frequency of power system waveform and is considered as an AMI. The PMU, adopted by authors in this prototype is synchrophasor estimation algorithm \cite{romano2014enhanced} which improves the performances of interpolated-DFT method using a specific compensation method for the spectral interference generated by negative image of the spectrum. For interfacing the PMU analog input modules with the power system waveform, the high current/voltage signals have to be transformed to low voltage signals first with the minimum possible phase/amplitude distortion. For this purpose, Altea CVS-24 sensor \cite{altea} was chosen which are 0.1-class compliant. The connection between the PMU and sensors output was established through dedicated shielded cables for all the substations. For each substation, GPS was used for PMU synchronization and the connection between the PMU and GPS antenna was established through RG213 cables.

The communication network was built on IPv6 by reusing existing twisted pair cables over which the communication takes place using the single-pair high-speed digital subscriber line (SHDSL) technology. The traffic from all PMUs is centered at the SHDSL concentrators, located in the PBX room. The communication between Phasor Data Concentrator (PDC) and PBX room was established using optical fibres at 100Mb/s because the bitrate of SHDSL is 2 Mb/s which is not sufficient. The whole network is durable up to 8 hours of power outage. The authors of this prototype developed an IP version of the parallel redundancy protocol, known as iPRP \cite{popovic2015iprp} which handles the duplication of UDP packets at PMUs and removes the duplicates at PDC. The iPRP is implemented as a transport layer solution in Linux OS of PDC and PMUs, which offers the advantage of not needing any modification to any network devices or PDC/PMU applications.

The PDC collects the data of synchrophaser, nodal absorbed/injected powers, frequency and rate of change of frequency which is estimated by PMUs and forwards it to other applications (e.g., state estimator or visualization tools). The PDC communicates with PMUs and decapsulates their datagrams. The synchrophasers are time aligned and aggregated in a circular buffer. Subsequently, a subset of these measurements is forwarded to the state estimator. While handling with communication networks and devices, the non-deterministic nature of network traffic, varying reporting rates among devices and data streaming can cause the measurements to be delayed which make it difficult to identify the accurate amount of reception time of the data packets. Hence, the authors use adaptive algorithm in this prototype that identifies the time each dataset need to wait for the rest of phasors based on an event timeout with the same timestamp. The arrival of more recent measurement identified by its timestamp expires the timeout. On event triggering, the current dataset is transmitted to the state estimator application. This approach guarantees that the available measurements will be forwarded in an acceptable time range (e.g., around 20ms for PMUs streaming at 50fps), increasing the determinism of the process. Since machine hosting the PDC and state estimator are GPS synchronized, the bottlenecks in the whole chain causing higher delay can be identified and solved.
The state estimator receives data from the PDC and estimates the system state. This prototype adopted discrete Kalman Filter as it provides more accurate estimations as compared to the weighted least squares \cite{zanni2014probabilistic}. The discrete Kalman Filter with no control input solves the problem of estimating the state of a discrete-time system process. It is suitable for 3-phase systems and solely relies on nodal synchrophasor measurement offered by PMUs.

The performance of this prototype seems compatible for real-time protection and control functionalities, which are expected to be developed for active distribution networks.

\subsection{Summary and Insights}
\begin{figure}[ht]
\centering
\includegraphics[width=0.5\textwidth]{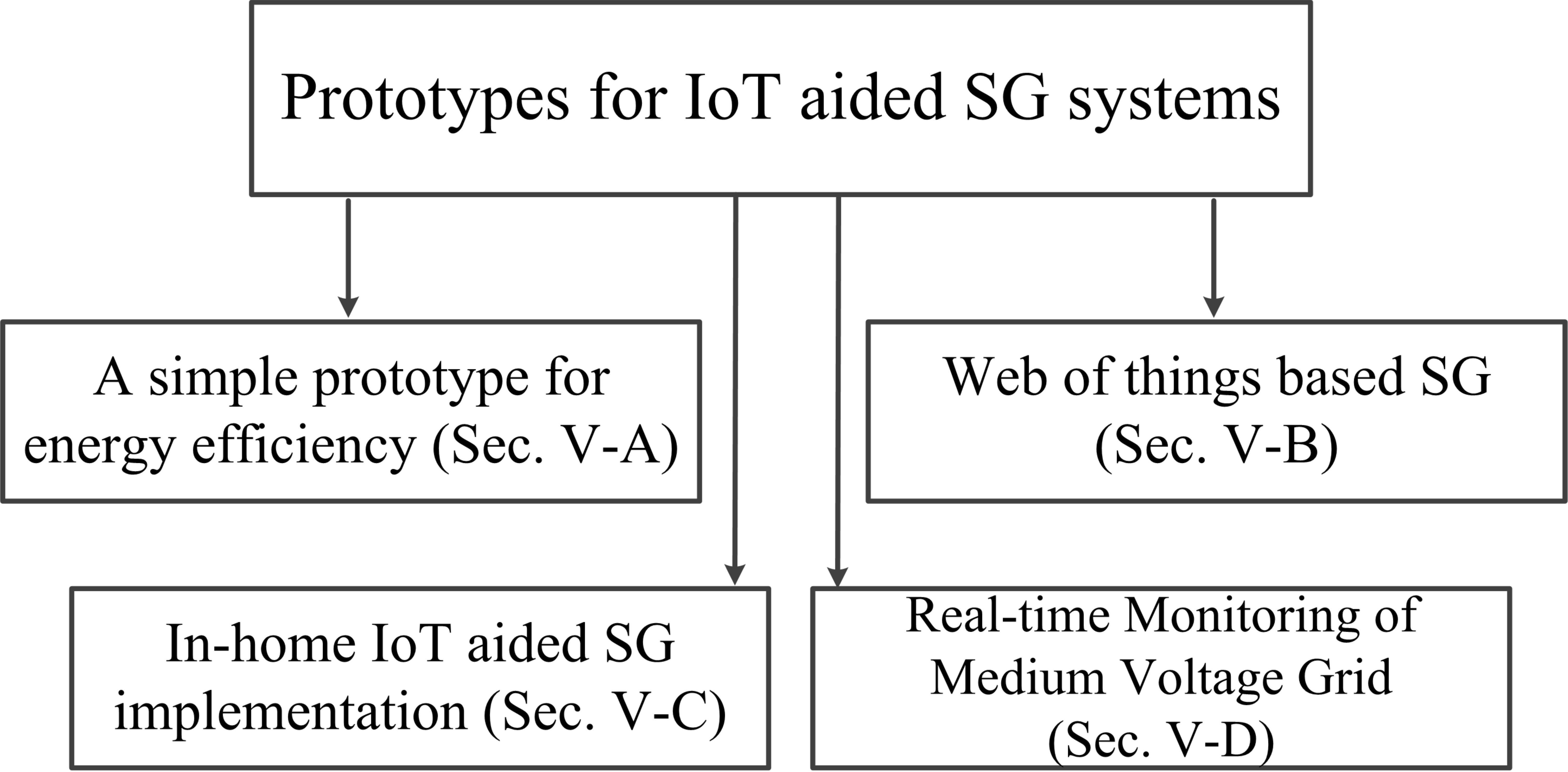}
\caption{Prototypes for IoT-aided SG systems.}
\label{fig:prototypes_iot_sg}
\end{figure}

As presented in Fig. \ref{fig:prototypes_iot_sg}, there are not many published prototypes in this area, and they tend to be very simple. For instance, the prototype for energy efficiency (see Section \ref{sec:prototype_energy_efficiency}) is a very simple prototype involving one user controlling home appliances in two locations (i.e., at their home and in an office building). Similarly, the integration of renewable and non-renewable energy sources prototype (see Section \ref{sec:prototype_web_of_things}) and in-home appliance monitoring prototype (see Section \ref{sec:prototype_in_home_iot_sg}) associate the home appliances with the web through web services. Consumers can access these web services through any browser and use them to control their home appliances, to track energy consumption and to switch the energy sources (e.g., renewable and non-renewable). Additionally, the real-time monitoring of medium voltage grid is prototyped for estimating the network state for a 20kV active distribution system through PMU in order to achieve low latency and high state estimation frame rates (see Section \ref{sec:prototype_real_time_state_estimation}). More challenging prototypes would require cooperation of HAN, NAN and WAN, affecting the various subsystems (i.e., power generation, transmission, distribution and utilization) of SG. In addition, combining different IoT knowledge systems with different sensor input systems will yield further intelligent functionality than available today. 

From this viewpoint of the prototype literature, it becomes apparent that there are no easily available open-source testbeds and simulation tools to enable developing experimentations and performance evaluation of IoT aided SG systems. For this purpose, \cite{tsg7} is a good source of guidance.

\section{Big Data Analytics and Cloud for IoT-aided SG Systems}
\label{sec:big_data_mgmt_iot_sg}

\subsection{Need of Big Data Management in IoT-aided SG Systems}
The integration of IoT technology with SG comes with a cost of managing huge volumes of data, with frequent processing and storage. Such data includes consumers load demand, energy consumption, network components status, power lines faults, advanced metering records, outage management records and forecast conditions. This means that the utility companies must have hardware and software capabilities to store, manage and process the collected data from IoT devices efficiently and effectively \cite{is2012managing}.

Big data is defined as data with huge volume, variety and velocity (three V's). The high frequency of data collection by IoT devices in SG makes the data size very large. The variety is represented by the different sensors that produce different data. The data velocity represents the required speed for the data collection and processing. Hence, IoT-aided SG systems can apply the techniques of big data management and processing (such as hardware, software and algorithms). 

The frequency of data processing and storage for IoT-aided SG systems varies from application to application. For instance, some applications perform their tasks during a specific time of a day, such as weather forecasting, which can be performed daily at the night time. Other applications perform their tasks all the times, such as real-time online monitoring of transmission power lines, so these requirements need consideration in managing and processing their data \cite{jaradat2015internet,tsg2}. Big data analytics can help to manage real-time and huge data \cite{sun2016internet}. 

In SG, the Supervisory Control and Data Acquisition (SCADA) system is the main element of decision making. It collects data from IoT devices which are distributed over the grid and provides real-time online monitoring and controlling. Additionally, it helps to manage the power flow throughout the network in order to achieve consumption efficiency and power supply reliability. Generally, it is located on local computers at various sites of the utility companies. With the growing size of SGs, utility companies face a challenge in keeping SCADA systems updated and upgraded. In order to solve this problem, cloud computing is a good solution to host SCADA systems. Cloud computing enables on-demand access to a shared pool of computing resources, such as storage, computation, network, applications, servers and services \cite{jararweh2014cloudexp,tsg1,tsg4}. 

\begin{figure*}[!t]
\centering
\includegraphics[width=1.0\textwidth]{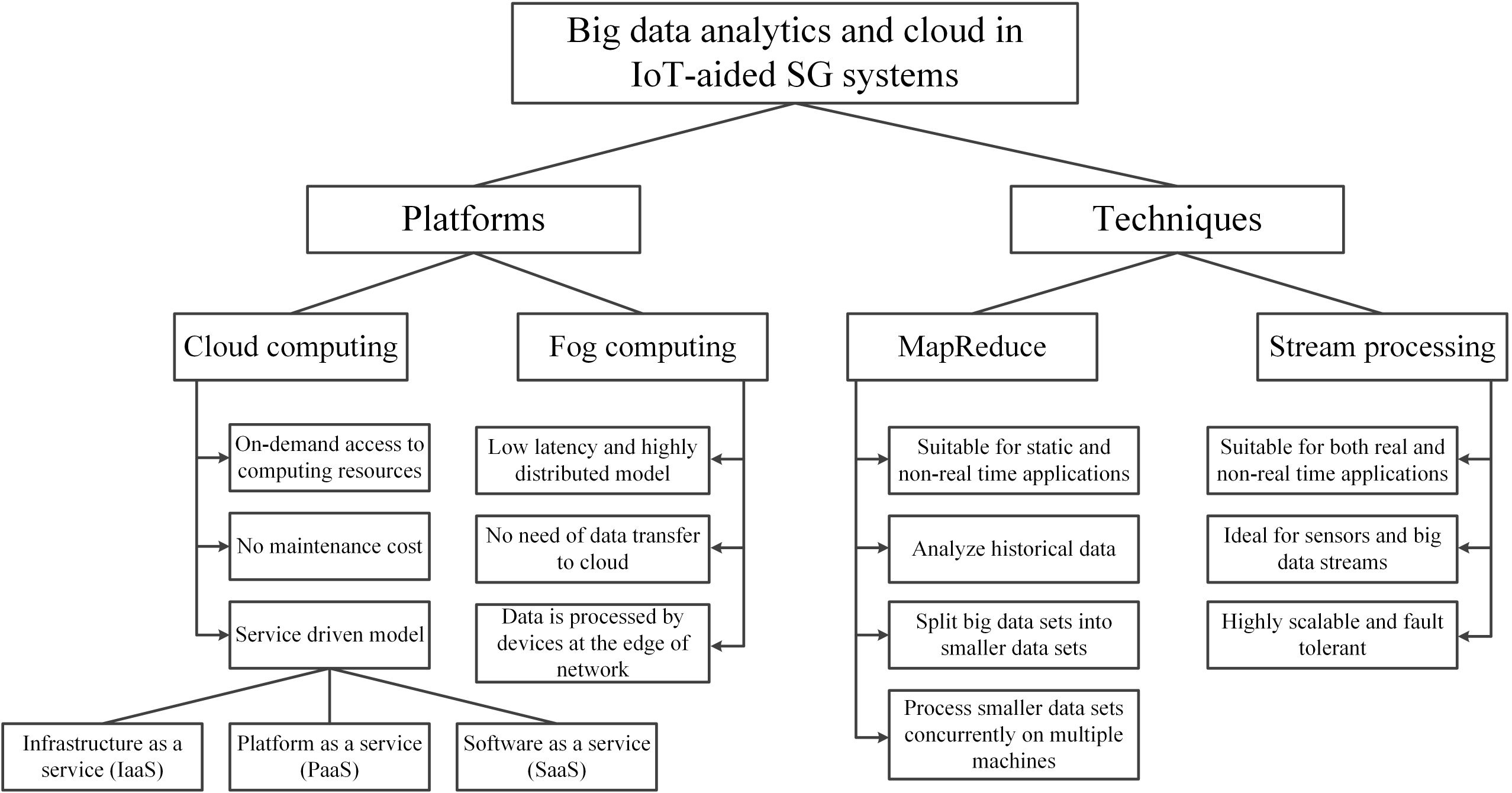}
\caption{Classification of big data management in IoT-aided SG systems into platforms and techniques. The platforms include cloud computing and fog computing while the techniques include MapReduce and stream processing.}
\label{fig:big_data_management_iot_sg}
\end{figure*}

\subsection{Platforms}
Cloud computing provides a service-driven model which can be categorized into three types, namely Infrastructure as a Service (IaaS), Platform as a Service (PaaS) and Software as a Service (SaaS) \cite{saleem2014integration}. IaaS offers the provisioning of computer infrastructure as a service, generally via Virtual Machines (VMs), such as Amazon EC2, Flexiscale, GoGrid and Joyent. PaaS provides the support of operating system and software development to the developers to deploy their applications within the Cloud, such as Microsoft Windows Azure, Force.com and Google app engine. SaaS is a multi-tenant platform and can be referred as on-demand services over the Internet. It provides a ready to use application for the end users, such as IBM, Microsoft, Oracle and SAP. Hence, the utility companies can transfer their SCADA systems entirely to the Cloud and utilize the storage, processing and other services offered by Cloud computing. In this manner, the utility companies can get many advantages, such as no maintenance cost and overhead, better collaboration, pay per usage and reduced energy cost \cite{markovic2013smart}. Meloni et al. \cite{meloni2016cloud} proposed a cloud-based architecture for SG based on REST API by considering interoperability, reusability and security in their architecture. The authors also presented a case study to highlight the benefits of this architecture.

However, there is a security risk of using Cloud computing for SCADA systems due to the shared storage among several users, which makes it vulnerable to various attacks. Such security concerns are discussed in detail in \cite{sajid2016cloud}. Fog computing, a low latency and highly distributed model, extends the paradigm of Cloud computing to network 'edge' to overcome this issue. In fog computing, there is no need to transfer data to the Cloud, and locally stored data is processed by devices located at the edge of the network \cite{bonomi2012fog}. Hence, fog computing is a good alternative to Cloud computing for IoT-aided SG systems. Jalali et al. \cite{jalali2016interconnecting} explore and evaluate the usage of Fog computing and microgrids together for reducing the energy consumption of IoT-aided SG systems.

\subsection{Techniques}

There are two main techniques for big data management, namely MapReduce and stream processing \cite{jaradat2015internet}. MapReduce \cite{mapreduce} is suitable for static and non real-time applications of IoT-aided SG systems, such as weather forecasting, while stream processing is suitable for real-time applications, such as online monitoring, self-healing and fraud detection, as well as non-real-time applications. The MapReduce technique is a tool for analyzing large historical data and is frequently used. It splits big data sets into smaller data sets and processes these smaller data sets concurrently on multiple machines using the same code. Stream processing \cite{gartner} is considered ideal for processing sensor data and big data streams. Its architecture is designed to handle big data in real-time with high scalability and fault tolerance. Hence, stream processing has a great potential for big data management in IoT-aided SG systems \cite{jaradat2015internet}.

\subsection{Summary and Insights}
IoT-aided SG systems involve processing data that have the same characteristics as data that requires Big Data techniques. Fig. \ref{fig:big_data_management_iot_sg} presents the classification of big data management in IoT-aided SG systems. Big data management is highly suitable for IoT-aided SG systems due to its huge volume of data and its 'velocity' - the need for real-time processed information. The operation platforms for big data management in IoT-aided SG systems include cloud computing and fog computing. Cloud computing provides on-demand access to a shared pool of computing resources. It is a service driven model which is categorized by IaaS, PaaS and SaaS. Sending and storing data on the Cloud raises security risk issues, due to the shared storage among several users, which makes it vulnerable to attacks. Hence, fog computing is used to solve this security risk. Fog computing does not require data transfer to the cloud and the data is processed by devices that are located at the network edge. The techniques for big data processing include MapReduce and stream processing. MapReduce is suitable for static and non-real time applications and it analyzes large historical data. It splits big data sets into smaller data sets and processes these smaller data sets concurrently on multiple machines. Stream processing is suitable for both real and non-real time applications and is ideal for sensors and big data streams. It is highly scalable and fault tolerant, and it has a great potential for big data management in IoT-aided SG systems. So far, IoT-aided SG systems do not present new issues for Big Data processing. However, different SG Big Data applications with different data sampling frequencies and data types should be further investigated. Additionally, IoT-aided SG system is prone to various security vulnerabilities caused by Internet-based protocols and public communication infrastructure. To cope with such security vulnerabilities, it is important that the big data produced by IoT-aided SG systems need to be stored and processed intelligently for extracting critical data and develop a mechanism to trigger security warnings in an early stage. For this purpose, Chin et al., \cite{chin2017energy} present research challenges for big data security threats in IoT-aided SG systems which can be considered while developing security solutions for Big Data in IoT-aided SG system.

\section{IoT and Non-IoT Communication Technologies for SG}
\label{sec:iot_communication_tech_sg}

There are many IoT and non-IoT communication technologies available for SGs and the users have to select among them, suitable to their needs. There is a lack of proper guidelines for the selection of communication technologies for SG. Therefore in this section, we have listed and discussed various IoT communication technologies together with non-IoT communication technologies including their characteristics, advantages and disadvantages. Table \ref{tab:iot_comm_technologies} provides a summary of each communication technology which is discussed in detail in each sub-section \cite{gungor2011bsmart, parikh2010opportunities, jain2014survey}. From this section, users will have a broader view of available communication technologies for SG and subsequently, they can make better decision for the selection of most suitable communication technology based on their needs. In the rest of this section, we first briefly explain the role and need of communication technologies for SG and subsequently provide discussion on each IoT and non-IoT communication technology.

The IoT and non-IoT communication technologies used in the SG for data transmission between smart meters and electric utilities are categorized into two types, wireless and wired technologies. Wireless technologies have some advantages over wired technologies in some cases, such as their ease of connection to otherwise unreachable or difficult areas, as well as their low cost infrastructure. However in wireless technologies, signal attenuation may occur due to the nature of the transmission path. On the other hand, wired technologies do not have such interference issues and, unlike wireless technologies, their functions do not depend upon batteries. 

An SG requires two types of information flows. Firstly, between smart meters and IoT devices, sensors and home appliances, and secondly, between smart meters and utility control centers. The first data flow can be achieved through powerline communications or wireless communications, such as by using 6LowPAN, ZigBee and Z-wave \cite{luan2010smart}. The second data flow can be achieved by using cellular communications or via the Internet. However, there are various key limiting factors that should be considered in the deployment process of smart metering, such as operational costs, deployment time, availability of technology, and the operating environment (such as rural, urban, indoor, outdoor). Hence, the choice of technology that suits one environment may not be applicable for another. We discuss communication technologies for SGs in the rest of this section.

\begin{table*}[!thpb]
\caption{IoT and non-IoT communication technologies for the smart grid}
\label{tab:iot_comm_technologies}
\centering
\scriptsize

\resizebox{\textwidth}{!}{
\begin{tabular}{|p{0.5cm}|p{1.75cm}|p{3.25cm}|p{3.25cm}|p{1.5cm}|p{3cm}|p{1.25cm}|p{1.25cm}|} 
\hline
Tech. & Protocol & Advantages & Disadvantages & Applicability & SG Application Areas & Data rate & Coverage range \\
\hline

\multirow{8}{*}{\rotatebox[origin=c]{90}{IoT Wireless Technologies~\hspace{30px}}}

& 5G
& Low latency; high reliability; high bandwidh; high speed; capablity of handling large number of devices
& Low range; high infrastructure set up cost; security and privacy issues
& NAN, WAN
& distributed monitoring and control
& Up to 20Gbps
& 10s-100s meters (depending upon cell size)
\\ \cline{2-8} 

& Z-Wave
& No interference with other wireless technologies; reliable; low latency; scalable
& Not suitable for NAN and WAN; only one propriety company for making chips; short range
& HAN
& Home automation
& 100 Kbps
& 30 m
\\ \cline{2-8} 

& 6LoWPAN
& Robust; low-power; support large mesh network topology; can be applied across various communication platforms
& Require extensive training and knowledge; short range; low data rate
& HAN
& Smart metering; home automation
& 250 Kbps
& 10-100 m
\\ \cline{2-8} 

& LoRaWAN
& Low-power; long range; no interference with different data rates; enhance gateways capacity by creating virtual channels; low-cost secure bi-directional communication
& --
& NAN; WAN
& Management of operation and equipment; online monitoring of power transmission lines and tower
& 0.3-50 Kbps
& 2-5km (urban environment); 15km (suburban environment)
\\ \cline{2-8} 

& ZigBee 
& 16 channels each with 5 MHz of bandwidth in 2.4 GHz spectrum; low power usage; low complexity; low deployment cost
& Low data rates; limited energy of battery; low processing capabilities; short range
& HAN
& Energy monitoring; smart lightning; home automation; automatic meter reading
& 250 Kbps
& 10-100 m
\\ \cline{2-8} 

& WirelessHART
& Backward compatibility; reliability; robustness
& Low data rate; short range; lack of security specifications
& HAN; WAN
& smart meters; power generation
& 115 Kbps
& 200 m
\\ \cline{2-8} 

& Bluetooth
& Low power consumption
& Low data rate; short range; weak security; prone to interference from IEEE 802.11 WLANs
& HAN
& Home automation
& 721 Kbps
& 1-100 m
\\ \cline{2-8} 

& Bluetooth Low Energy (BLE)
& Ultra-low power consumption; low cost; low complexity
& Short range; handling lower amount of data; high setup time
& HAN
& Home automation
& 25 Mbps
& 5-10 m
\\ \cline{2-8} 

& Narrowband IoT (NB-IoT)
& Crowd-free unlicensed bands, ultra-low power consumption; low cost; low complexity
& Latency
& HAN; NAN
& Home automation; AMI
& 250 Kbps for uplink; 230 Kbps for downlink
& 35 km

\\ \hline

\multirow{5}{*}{\rotatebox[origin=c]{90}{Non-IoT Wireless Technologies~\hspace{55px}}}

& Cellular communications 
& Wide area coverage; improved QoS
& Network congestion due to high density; critical for emergency applications; non-guaranteed service in unfavorable conditions (e.g., wind storms) 
& HAN; NAN; WAN
& Monitoring and management of DERs; SCADA
& 60-240 Kbps
& 10-50 km
\\ \cline{2-8} 

& Wireless mesh 
& Low cost; self healing; self organization; high scalability; high data rate
& Prone to interference and fading; network coverage problem in rural areas due to low density of smart meters; prone to loop problem due to inclusion of multiple relay nodes 
& HAN; NAN
& Monitoring and controlling of DERs; automation and protection of substation automation
& Depends upon protocol
& Depends upon deployment
\\ \cline{2-8} 

& WiMAX
& Long range; high data rate
& Trade-off between performance and distance; costly RF hardware; low frequencies are already licensed and require leasing; high frequencies do not penetrate obstacles
& HAN
& Real time pricing; automatic meter reading; outage detection and restoration
& 75 Mbps
& 10-50 km (LOS); 1-5 km (NLOS)
\\ \cline{2-8} 

& Mobile broadband wireless access
& Low latency; high mobility; high bandwidth
& Communication infrastructure not readily available; high cost; moderate data rate
& NAN; WAN
& Broadband communication for electric vehicles; SCADA system; wireless backhaul for SG monitoring
& 20 Mbps
& Vehicular standard (up to 240 km/h)
\\ \cline{2-8} 

& Digital microwave technology
& Long distance coverage; high data rate; high bandwidth
& Prone to multipath interference and precipitation
& HAN; NAN
& Transfer trip between DER and distributed substation
& 155 Mbps
& 60 km
\\ \hline

\multirow{3}{*}{\rotatebox[origin=c]{90}{Non-IoT Wired Technologies~\hspace{20px}}}

& Powerline communication
& Cost effective; low installation cost; wide availability; utility's own ownership and control; dedicated network
& Noisy and harsh medium; sensitive to disturbances; signal quality gets affected by the type and number of devices, wiring distance between nodes and network topology; cost of ownership; complexity of management
& HAN; NAN
& Low voltage distribution; automatic meter reading
& 2-3 Mbps
& 1-3 km
\\ \cline{2-8} 

& Digital subscriber lines (DSL)
& High speed; low latency; low installation cost; high data rate, high capacity; long range; wide availability
& Quality is distance dependent; high installation cost in low density (rural) areas; unreliable
& HAN; NAN; WAN
& Smart metering
& 1-100 Mbps
& 5-28 km
\\ \cline{2-8} 

& Optical communications
& Long distance communication; ultra-high bandwidth; robustness against radio and electromagnetic interference; high reliability
& High deployment cost of fiber installation; high cost of terminal equipment; difficult to upgrade; not suitable for metering applications
& WAN; NAN
& Physical network infrastructure
& Up to 100 Tbps
& 10-60 km
\\ \hline

\end{tabular}}
\end{table*}

\subsection{IoT Technologies}

\subsubsection{Wireless Technologies}

\paragraph{5G}
5G is the next generation cellular communications network and is the successor of 4G, 3G and 2G having the objectives of significantly increasing the speed and responsiveness of wireless networks. It is built mainly by considering the requirements of emerging M2M applications, such as low latency, noise immunity, reliability and traffic safety \cite{moongilan20165g,garau20175g},  as well as for controlling and monitoring of sensitive infrastructures, such as the SG \cite{moongilan20165g}. 5G improves LTE-A by adding multicast transmissions and the IoT perspective. For SG, 5G Radio Access Technology (RAT) is designed by considering the applications requirements of SG \cite{moongilan20165g,garau20175g}. Hence, 5G is a very suitable technology for SG mainly for distributed monitoring and control functionality \cite{cosovic20175g,garau20175g}. 5G enables utility centers to remotely connect to the assets of whole distribution grid.

5G operates on various licensed and unlicensed frequency bands, e.g., below 1GHz, below 6GHz and above 6GHz. The coverage of 5G depends upon the small cells (e.g., femtocells, picocells, microcells) having coverage range from 10s of meters up to 100s of meters \cite{5g_rf_world}. It has data rates of up to 20 Gbps and latency of 1ms \cite{rouse20195g_online}. In SG, 5G is mainly suitable for NAN, such as for distributed controlling and monitoring \cite{cosovic20175g,garau20175g}.

\textit{Advantages:}
5G offers low latency, reliability, high bandwidth and high-speed. Since 5G communicates with a large number of devices, capable of sensing and performing actions, therefore it is a very good candidate for IoT-aided SG systems which includes a large number of sensors and meters.

\textit{Disadvantages:}
Since 5G offers high bandwidth, therefore it does not offer high coverage range. Additionally, it requires high set up cost of creating the infrastructure.  Also, the security and privacy issues still need to be solved.

\paragraph{Z-Wave}
Z-Wave \cite{zwave, samuel2016review} is a low-power radio frequency (RF) IoT communication technology for short range communication which is mainly suitable for home automation \cite{kuzlu2015review, withanage2014comparison}. It operates on 1 GHz band and offers a data rate of 100 Kbps with coverage range of 30m. Z-Wave is developed by Sigma Designs \cite{sigmadesign} and it applies 128-bit AES encryption for ensuring security \cite{huq2010home}.

\textit{Advantages: }
The main advantage of Z-Wave is that it does not interfere with WiFi and other wireless technologies operating in 2.4 GHz band, such as ZigBee and Bluetooth \cite{huq2010home, 11_iot_protocols}. It is optimized for reliable and low-latency communication of small data packets. It is also very scalable and can control up to 232 devices \cite{11_iot_protocols}. It can also support whole mesh networks without needing any coordinator.

\textit{Disadvantages: }
The main disadvantage of Z-wave is that since it is mainly designed for small data packets, it is not much suitable for NAN and WAN because NAN and WAN have large aggregated data packets. Additionally, it is a short range and low data transmission rate communication technology. Moreover, there is only one company, Sigma Designs, for developing the chips of Z-Wave, unlike multiple sources for other wireless technologies (e.g., ZigBee) \cite{11_iot_protocols}.

\paragraph{6LowPAN}

6LoWPAN (IPv6 over Low-power Wireless Personal Area Networks) \cite{rfc4944} is a key IP-based communication technology. It enables IPv6 packets to be carried within small link layer frames efficiently which are defined by IEEE 802.15.4. The key attribute of 6LoWPAN is the IPv6 (Internet Protocol version 6) stack, which has the main role in recent year for enabling IoT. IPv6 provides around 5x1028 addresses for each person in the world, which allows any device or embedded object to have its own unique IP address and to connect to the Internet. It is mainly designed for home automation (including smart metering). Initially, 6LoWPAN was designed to support IEEE 802.15.4 low-power wireless networks in 2.4 GHz band, however it is now adapted to support a number of other wireless newtorking media, including sub-1 GHz low-power RF \cite{kuzlu2015review}. Unlike Bluetooth and ZigBee, 6LoWPAN provides encapsulation and header compression mechanisms \cite{11_iot_protocols}. Similar to Z-Wave, it also applies strong 128-bit AES encryption for link layer security defined in IEEE 802.15.4. It offers a data rate of 250 Kbps and coverage range of 10m to 100m.

\textit{Advantages: }
The characteristics of 6LoWPAN make it ideal for SG HAN including home automation and smart metering \cite{kuzlu2015review}. It is very robust and low-power communication technology which can also support a large mesh network topology \cite{kuzlu2015review}. Additonally, its standard has the liberty of physical layer and frequency band, and it can be applied across various communication platforms (such as IEEE 802.15.4, WiFi, Ethernet and sub-1 GHz band).

\textit{Disadvantages: }
The main disadvantage of 6LoWPAN is the extensive training required to work with IPv6 protocol. The end users cannot deploy this protocol easily because it requires extensive knowledge of IPv6 stack and its functionality. Moreover, one of its disadvantages are similar to Z-Wave, i.e., it is designed for small link layer frames, and it has short range and low data transmission rate communication technology, therefore it is not much suitable for NAN and WAN.

\paragraph{LoRaWAN}

LoRAWAN \cite{lorawan} is a Low Power Wide Area Network (LPWAN) communication technology which is designed for wireless battery-operated devices in Internet of Things \cite{schroder2016adequacy}. It is mainly designed for wide area network coverage in IoT and it can be applied in NAN and WAN of SG. It provides mobility, secure bi-directional communication and localization services, as well as interoperability. It offers a data rate of 0.3 Kbps to 50 Kbps and coverage range of 2-5km in urban environment and 15km suburban environment. Most of the existing wireless communication technologies have been designed to maximize their peak data rate, however LoRa is mainly designed for low-power consumption and long communication range by sacrificing its peak data rate \cite{petajajarvi2015coverage}. Furthermore, LoRaWAN network server manages the RF output and data rate for each IoT device through adaptive data rate (ADR) approach in order to enhance network capacity and battery life of each IoT device \cite{lorawan}. It is comprised of wide spectrum bands, there is no interference with different data rates during the communication. Furthermore, it creates virtual channels which enhances the capacity of gateways. Additioanlly, it provides low-power WAN coverage with features required to support low-cost secure bi-directional communication \cite{11_iot_protocols}.

\paragraph{ZigBee}
ZigBee is a cost effective, reliable, low-power, low complexity and low data rate wireless communication technology, developed by the ZigBee Alliance. It is suitable for home automation, energy monitoring, smart lightning and automatic meter reading in SGs. ZigBee has been considered as the most suitable communication technology for HANs in SGs by the National Institute for Standards and Technology (NIST) \cite{yi2011developing}. Many AMI vendors, such as Elster, Itron and Landis Gyr, integrate ZigBee with smart meters for communication with IoT devices that control the home appliances in HANs \cite{gungor2011smart}. ZigBee is also used to inform consumers about their real-time energy consumption.

\textit{Advantages: }
ZigBee operates on the 868 MHz, 915 MHz bands and also uses 16 channels in the 2.4 GHz band (each having 5 MHz of bandwidth). It provides compatibility with the IEEE 802.15.4 standard and offers a data rate of 250 Kbps with coverage of 10-100m. It supports tree, mesh and star topologies and uses 128-bit AES encryption for security. Due to its lightweight nature, ZigBee is a good choice for metering management, it is considered as a good choice for the SG implementation in HANs due to its simplicity, robustness, mobility, low operational cost and low bandwidth requirement. 

\textit{Disadvantages: }
ZigBee has some limitations for practical implementations, such as low memory and processing capabilities, high latency and interference with other devices due to its wireless medium because multiple devices operate on the same spectrum \cite{yi2011developing}. Such interference in the vicinity of ZigBee can damage the communication in the whole SG network \cite{lewis2009assessment}. Therefore, interference avoidance and detection schemes must be designed, as well as energy efficient routing protocols in order to provide reliable performance and to maximize the network lifetime. 

\paragraph{WirelessHART}
WirelessHART is an extension of the HART (Highway Addressable Remote Transducer) protocol that is a hardwire control protocol designed for industrial automation \cite{chhaya2017wireless, howell2009update}. HART is a global standard that is widely installed. WirelessHART is a centralized and real-time protocol for industrial process monitoring and control \cite{gungor2009industrial}. It is a low power communication protocol that is based on IEEE 802.15.4 and operates on 2.4GHz band \cite{chhaya2017wireless, gungor2009industrial}. It has a range of 200m, offers 115 Kbps data rate \cite{digades2018lokal} and it uses 128-bit AES encryption for security \cite{chhaya2017wireless}. WirelessHART was included in the HART protocol suite in June 2007 after approval by the HART communication foundation as part of the HART 7 Specification \cite{wireless_hart}. In SG, WirelessHART can be used in power generation for sensors deployment \cite{liu2012wireless}. Emerson introduced WirelessHART power meter that overcomes the limitations of wired meters that require a power source and handwiring to the monitoring and control system \cite{emerson2017wirelesshart}. The required power for WirelessHART power meter is provided from the main electrical supply to the monitored equipment that eliminates the requirement of a dedicated power source. Emerson white paper on WirelessHART power meter discusses improving the energy management and equipment reliability through monitoring the power use, the interconnection of WirelessHART and wireless network, as well as some examples on the organizations improving their operations using the WirelessHART power meter \cite{emerson2017wirelesshart}.

\textit{Advantages:} 
WirelessHART is backward compatible, e.g., systems operating on wired protocol can be connected to WirelessHART network using simple adapters by offering the least resistance path while updating to the wireless medium \cite{howell2009update}. It overcomes the limitations of ZigBee, such as reliability, robustness, message delivery \cite{chhaya2017wireless}. Lennvall et al. \cite{lennvall2008comparison} have made a comparison of ZigBee and WirelessHART for industrial applications and discussed how WirelessHART can outperform ZigBee. 

\textit{Disadvantages:}
Although WirelessHART overcomes some of the limitations of ZigBee and it is well organized in all the manners, however, it still has the main limitation of security. WirelessHART has no dedicated specifications of security requirements. It limits the implementation and development of applications because the developer must need to first acquire the knowledge of all the core specifications. Raza et al. \cite{raza2009security} present a detailed overview of WirelessHART security which includes analyzing the security mechanisms provided by WirelessHART against main threats in wireless medium, as well as providing suggestions on how to overcome such limitations.

\paragraph{Bluetooth and Bluetooth Low Energy (BLE)}
Bluetooth is a part of the Wireless Personal Area Network (WPAN) standard, known as IEEE 802.15.1, which is a low power, short range radio frequency communication standard. It provides a data rate of 721 Kbps and operates on the 2.4 - 2.4835 GHz unlicensed Industrial, Scientific and Medical (ISM) band. Bluetooth can facilitate point-to-point and point-to-multipoint communication configurations. It offers coverage between 1m - 100m, depending on the communication configuration. Furthermore, Bluetooth-configured devices are comprised of the whole seven-layered OSI communication stack. 

BLE is the latest version of Bluetooth which provides short range, low cost and low complexity communication. It overcomes the limitations of Bluetooth, such as insufficient power requirements and high complexity. It is an ultra-low power technology for devices having very low battery capacity. BLE specification is part of the Bluetooth specification. BLE offers data rates of upto 25 Mbps \cite{bluetooth2018allurwardi} with a range of 5-10 m in 2.45 GHz band \cite{gungor2009industrial}. It has a variable packet length as compared to fixed packet length in Bluetooth and BLE uses a different modulation scheme \cite{gungor2009industrial}. Bluetooth silicon vendors have now added BLE in the latest chips. Almost all the smartphones launched during the last couple of years are equipped with BLE hardware having various levels of capability based on the operating system support \cite{want2015enabling}.

The main advantage of Bluetooth is its low power usage, which is useful for HANs, as well as for online monitoring of substation automation \cite{zhang2007design}. Its main disadvantages are its low data rate, short range and weak security. Furthermore, Bluetooth configured devices may interfere with WLANs (Wireless Local Area Networks) due to their high influence on the surrounding communication links. The main advantages of BLE are ultra-low power consumption, low cost and low complexity. However, its main disadvantages are short range, handling lower amount of data as compared to Bluetooth and high setup time \cite{bluetooth2018allurwardi}.

\paragraph{Narrowband IoT (NB-IoT)}
Narrowband IoT (NB-IoT) is a licensed LPWAN technology which is developed on the existing LTE specifications, providing cellular-level QoS facilities \cite{li2018smart}. Since NB-IoT uses existing cellular infrastructure rather than a new one, it saves significant investments, time and efforts for developing a utility-dedicated communication infrastructure and the deployment of applications \cite{li2018smart}. NB-IoT is standardized by 3rd Generation Partnership Project (3GPP) in LTE Release 13 \cite{reininger20163gpp} and has attracted strong support from Qualcomm, Ericsson and Huawei. The main objectives of NB-IoT include: a device should cost less than 5 US\$ having 164 dB coupling loss, uplink latency should be below 10s, each household should be able to support 40 connected devices, and a device battery life should last for 10 years if it transmits 200 bytes of data per day \cite{tr45820, ratasuk2016nb}. NB-IoT operates on 900-1800 MHz frequency bands with coverage of upto 35 km \cite{li2018smart}. It has data rates of upto 250 Kbps for uplink and 230 Kbps for downlink communications \cite{wang2017primer}. In SG, NB-IoT is mainly suitable for HAN and NAN, such as home automation and AMI \cite{li2018smart}. Since NB-IoT offers ultra-low cost (i.e., a device costs $<$ 5 US\$), therefore it is a very good choice for large-scale deployment in the SG \cite{li2018smart}.

\textit{Advantages:}
NB-IoT offers ultra-low power, low cost and low complexity. It also mainly overcomes the limitation of unlicensed LPWAN technologies of crowded license-free frequency bands for mission critical and Quality of Service (QoS)-aware SG requirements. 

\textit{Disadvantages:} 
NB-IoT incurs latency, so it is not suitable for delay-tolerant applications, such as distributed automation and Distributed Energy Resources (DERs) \cite{li2018smart}.




\subsection{Non-IoT Technologies}

\subsubsection{Wireless Technologies}

\paragraph{Cellular Communications}
Cellular networks can be a good choice for communication between smart meters and utility centers in an SG. The use of existing cellular communication infrastructures save the costs and time of the utility centers to build a private and dedicated communication infrastructure for a SG. A number of cellular communication technologies, including 2G, 3G, LTE and WiMAX are available to utility centers for smart metering deployments in SG. Currently, cellular networks have been used in SG by a number of companies. For instance, Echelon's networked energy services system uses T-Mobile's Global System for Mobile (GSM) communications for its deployment by integrating T-Mobile's SIM (Subscriber Identity Module) into Echelon's smart meters to enable communication between smart meters and the backhaul servers \cite{gungor2011bsmart}. Since T-Mobile's GSM network handles all the communication requirements of smart metering in the SG, investments in a new dedicated communication network are not required by the utility centers. Similarly, Telecom Italia, Telenor, Vodafone and China Mobile have also offered their GSM networks for communication in SG. 

\textit{Advantages: }
Since cellular networks have existing infrastructure, utilities do not have to incur the additional costs of building the dedicated communication infrastructure for SG. Due to their ubiquity and cost effectiveness, cellular network is the leading communication technology for the SG. Furthermore, since GSM capacity is now progressively freed up due to the migration to VoLTE (Voice over LTE), mobile operators can offer such solutions for a relatively low cost. In SG, the data collection by IoT devices is performed at short intervals, which generates a huge amount of data in short bursts, and so cellular networks provide sufficient bandwidth to handle such data. Moreover, cellular networks also ensure the security of data transmission security in SG systems. Since the cellular network coverage has reached almost 100\% in the developed world, cellular networks can provide the best coverage for SGs. Additionally, lower maintenance and deployment costs, and rapid installation make cellular networks a better choice for SG applications, such as AMI, HAN, outage management and demand response management.

\textit{Disadvantages: }
The cellular networks' services are shared among multiple customers which may cause network congestion and performance deterioration in emergency situations, presenting a severe challenge for mission critical applications of SG which require continuous communications. In unfavorable climate conditions (e.g., wind storms), the cellular networks may be unable to provide guaranteed service, compared to wired networks. The utilities therefore, have to build their own private communication networks in order to fulfill the communication needs of mission critical SG applications. 

\paragraph{Wireless Mesh}
A wireless mesh network comprises of a group of wireless nodes where new nodes can join the network and each node can act as an router. If any node leaves the network, the self healing characteristic of the network allows it to establish an alternative route through the active nodes. A wireless mesh network was applied in a SG by the Pacific Gas, SkyPilot Networks and Electric Company in a smart metering system. In such a system, each smart device is equipped with a radio module and each device routes the metering data through neighboring meters. In this manner, each meter acts as a relay until the collected data reaches the electric network access point. The collected data are then tranmitted to the utility.

\textit{Advantages: }
The wireless mesh networking is an effective solution with low cost, self healing, self organization, self configuration and high scalability properties. It enhances the network performance, balances the network load, and provides scalability, and network coverage range \cite{yarali2008wireless}. Urban and suburban areas can be well-covered with the help of a multi-hop routing technique. Since smart meters act as relay nodes, a higher number of relay nodes maximizes the network coverage and capacity. Wireless mesh networking is highly suitable for home energy management and AMI applications of SGs. 

\textit{Disadvantages: }
The major challenges of wireless mesh networking are their network capacity, fading and interference. Wireless mesh networks are difficult to implement in rural areas because the smart meters' density is not sufficient to cover the whole communication network. Moreover, as data travels though multiple relay nodes, there could be a routing looping problem which would cause additional overhead and affect the available bandwidth \cite{lewis2009assessment}. 

\paragraph{WiMAX}
WiMAX (worldwide interoperability for microwave access) is based on the IEEE 802.16 standard for Wireless Metropolitan Area Networks (WMAN) with the main goal of achieving worldwide interoperability for microwave access. WiMAX has its own spectrum bands for interoperability. It has been assigned the 3.5 GHz and 5.8 GHz bands for fixed communication, while the 2.3 GHz, 2.5 GHz and 3.5 GHz bands are reserved for mobile communications. The spectrum bands 2.3 GHz, 2.5 GHz and 3.5 GHz are licensed spectrum, while the 5.8 GHz spectrum band is unlicensed. It provides up to 70 Mbps data rate and 48km coverage. The licensed spectrum bands are more suitable for long distance communications as they allow for long range transmission and high power. 

WiMAX's main advantages are the long range and the high data rate. It is mainly applied in HANs and AMI, for the real time pricing, automatic meter reading, and outage detection and restoration. Its two main disadvantages include costly Radio Frequency (RF) hardware for WiMAX towers and a trade-off between network performance and distance. Furthermore, high WiMAX frequencies cannot penetrate through obstacles and lower frequencies have been already assigned, and so require leasing from third-parties.  

\paragraph{Mobile Broadband Wireless Access}
Mobile Broadband Wireless Access (MBWA) technology, also known as MobileFi, is based on the IEEE 802.20 standard. It provides high mobility, high bandwidth and low latency. It operates on the 3.5 GHz licensed band by exploiting the features of the IEEE 802.11 WLANs as well as the IEEE 802.16 WMANs. It provides a real time data rate of 1 Mbps up to a high speed data rate of 20 Mbps. 

MBWA's main advantages are its high mobility support, high bandwidth and low latency. It can be used in NANs and WANs for SG applications, such as broadband communication for electric vehicles, SCADA systems and wireless backhaul for SG monitoring. Its main disadvantage is that since it is a new technology, there is no communication infrastructure readily available for this technology. Hence, the use of this technology would be costly, compared to other available technologies. 

\paragraph{Digital Microwave Technology}

Digital microwave technology operates on the 2 - 40 GHz licensed bands and provides a data rate of up to 155 Mbps. It offers long distance and wide coverage of up to 60 km, and accepts data from ATM (Asynchronous Transfer Mode) or Ethernet ports and forwards it as microwave radio. 

The main advantages of digital microwave technology are its long distance coverage and high bandwidth. It can support point-to-point communications for SG applications, such as transfer trips between DERs and a distribution substation \cite{parikh2010opportunities}. Its main disadvantage is that it is prone to two types of signal fading, namely multipath interference and precipitation. Moreover, it incurs additional latency due to its encryption of messages for security, which makes the messages larger.

\subsubsection{Wired Technologies}

\paragraph{Powerline Communication (PLC)}
PLC uses existing powerlines to transmit high speed data of about 2-3 Mbps from one device to another. PLC has a direct connection to the meter, and therefore, it was the first communication technology for electricity meters \cite{lewis2009assessment}. In a PLC network, the data from smart meters is transmitted through a PLC to a data concentrator, as they are directly connected to each other. Subsequently, the data from data concentrators is transmitted to a utility control center using other wireless communication technologies (such as cellular networks), as they are not directly connected to each other through powerlines \cite{gungor2011bsmart}. 

\textit{Advantages: }
PLC reduces the installation cost by using the existing powerline infrastructure instead of developing a private communication infrastructure. The strengths of PLC are its cost-effectiveness (by reusing existing infrastructure), widely available infrastructure, standardization efforts and ubiquitous nature. HAN is one of the largest applications of PLC technology. Since the PLC infrastructure broadly covers almost all the urban areas in the range of a utility company's service territory, it can be suitable for SG applications in urban areas. 

\textit{Disadvantages: }
The powerline transmission medium of the PLC technology is noisy and harsh, increasing the challenges of channel modeling \cite{gungor2011smart}. Moreover, the quality of signals are adversely affected by using powerlines in SGs due to various parameters, such as the number and types of IoT devices connected through powerlines, the wiring distance between transmitters and receivers, and the network topology \cite{gungor2011bsmart}. 

\paragraph{Digital Subscriber Lines (DSL)}
DSL is a high speed digital data wired transmission technology which uses telephony networks land wires. A number of SG projects have selected DSL as their communication technology. As an example, an SG project was carried out at Deutsche Telekom for the Stadtwerke Emden municipal utility in Germany in which Deutsche Telekom was responsible for providing data communications to electric and gas meters. A communication box was installed at each consumer's premises, which then transmitted the consumption data to Stadtwerke Emden utility over DSL. Deutsche Telekom offered a number of services in this project, such as installation and operation, analyzing consumption data and data transmission. 

\textit{Advantages: }
DSL reduces the installation cost by exploiting the existing broadband infrastructure. The main strengths of DSL technology are its wide availability (in the developed world), high bandwidth data transmission and low cost. These strengths make it suitable for electric utilities to implement SGs with smart metering and data transmission applications.

\textit{Disadvantages: }
DSL technology requires installation and maintenance of communication cables which makes it difficult to implement in rural areas due to the high cost of installation and maintenance of fixed infrastructure for low density areas. Additionally, the quality of the DSL connection depends upon the distance between the subscriber and the serving telephone exchange, which makes it difficult to characterize the performance of DSL technology. Also, due to its unreliability and down time, it is not appropriate for mission critical applications.

\paragraph{Optical Communications}
Optical communication technologies have been widely used by electric utilities in the last decade by building the communication backbone for connecting substations with control centers. Fiber optic communications play a significant role in SGs. Recently, the usage of optical fibers in SGs has been proposed to provide SG services directly to consumers \cite{maier2011fiber, jianming2011smart}. Furthermore, the Ethernet Passive Optical Network (EPON) technology enables the use of a standard Ethernet communication protocol over an optical network, and it is therefore gaining considerable attention from grid operators because of its interoperability with existing IP-based networks \cite{ancillotti2013role}.

\textit{Advantages: }
There are three main advantages of optical communication technology. Firstly, it can transmit data packets over long distances (about several kilometers) by providing up to tens of Gbps. Secondly, it is robust against radio and electromagnetic interference, which makes it a desirable candidate for high-voltage environments. Thirdly, a special type of optical cables, known as Optical Power Ground Wire (OPGW), combines the features of optical and grounding communications which allows long distance transmission at high data rates. Hence, OPGW can be used for building transmission and distribution lines \cite{ancillotti2013role}. 

\textit{Disadvantages: }
Although optical communications are very favorable for SG, they do have some limitations. The main disadvantage of optical communications is the high deployment cost of the fiber installation and the terminal units. Furthermore, they are very difficult to upgrade. Optical networks, which provide high quality wideband, are not suitable for metering applications, because metering devices are memory-constrained, requiring only narrow band communication. Therefore, these devices are not likely to be connected to broadband networks \cite{ancillotti2013role}.

\subsection{Summary and Insights}
In this section, we have discussed several IoT and non-IoT communication technologies for SG. The main motivation of discussing these communication technologies for SG is to provide some guidelines for the selection of communication technologies of SG based on the requirements. For this purpose, we have discussed which communication technologies are preferred for which scenarios, along with their characteristics, advantages and disadvantages. We came to know that most of the communication technologies are designed by focusing HAN. However, there is a new IoT communication technology, named LoRaWAN, which is a very good candidate for NAN and WAN and it is a long range and low-power communication technology. Furthermore, it is important to note that there is no overall the best technology, but certain ones are more suitable to particular SG applications than others. In general, wired technologies such as DSL, PLC and optical fiber are expensive for wide area deployments, especially in rural areas. However, they can maximize both communication capacity and security. Wireless technologies, on the other hand, can reduce installation costs but they have bandwidth and security limitations. 

\begin{figure*}[!th]
\centering
\includegraphics[width=1.0\textwidth]{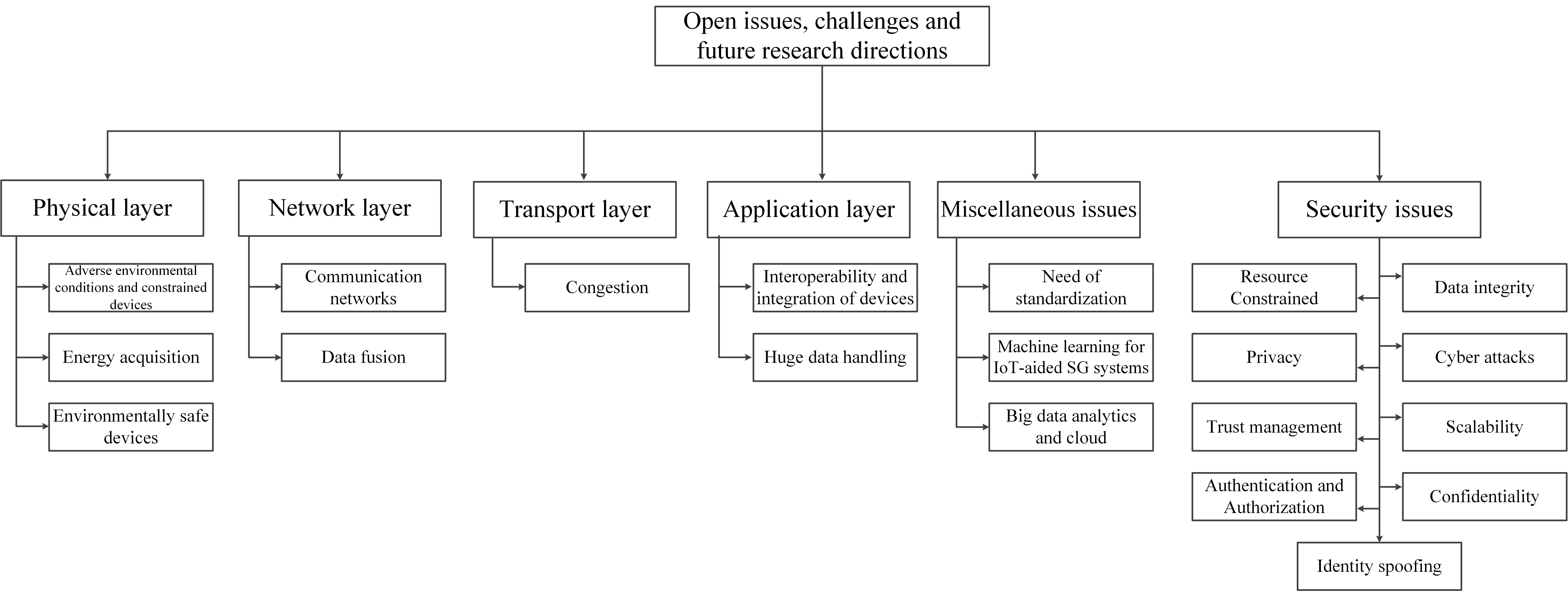}
\caption{Layered-wise classification of open issues, challenges and future directions for IoT-aided SG systems.}
\label{fig:issues_challenges_future_directions}
\end{figure*}



\section{Open Issues, Challenges, and Future Research Directions}
\label{sec:open_issues}

In this section, we highlight open issues, challenges and future research directions which are also categorized in Fig. \ref{fig:issues_challenges_future_directions}.

\subsection{Physical Layer}

\subsubsection{Adverse Environmental Conditions and Constrained Devices}
IoT-aided SG systems operate under different environments including some very severe conditions, such as the monitoring of power transmission lines. Therefore, it is important to consider requirements for reliability, availability, compatibility for hybrid communication technologies and signaling coverage at adverse environmental conditions \cite{ou2012application}. IoT solutions for self-healing and self-organization should also be considered. As an example, when a set of IoT devices fail, an alternative route should be selected by the self-healing capability, so that the reliability of IoT-aided SG systems should not be compromised. 

The limitations of IoT-aided SG systems also include the constrained devices that are used. Such devices may be ruggedized for adverse conditions, but they lack memory and processing power that limits their ability to perform local functions. The lifecycle for such devices tends to be rather long (up to 10 years), so backward compatibility remains a serious issue.

\subsubsection{Energy Acquisition}
IoT end devices and sensors operate on batteries in many applications of IoT-aided SG systems. For example, the online monitoring of power transmission lines includes various sensors, video cameras and backbone nodes installed on the transmission towers and transmission lines, which normally operate on batteries. Therefore the energy acquisition for power consumption of these devices is a serious problem in realizing the application of IoT technology in SG \cite{liu2011applications, ou2012application}. For this purpose, efficient energy storage sources for IoT devices, and energy generation devices coupled with energy harvesting using energy conversion need to be used designed \cite{shu2011research}. Already, the new generation of batteries can boast over ten years of life \cite{estimating2014steiner} at reasonable usage. However, there are still limitations on the level of power usage at the device level.

\subsubsection{Environmentally Safe Devices}
IoT devices in some SG applications are deployed outdoors and at severe electromagnetic conditions, such as power generation, transmission and distribution substations. Therefore, it is very crucial to protect the instruments and to consider and embed new technologies, such as dustproof, waterproof, anti-electromagnetic, anti-vibration, low temperature and high temperature in the manufacturing of IoT devices and their chips that will prolong their lifetime under such severe environments \cite{liu2011applications}.

\subsection{Network Layer}

\subsubsection{Communication Networks}
Information and communication networks are very important for the realization of transmission and collaboration of IoT devices in IoT-aided SG systems. The information and communication networks can be divided into two main categories based on the range of transmission \cite{ou2012application}, namely wide area and short range communication networks. In wide area communication network, the long distance information transmission is achieved through IP-based Internet, PLC, 2G/3G mobile networks, LTE and satellite networks. In short range communication network, Bluetooth, ZigBee (IEEE 802.15.4) and Ultra-Wideband (UWB) are used. As an example, the wireless sensor nodes in IoT-aided SG systems have the characteristics of low power, low rate and short distance with the limitations of storage and processing capabilities. Hence, ZigBee is an appropriate communication network in this scenario. As IoT-aided SG applications progress from merely informative facilities and appliance scheduling towards automated power management and mission-critical power supply, the reliability and speed of the communication become paramount. As Section \ref{sec:iot_communication_tech_sg} discusses several IoT communication networks, IoT-aided SG systems utilize a hybrid combination of communication networks at different stages in the same operation, a situation that is not common in other fields. This communication path goes through several `legs', from devices to local network, then to gateways, then to core servers and potentially to the Cloud too, using different protocols each time. This means that more elaborate applications of IoT-aided SG systems need to be assured of solid network support, despite fragmented communication phases, and multiple network providers. 

\subsubsection{Data Fusion}
Data fusion is the process of combining data from multiple sources. The IoT devices in IoT-aided SG systems are resource constrained and have limited battery life, processing, bandwidth and storage capabilities. Therefore, it is not efficient for IoT devices to transmit all the data to a gateway in the process of data collection, as it would consume significantly high energy and bandwidth. It is desirable to use data fusion technologies to filter and aggregate only useful data from multiple IoT devices, which will enhance the efficiency of data collection, as well as save energy and bandwidth \cite{ou2012application}. The techniques of identifying the significant data, such as smart aggregation, is a new field that will undoubtedly impact IoT-aided SG systems. 

\subsection{Transport Layer}

\subsubsection{Congestion}

Congestion causes delay and packets loss, which are the important performance parameters for SG \cite{hauser2008security}. An input from a NAN gateway can be delayed or missed by the control center due to congestion, which may affect important decision making by the control center, thus causing performance degradation or non-fulfillment of the user requirements. It may also be possible that a message could be dropped by a HAN gateway due to congestion and its memory overload if multiple messages arrive to the HAN gateway from a higher number of IoT devices simultaneously. In such scenarios, the IoT devices have to retransmit the packets on the request of the HAN gateway upon the expiration of the acknowledgement interval at HAN gateway, which contributes to higher delay. Since some SG applications are delay sensitive, it is necessary to minimize the communication delay in IoT-aided SG systems. For many IoT-aided SG systems, it is essential to accommodate a higher number of simultaneous messages from multiple IoT devices without causing a major impact on delay and packets loss. This means not only high bandwidth, but also minimizing the number of messages from multiple IoT devices to each HAN gateway, i.e., careful network design and optimal numbers of nodes and gateways \cite{fadlullah2011toward}.

\subsection{Application Layer}

\subsubsection{Interoperability and Integration of Devices}
Interoperability is defined as the ability of two or more heterogeneous networks/devices to exchange information between them, and to use the exchanged information in a common function \cite{shu2011research}. The IoT-aided SG system is comprised of a large number of different types of IoT devices and gateways which vary in characteristics, operation, resources (such as computation power, memory, energy, bandwidth, time sensitivity), as well as their implemented communication stacks and protocols (for non IP-based devices). The lack of device interoperability and integration imposes a serious constraint on the development of IoT-aided SG systems. One possible solution proposed in \cite{collier2017emerging} is to convert the networks based on proprietary protocols into IP-based networks for the realization of IoT-aided SG systems. This will enable the SG to get benefits from the seamless integration of various types of networks/devices for achieving interoperability. Moreover, future IoT devices should integrate different communication protocols and standards which operate at different frequencies and allow different architectures to communicate with other networks \cite{shu2011research}. Additionally, the interoperability issues need to be differentiated at different levels, such as in the communication layer, the physical layer or the application layer. A holistic approach to services, devices and semantics should be addressed for solving the interoperability in IoT-aided SG systems \cite{shu2011research}. 

\subsubsection{Communication Interoperability}
SG interconnects a large number of power generating sources, energy distribution networks and energy consumers. Each element of these systems require a communication medium independent of the physical medium, as well as manufacturers and the type of devices. The communication architecture of IoT-aided SG systems should have coexistence of multiple communication technologies and standards. An example of such communication architecture is presented in \cite{fan2013smart, souryal2011analysis, iyer2011performance}. For instance, for interfacing between smart meter and end devices, Ultra-wideband (UWB) or Bluetooth can be used. For interfacing smart meters in HAN, IEEE 802.11 (WiFi) and IEEE 802.15.4 (ZigBee) can be used. For interfacing between smart meters and central system, cellular wireless technologies (such as 4G, 5G, UMTS and GPRS) can be used \cite{souryal2011analysis, iyer2011performance}. The importance and need of interoperability for communication architectures, smart metering devices and systems has also been highlighted by EU M/441 standardization Mandate on smart meters \cite{m441mandate}.

\subsubsection{Cross-domain Interoperability}
IoT is comprised of various heterogeneous devices, communication protocols and applications. Therefore, interoperability is a major challenge in IoT. When IoT is integrated with SG, IoT-aided SG systems also face the challenge of interoperability, i.e., the inherent issue of IoT. SG is comprised of various systems (such as power generation, transmission, distribution and consumption) and sub-systems (such as bulk generation, renewable energy, storage and smart home). Hence, IoT-aided SG systems have to cope with two types of cross-domain interoperability issues. The first cross-domain interoperability is required between each system and sub-system of SG. The second cross-domain interoperability is required between various vertical IoT applications, e.g., the interoperability of SG with smart cities, smart homes etc., in order to fully exploit the benefit of recent developments of IoT. Therefore, the cross-domain interoperability in IoT-aided SG system requires significant investigation and development, which is one of the important future research directions. 

\subsubsection{Huge Data Handling}
The integration of IoT technology with SG comes with the cost of more frequent processing and storing the huge volume of data which would impose a higher load on the IoT communication networks. Such data includes energy consumption, consumers load demand, advanced metering records, power lines faults etc. Using high bandwidth and data rates, such as the ones offered by LTE, increases the ability to transport such data but creates bottleneck elsewhere. Consequently, the utility companies should need to design systems with enhanced capabilities to store, manage and process the collected data efficiently and effectively \cite{is2012managing}.

\subsection{Miscellaneous Issues}

\subsubsection{Need of Standardization}
Standardization is important for interoperability, compatibility, reliability and security. Although there have been separate investigations on the standardization activities of IoT \cite{palattella2013standardized, sheng2013survey} and SG \cite{bhatt2014instrumentation}, there is no standardization activity specifically for IoT-aided SG systems. However, standardization of IoT data collection under OneM2M \cite{swetina2014toward, onem2m2016} is in full swing, but the energy industry regards it as an overkill and unsuitable for constrained devices. OMA's standard for LWM2M (Lightweight M2M) \cite{oma2016} is gaining greater popularity due to its greater simplicity, which is required for such applications. 

De facto standards emerge 'organically' in the web world by wide adoption of what is usually open source or freely available software components. This may eventually occur for IoT-aided SG systems. However, the security requirements need more urgent solutions.

\subsubsection{Machine Learning for IoT-aided SG Systems}
Machine learning can help IoT-aided SG systems to learn from the past actions and improve the decision making. For example, deep learning has been applied in IoT-based electrical load prediction in SG \cite{li2017when}. Electrical load prediction is a complex problem because of variable factors, such as weather and time. The latest development in IoT and smart meters can store relevant information at a large scale. Hence, deep learning has been applied in electrical load prediction to automatically extract the information from captured data, process the extracted information and subsequently offer an estimation of future load value \cite{li2017when}. Similarly, it is a very interesting area to investigate the application of machine learning to IoT-aided SG in other systems which can significantly improve the performance.

\subsubsection{Big Data Analytics and Cloud for IoT-aided SG Systems}
In Section \ref{sec:big_data_mgmt_iot_sg}, we highlighted the need of Big Data Analytics and Cloud in IoT-aided SG systems, as well as presented the platforms and techniques. However, there is not much work done in this domain for IoT-aided SG systems, therefore it is very important to analyze whether there are any differences for Big Data and Cloud in IoT-aided SG systems as compared to the classical methods. Additionally, it is equally important to analyze the special requirements for Big data Analytics and Cloud in IoT-aided SG systems.

\subsection{Security Issues}
Although IoT technologies have been widely applied in SG, these could lead to various security vulnerabilities. Since the monitoring and controlling in IoT-aided SG systems is performed over the open Internet, it has no more than the Internet-based security, which has inferior security than the managed mobile and fixed networks. The Internet is far more vulnerable to cyber attacks. By manipulating the data either generated by smart objects or sent from the utility, an attacker can affect the real-time balance between energy production and consumption, and cause considerable financial loses to the utility and power assets \cite{bekara2014security}. For instance, very recently, the management of contingencies in smart grid through Internet of Things have been proposed \cite{tsg6}. Therefore, security considerations for IoT-aided SG systems is a high priority. Such considerations consist of several aspects which are discussed below and some calibrated security measures for centralized IoT-aided SG systems are discussed in \cite{koundinya2016calibrated}. Additionally, there are two recent survey on security and cyber threats in IoT-aided SG systems \mbox{\cite{gupta2019prevailing, de2019implementation}} which are a very good resource to be considered while dealing with security in IoT-aided SG systems.

\subsubsection{Resource Constrained}
In IoT-aided SG system, several IoT devices and smart objects are resource constrained, specially those that are deployed in great numbers. Since they have limited computational and storage capacities, they cannot run complex security algorithms. This constraint makes the application of classical security solutions (such as public key infrastructure (PKI) and public key cryptography) more challenging \cite{bekara2014security}. Hence, there is a need to take special care while developing security solutions for IoT-aided SG systems to ensure that the resource constrained IoT devices are able to accommodate the proposed security solution.

\subsubsection{Privacy}
The smart meters and the appliances in the houses could provide information much more than just energy consumption, such as the habits of consumers (wake up, sleeping, lunch and dinner timings), whether they are away from premises, or on vacation. Such information can be used for marketing (e.g., timing of adverts) but in the wrong hands, it can be used for burglaries, for example. Utilities must guarantee that such private user data should not be obtained without the users' approval and that such data would only be used for the intended purpose \cite{tsg8}. A secure and differentiated access to data for IoT-aided SG systems is proposed in \cite{spano2015last} to ensure privacy. In this design, the consumers have complete fine-grained access to their data while the distributors and energy utilities receive only aggregated statistical data and have coarse-grained access to consumers data. Addtionally, some other privacy considerations for IoT-aided SG systems are discussed in \cite{dalipi2016security}.

\subsubsection{Trust Management}
A certain level of trust establishment is required for two devices to communicate. It is easy to establish a trust relationship between devices which are owned/managed by the same entity, but not where they are managed by different entities, such as the consumers for appliances and operators for smart meters. In a large scale IoT-aided SG system, it is very challenging to establish trust between IoT devices which are owned/managed by different entities \cite{bekara2014security}.

\subsubsection{Authentication and Authorization}
The energy provider has to authenticate each smart meter in order to bill the corresponding consumer. The identity of the IoT device in an SG is also authenticated in order to avoid any misuse of the system. Only the authenticated user or device should be authorized to accomplish the tasks or is granted the required privileges to access the resources. For example, the configuration of a smart meter can only be done by the field agent, who should be authorized and granted privileges \cite{bekara2014security}.

\subsubsection{Data Integrity}
Data integrity is also very important to ensure that the received data from IoT devices (such as smart meters) cannot be modified by an unauthorized party. In IoT-aided SG systems, the IoT devices generally communicate using the open Internet. Therefore, data exchanges can be easily comprised by the attacker. Energy consumption of households can be altered by an attacker and can also modify the data exchanges in SG systems to lower the prices during the peak hours, for example. This would significantly increase the energy consumption of households (such as charging cars) instead of minimizing it which will eventually result in an overloaded power network. Therefore, data integrity in IoT-aided SG systems ensures that consumers are charged according to their exact energy consumption \cite{bekara2014security}.

\subsubsection{Cyber Attacks}
SG involves various physical objects, such as smart meters, cables and transformers that are managed by IoT. Hence, SG is vulnerable to cyber attacks which could subvert the management and cause indirect damage to these assets. Attacks can prevent smooth operations, and interfere with maintenance procedures. Attacks can also prevent the billing process to proceed, and the energy supply to be balanced. The more IoT-aided SG applications are developed, the greater the need to prevent cyber attacks. 

\subsubsection{Scalability}
The IoT-aided SG systems consist of a large number of IoT devices and smart objects which are installed over large areas which may comprise of few cities in a country. Hence, scalability is a serious challenge.

\subsubsection{Confidentiality}
SG data should be protected from prying eyes of unauthorized parties. Confidentiality means that the stored and transmitted data is only accessible to the concerned persons. As an example, the energy consumption data of consumers should not be accessible to anyone except the SG's operator and the energy providers' appropriate departments.

\subsubsection{Identity Spoofing}
In identity spoofing, an attacker takes an identity of a legitimate user/thing and uses it for communication on its behalf in an unauthorized manner. In IoT-aided SG sytems, the identity of a smart meter at someone's home can be spoofed by the attacker and in this manner, the legitimate user would be charged for energy consumption of the attacker.

\section{Conclusion}
\label{sec:conclusion}
SG is the future grid which solves the problems of uni-directional information flow, energy wastage, growing energy demand, reliability and security in the traditional power grid. The IoT technology provides connectivity anywhere and anytime. It helps SG by providing smart devices or IoT devices (such as sensors, actuators and smart meters) for the monitoring, analysis and controlling the grid, as well as connectivity, automation and tracking of such devices. This realizes the IoT-aided SG system which supports and improves various network functions at the power generation, transmission, distribution, and utilization. In this paper, we have presented a comprehensive survey on IoT-aided SG systems and we found a number of issues that needs to be addressed. Firstly, from the perspective of standardization, there are various standards for IoT and SG, however, for the IoT-aided SG system as a whole, there is a lack of standardization. Hence, it is very important to take into account the standardization of IoT-aided SG systems. Secondly, from the perspective of applications, very less work has been done on the applications of IoT-aided SG systems. In this paper, we have identified twenty-seven applications of IoT-aided SG systems, while there are only nine applications that have been developed so far. So, there is a need to consider the fast development of rest of the applications for the better realization of IoT-aided SG systems. Thirdly, from the perspective of architectures, a major focus of the existing architectures is on generic layered architecture and HAN architectures (mainly for remote controlling and managing home appliances). The layered architectures are generic architectures that do not cover many important aspects of SG, such all the networks (HAN, NAN and WAN) and systems (power generation, transmission, distribution and consumption). Therefore, there is a need of designing a new reference architecture for IoT-aided SG systems and consider these specific aspects in that architecture. Fourthly, there are not many published prototypes for IoT-aided SG systems, and the available ones are very simple. Also, there are no easily available open-source testbeds and simulation tools to enable the experimentations and performance evaluation of IoT-aided SG systems. Both these aspects of prototyping needs to be considered.

Apart from applications, architectures and prototypes, there are many other issues in IoT-aided SG systems that require serious attention and consideration, such as interoperability and integration of devices, Big Data analytics, huge data handling, security (privacy, data integrity, confidentiality, trust etc.), communication networks, congestion etc. In summary, IoT-aided SG systems is an integration of two emerging and promising worlds: IoT and SG. There is already been significant work done in the domain of IoT-aided SG systems which we presented in this paper, but there is much more yet to materialize for the better and complete realization of IoT-aided SG systems.

Table \ref{acronym} presents the list of acronyms used in this paper.

\begin{table}[!thbp]\scriptsize
\centering
\caption{List of acronyms and corresponding definitions.}
\label{acronym}
\label{tb1}
\begin{tabular}{|p{2cm}|p{5cm}|}
\hline
\bfseries Acronyms & \bfseries Definitions \\
\hline
3GPP & 3rd Generation Partnership Project \\
\hline
5G & Fifth-Generation \\
\hline
ADR & Adaptive Data Rate \\
\hline
AMI & Advanced Metering Infrastructure \\
\hline
ATM & Asynchronous Transfer Mode \\
\hline
CGI & Common Gateway Interface \\
\hline
DER & Distributed Energy Resource \\
\hline
DSL & Digital Subscriber Lines \\
\hline
EPON & Ethernet Passive Optical Network \\
\hline
FAN & Field Area Network \\
\hline
GPS & Global Positioning System \\
\hline
GSM & Global System for Mobile \\
\hline
GUI & Graphical User Interface \\
\hline
HAN & Home Area Network \\
\hline
HART & Highway Addressable Remote Transducer \\
\hline
IaaS & Infrastructure as a Service \\
\hline
IoT & Internet of Things\\
\hline
IPv6 & Internet Protocol version 6 \\
\hline
ISM & Industrial, Scientific and Medical \\
\hline
LTE & Long Term Evolution \\
\hline
MAC & Medium Access Control \\
\hline
M2M & Machine-to-Machine \\
\hline
MBWA & Mobile Broadband Wireless Access \\
\hline
NAN & Neighborhood Area Network \\
\hline
NAT & Network Address Translation \\
\hline
NIST & National Institute for Standards and Technology \\
\hline
NMEA & National Marine Electronics Association \\
\hline
LWM2M & Lightweight Machine-to-Machine \\
\hline
OPGW & Optical Power Ground Wire \\
\hline
PaaS & Platform as a Service \\
\hline
PDC & Phasor Data Concentrator \\
\hline
PLC & Power Line Communications \\
\hline
PMU & Phasor Measurement Unit \\
\hline
PKI & Public Key Infrastructure \\
\hline
PSTN & Public Switched Telephone Network \\
\hline
RAT & Radio Access Technology \\
\hline
RF & Radio Frequency \\
\hline
RFID & Radio Frequency Identification \\
\hline
SaaS & Software as a Service \\
\hline
SCADA & Supervisory Control and Data Acquisition \\
\hline
SG & Smart Grid\\
\hline
SHDSL & Single-pair High-Speed Digital Subscriber Line \\
\hline
SIM & Subscriber Identity Module \\
\hline
WPAN & Wireless Personal Area Network \\
\hline
UPnP & Universal Plug and Play \\
\hline
WiFi & Wireless Fidelity \\
\hline
WiMAX & Worldwide inter-operability for Microwave Access  \\
\hline
WMAN & Wireless Metropolitan Area Network \\
\hline
VoLTE & Voice over Long Term Evolution \\
\hline
VM & Virtual Machine \\
\hline
UAV & Unmanned Aerial Vehicles \\
\hline
UWB & Ultra-Wideband \\
\hline
V2G & Vehicle-to-Grid \\
\hline
V2H & Vehicle-to-Home \\ 
\hline
V2V & Vehicle-to-Vehicle \\
\hline
WAN & Wide Area Network \\
\hline
WLAN & Wireless Local Area Networks \\
\hline
WWW & World Wide Web \\
\hline
\end{tabular}
\end{table}

\end{document}